\documentclass[a4paper,11pt]{article}
\pdfoutput=1 % if your are submitting a pdflatex (i.e. if you have
\usepackage{jheppub} % for details on the use of the package, please
\usepackage[T1]{fontenc} % if needed

\usepackage{hyperref}

\usepackage{amsmath}
\usepackage{amssymb}
\usepackage{amsthm}

\usepackage[table]{xcolor}

\usepackage{tikz-cd}

\newcommand{\RR}{\mathbb{R}}

\newcommand{\ZZ}{\mathbb{Z}}

\newcommand{\C}{\mathcal{C}}
\newcommand{\E}{\mathcal{E}}
\newcommand{\F}{\mathcal{F}}

\newcommand{\Hh}{\mathcal{H}}
\newcommand{\I}{\mathcal{I}}

\newcommand{\M}{\mathcal{M}}
\newcommand{\N}{\mathcal{N}}
\newcommand{\Q}{\mathcal{Q}}

\newcommand{\V}{\mathcal{V}}

\DeclareMathOperator{\Tr}{Tr}

\DeclareMathOperator{\im}{Im}

\numberwithin{equation}{section}

\def\be{\begin{equation}}
\def\ee{\end{equation}}

\allowdisplaybreaks[3]

\title{\boldmath Fricke S-duality in CHL models}

\author[1]{Daniel Persson}
\author[2,3]{Roberto Volpato}
\affiliation[1]{Fundamental Physics, Chalmers University of Technology,\\
412 96, Gothenburg, Sweden}
\affiliation[2]{Theory  Group,  SLAC  National  Accelerator  Laboratory,\\ Menlo  Park,
CA 94025, USA}
\affiliation[3]{Stanford  Institute  for  Theoretical  Physics,  Department  of  Physics,\\  Stanford  University,
Stanford, CA 94305, USA}

\emailAdd{daniel.persson@chalmers.se}
\emailAdd{volpato@slac.stanford.edu}
\abstract{We consider four dimensional CHL models with sixteen spacetime supersymmetries obtained from orbifolds of type IIA superstring on K3$\times T^2$ by a $\ZZ_N$ symmetry acting (possibly) non-geometrically on K3. We show that most of these models (in particular, for geometric symmetries) are self-dual under a weak-strong duality acting on the heterotic axio-dilaton modulus $S$ by a ``Fricke involution'' $S\to -1/NS$. This is a novel symmetry of CHL models that lies outside of the standard $SL(2,\mathbb{Z})$-symmetry of the parent theory, heterotic strings on $T^6$. For self-dual models this implies that the lattice of purely electric charges is $N$-modular, i.e. isometric to its dual up to a rescaling of its quadratic form by $N$. We verify this prediction by determining the lattices of electric and magnetic charges in all relevant examples. We also calculate certain BPS-saturated couplings and verify that they are invariant under the Fricke S-duality. For CHL models that are not self-dual, the strong coupling limit is dual to type IIA compactified on $T^6/\ZZ_N$, for some $\ZZ_N$-symmetry preserving half of the spacetime supersymmetries.}

\begin{document}
%\tableofcontents
\maketitle
\flushbottom

\section{Introduction and Summary}

CHL models are orbifolds of heterotic string theory on $T^6$ preserving $\mathcal{N}=4$ supersymmetry, or, equivalently, of type II string theory on $K3\times T^2$ \cite{Chaudhuri:1995fk,Chaudhuri:1995bf,Chaudhuri:1995dj,Chaudhuri:1995ee}. These models have been a remarkable arena for testing string dualities and analyzing black hole microstates (see \cite{Dabholkar:2004yr,Dabholkar:2005by,Dabholkar:2005dt,Dabholkar:2006xa,Dabholkar:2006bj,Dabholkar:2007vk,Sen:2007qy,Cheng:2008fc,Cheng:2008kt,Govindarajan:2008vi,Banerjee:2008yu,Dabholkar:2008zy,Govindarajan:2009qt,Cheng:2010pq,Govindarajan:2010fu,Sen:2010ts,Govindarajan:2011mp,Govindarajan:2011em,Dabholkar:2012zz,Dabholkar:2012nd} for a sample of references). In the non-orbifold case the $\mathcal{N}=4$ theory has duality group $SL(2,\mathbb{Z})\times SO(6,22;\mathbb{Z})$, where the first factor is the S-duality group which acts as a strong-weak duality on the coupling $S$, while the second factor is the T-duality group. When orbifolding this theory by an automorphism of order $N$ the S-duality group is  broken to $\Gamma_1(N)\subset SL(2,\mathbb{Z})$.

In this paper we show that the symmetry group of CHL models is in fact larger: generically there is an additional strong-weak duality transformation $S\to -1/(NS)$ which has so far gone unobserved in the literature. We call this new transformation \emph{Fricke S-duality}\footnote{In mathematics, a Fricke involution is a $PSL(2,\RR)$ fractional linear transformation acting on the upper half-plane by $\tau\to -1/N\tau$.}. This is a genuinely new duality of CHL models which lies \emph{outside} of the  $SL(2,\mathbb{Z})$-symmetry of the original theory. Our analysis is very general and includes in particular  orbifolds by non-geometric symmetries of the target space. We show that Fricke S-duality has novel and highly non-trivial implications for type II-heterotic duality and electric-magnetic duality. We also verify that the counting of 1/2 BPS-states is invariant under Fricke S-duality. In the following  introduction we shall give an overview of our main results and explain in some detail their connection with black hole microstate counting and Mathieu moonshine. 

\subsection{Fricke S-duality}

Consider heterotic string theory on $T^6$ and decompose the torus according to 
$T^6=T^4 \times S^1\times \tilde{S}^1$.
 We want to take the orbifold of this by a cyclic group $G$ of symmetries that preserve all the spacetime supersymmetries. This implies, in particular, that $G$ must act trivially on the left-moving (supersymmetric) sector. From the type IIA perspective, the target space is  $K3\times S^1\times \tilde{S}^1$ and the orbifold group must be a symmetry of the $\mathcal{N}=(4,4)$ superconformal sigma model $\mathcal{C}$ on $K3$. All such symmetries were classified in \cite{K3symm} and shown to correspond to elements of the Conway group $Co_0$. The group $G$ acts by an automorphism $g\in O(\Gamma^{4,20})$ of order $N$ on the $K3$-sigma model and by a shift on the circle $S^1$.

Consider the limit where the heterotic axio-dilaton $S_{\text{het}}$ is large (i.e. $\im S_{\text{het}} \to \infty$) so that the theory is weakly coupled. On the IIA side, this means that the radii of the circles $S^1\times \tilde{S}^1$ are large (but their ratio is fixed) since string-string duality interchanges $S_{\text{het}}$ with the type IIA K\"ahler modulus $T_{\text{IIA}}$. In this limit, the IIA theory decompactifies to a six dimensional theory, with the internal superconformal theory being the K3 sigma  model $\mathcal{C}$. What is the strong-coupling limit of the theory? On the heterotic side this means $\im S_{\text{het}} \to 0$ and it  corresponds to the small-volume limit of the torus $S^1\times \tilde{S}^1$ in type IIA. For this reason it is natural to perform a double T-duality: first dualizing on $S^1$ to type IIB and then back to type IIA by T-dualizing on $\tilde{S}^1$. Let us call the dual torus $S'^1\times \tilde{S}'^1$. We show that this has the following effect on the K\"ahler modulus:

\be
T_{\text{IIA}}\longrightarrow T^{\prime}_{\text{IIA}}= -\frac{1}{N T_{\text{IIA}}}.
\label{DoubleT}
\ee
Notice the unusual factor of $N$ in the denominator which is a consequence of the fact that we are T-dualizing on the `special circle' $S^1$ upon which the orbifold group acts. The T-dual type IIA theory in the $\im T'_{\text{IIA}}\to \infty$ limit is a six-dimensional theory described by a compactification $\mathcal{N}=(4,4)$ superconformal field theory $\mathcal{C}'$, which is simply the orbifold of $\C$ by $g$. For large but finite $\im T'_{\text{IIA}}$, the theory is an orbifold of type IIA on $\C'\times {S'^1}\times{\tilde {S}'^1}$ by a group $G'$ acting by a shift along ${S'}^1$ and by an automorphism of $\C'$ (the quantum symmetry).  As discussed in \cite{Gaberdiel:2012um}, the theory $\mathcal{C}'$ can be either a  non-linear sigma model on K3, or a sigma model on $T^4$. A similar construction was considered in \cite{Vafa:1997mh}, where it was used to provide a geometrical interpretation of Montonen-Olive duality for non simply-laced groups.

\begin{figure}[t]
\begin{center}
\includegraphics[width=14cm]{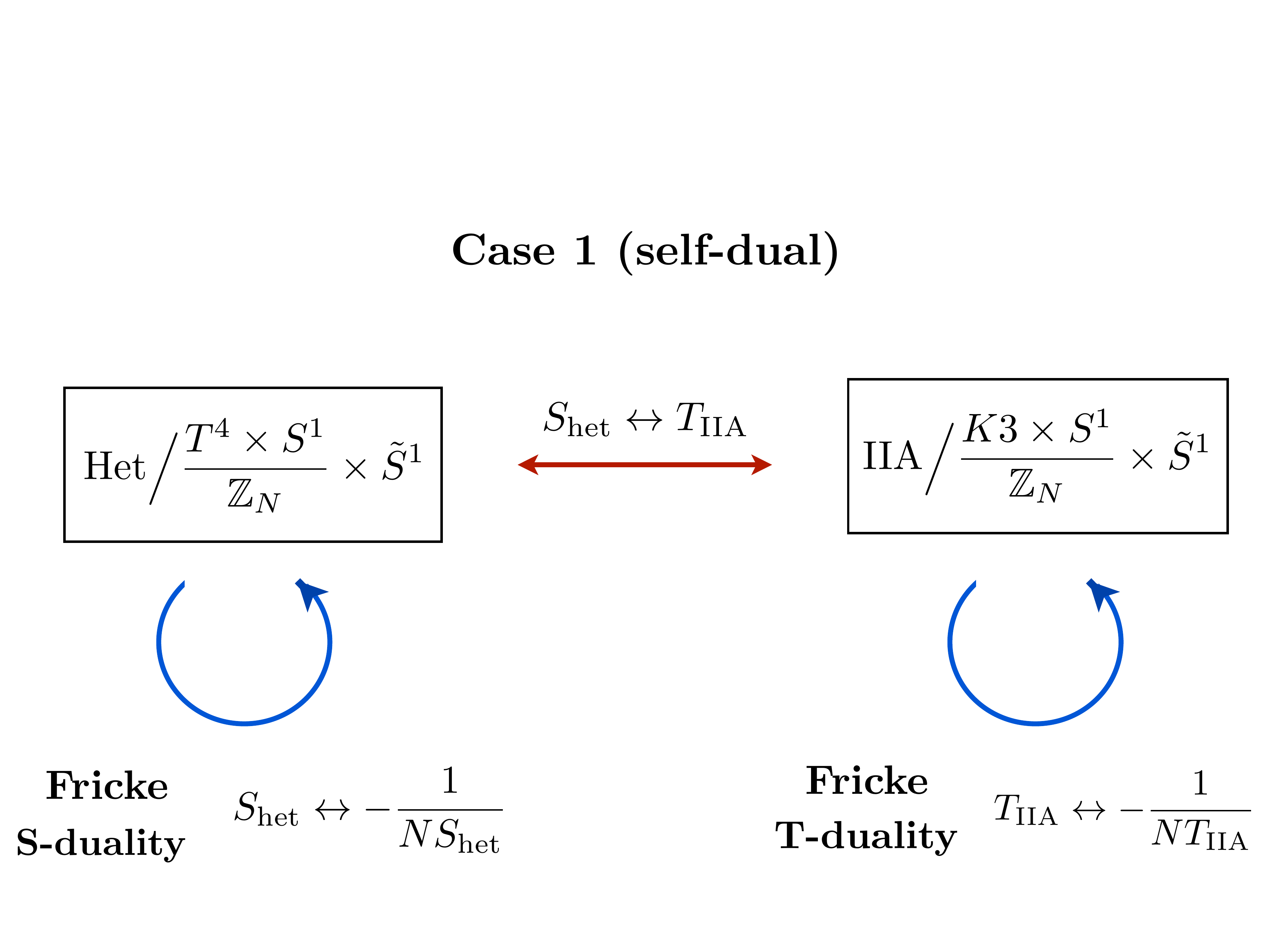}
\end{center}
\caption{\footnotesize{The  web of dualities in the case when the CHL model is self-dual under Fricke S-duality. The strongly coupled limit is given by the same CHL model at a different point in the moduli space.}}\label{fig:case1}
\end{figure}

\begin{figure}[t]
\begin{center}
\includegraphics[width=14cm]{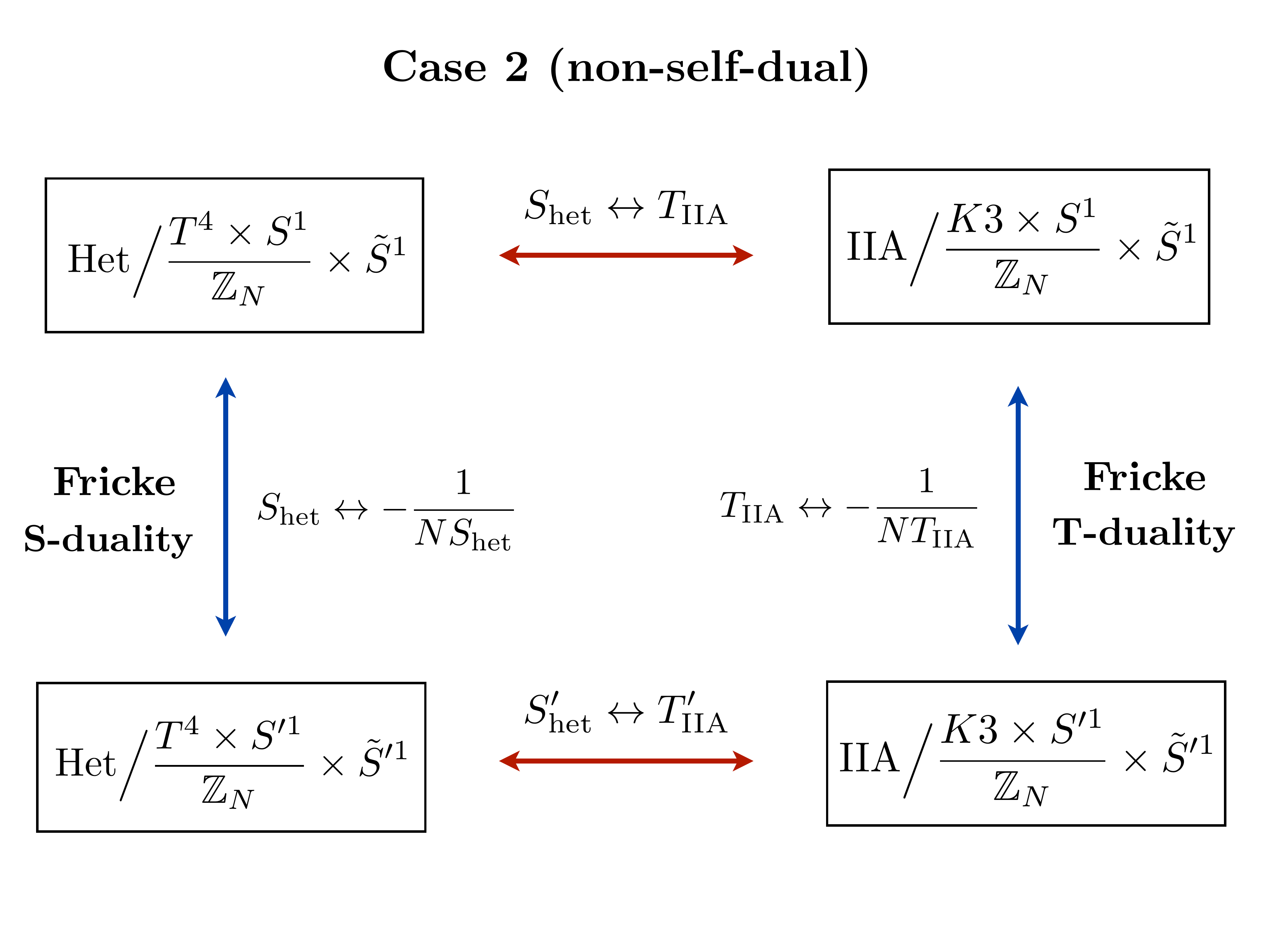}
\end{center}
\caption{\footnotesize{The  web of dualities in the case when the Fricke S-duality maps from a strongly coupled heterotic CHL model 
to the weakly coupled limit of a different model, i.e. a different connected component in the moduli space of $\N=4$ theories.} }
\label{fig:case2}
\end{figure}

Consider the case when $\mathcal{C}'$ is a K3-sigma model. We can then use string-string duality to map $T_{\text{IIA}}$ to $S_{\text{het}}$ and the map (\ref{DoubleT}) becomes 
\be
S_{\text{het}} \longrightarrow -\frac{1}{NS_{\text{het}}}.
\label{Fricke}
\ee
We call this transformation a \emph{Fricke S-duality}.  It relates the strongly coupled limit of a heterotic CHL model to the weakly coupled limit of another CHL model. The most interesting situation is when the dual models are just different points in the same connected component of the moduli space. In this case, we say that the CHL model is self-dual. We also show that  the symmetry under Fricke S-duality extends to the full Atkin-Lehner subgroup of $SL(2,\mathbb{R})$. See Figure \ref{fig:case1} for a pictorial overview of the web of dualities; the non self-dual case is depicted in Figure \ref{fig:case2}. 

This new type of S-duality yields non-trivial predictions for heterotic-type II duality. Let $\Lambda=\Lambda_e \oplus \Lambda_m$ be the electric-magnetic charge lattice of a \emph{self-dual} CHL model. In the non-orbifold case, these are simply the Narain lattices of the compactification: $\Lambda_e\cong \Lambda_m\cong \Gamma^{6,22}$ but in general they will be sublattices. Now, although Fricke S-duality relates two CHL models at different points in the moduli space, the associated lattices are moduli-independent and must therefore be isomorphic. For this to hold the lattices must be $N$\emph{-modular}, i.e. satisfy
\begin{eqnarray}
{} && \Lambda_e\cong  \Lambda_e^{*}(1/N), 
\nonumber \\ 
{} && \Lambda_m\cong \Lambda_m^{*}(N).
\end{eqnarray}
Here, $\Lambda^{*}_{e}$ and $\Lambda^{*}_{m}$ denote the dual lattices and the parentheses $(n)$ indicates that the  quadratic form of the lattice is rescaled by $n$. This is a very non-trivial 
prediction of Fricke S-duality that we verify explicitly by computing the charge lattices for several CHL models. 

A particular class of CHL models can be defined by considering a symmetry $g$ of the K3 sigma model induced a symplectic automorphism of the K3 target space. It  can be verified that all such `geometric' CHL models are self-dual. In these cases, $N$-modularity of the electric-magnetic charge lattice immediately translates into the $N$-modularity of the $g$-invariant sublattice $(H^{*}(K3,\ZZ))^g$ of the (even) integral homology of the K3 surface. While these sublattices have been studied in the mathematical literature \cite{GarbagnatiSarti2009}, their modularity properties have never been noticed before (to the best of our knowledge). It natural to wonder whether this observation admits a purely geometric interpretation, besides the present string theoretical analysis.

The third case occurs when the orbifold non-linear sigma model $\mathcal{C}'$, obtained after the double T-duality, is a torus model. In this case, the six dimensional type IIA theory compactified on $\mathcal{C}'$ has no heterotic dual. Thus, the standard heterotic-type IIA duality does not hold and the above argument cannot be used to give a weak-coupling description of the CHL model. However, there exists another theory which is dual to type IIA on $(\C'\times {S'^1}\times{\tilde {S}'^1})/G'$  under the interchange of the K\"ahler and axio-dilaton moduli: this is again an orbifold of type IIA on $T^4\times S^1\times \tilde{S}^1$, where the orbifold symmetry acts by a shift on $S^1$ and only on the right-moving oscillators of the sigma model. Thus, this breaks all spacetime supersymmetries in the right-moving sector of the type IIA superstring. Following \cite{Sen:1995ff} we call this a theory of \emph{(4,0)-type}. This should be contrasted with the previous case where the orbifold breaks half of the supersymmetries in both the right- and left-moving sector, giving rise to theories of \emph{(2,2)-type}.

\begin{figure}[t]
\begin{center}
\includegraphics[width=14cm]{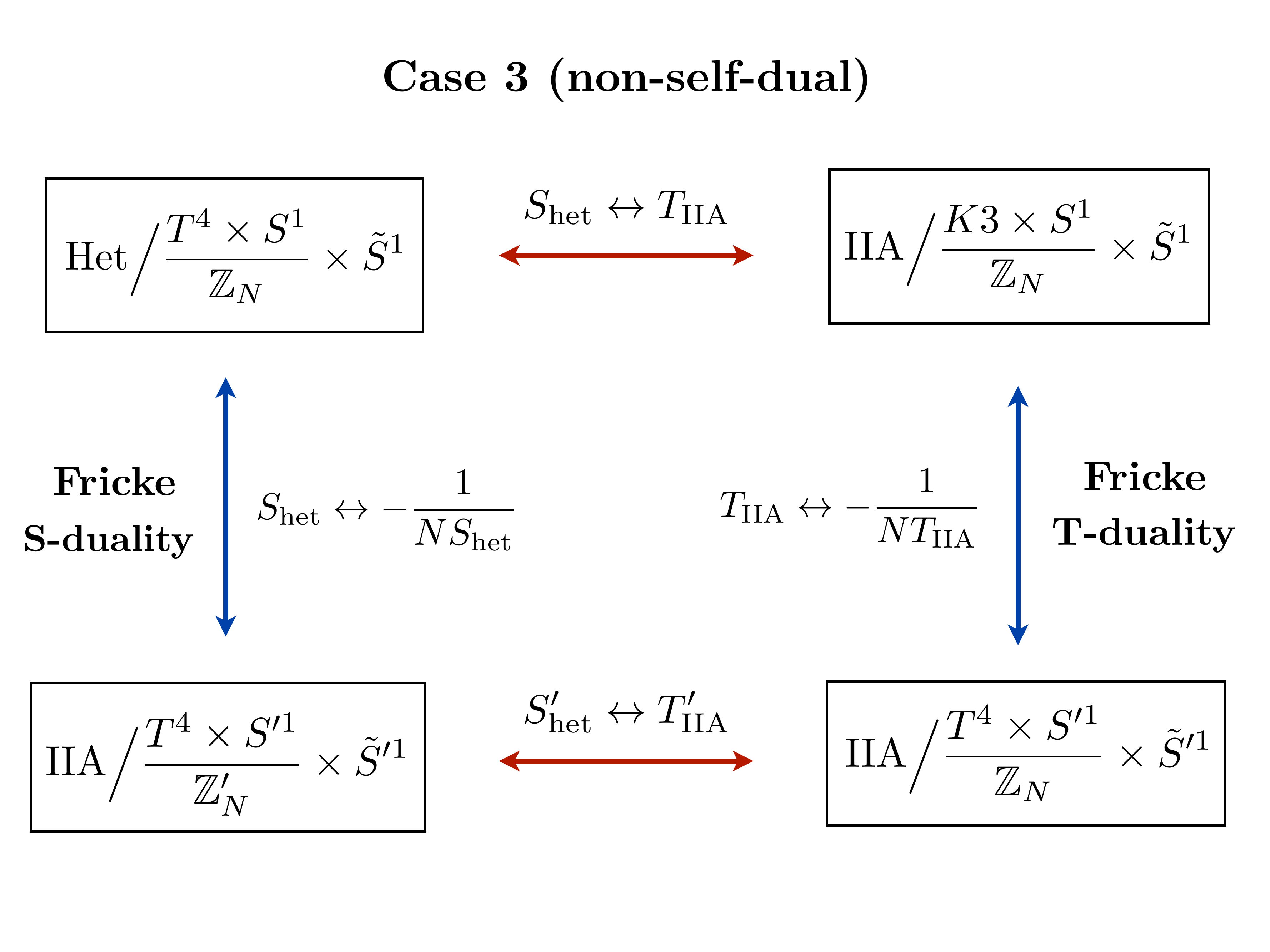}
\end{center}
\caption{\footnotesize{The web of dualities in the case when the CHL model is not self-dual under Fricke S-duality, but maps from a strongly coupled heterotic theory 
to a genuinely different, weakly coupled type IIA theory. Note in particular that the weakly coupled theory in the lower left corner is of (4,0)-type (i.e. all supersymmetries in the right-moving sector are broken) while the type IIA theory on the right hand side is   of 
(2,2)-type. } }
\label{fig:case3}
\end{figure}

To summarise, in the first case the heterotic description of the CHL model is self-dual under the Fricke S-duality, while the corresponding type IIA description is self-dual under Fricke T-duality (see Figure \ref{fig:case1}), i.e. relates different points in the same connected component of the moduli space. In the second and third case we have webs of dualities with 4 corners. In case 2, Fricke S-duality maps a strongly coupled CHL model to the weakly coupled limit of a different CHL model (see Figure \ref{fig:case2}), i.e. relates two different components of the moduli space. In case 3, on the other hand, the weakly coupled description of the heterotic CHL model is instead given in terms of a genuinely different theory, namely a type IIA torus orbifold of $(4,0)$-type (see Figure \ref{fig:case3}). 

%The topology of the target space of the orbifold theory $\C'$ can be distinguished by its Witten index
%\be
%I^{\mathcal{C'}}=\text{Tr}_{\mathcal{C}'}((-1)^{F+\tilde{F}} q^{L_0-c/24} \bar{q}^{\bar{L}_0-c/24})=\begin{cases}24 & \text{if $\C'$ is a K3 model}\\ 0 & \text{if $\C'$ is a torus model.}\ .\end{cases}
%\ee
Inequivalent CHL models can be characterised via the eigenvalues of $g\in O(\Gamma^{4,20})$ in the defining 24-dimensional representation. This fact allows to characterise the order $N$ orbifold element $g$ in terms of its \emph{generalised Frame shape}
\be
g \longleftrightarrow \prod_{a|N} a^{m(a)},
\label{frameshape}
\ee
where $m(a)$ are integers such that $\sum_{a|N} m(a) \, a=24$. For example, the identity element $e$ is represented by the Frame shape $1^{24}$ corresponding to a product of 24 identity permutations. 
The three different cases for the Fricke S-duality are easily distinguished by the Frame shape of $g$. For example, the Witten index of the non-linear sigma model $\C'$ (or equivalently, the Euler characteristic of its target space)  is given by:
\be
\I^{\C'}=\sum_{a|N} m(N/a)\, a. 
\ee
This reveals, in particular, that whenever the Frame shape is \emph{balanced}, i.e. such that 
\be
m(N/a)=m(a),
\ee
then $\mathcal{C}'$ is a K3-sigma model. This corresponds to the self-dual case, displayed in Figure \ref{fig:case1}. It turns out that it is also possible to have $\I^{\C'} =24$ even when the Frame shape is not balanced; this gives rise to case 2 (Figure \ref{fig:case2}). The third case then corresponds to $\I^{\C'} =0$ and yields the duality web in Figure \ref{fig:case3}.

\subsection{Black hole microstate counting}
The counting of BPS-states in CHL models have lead to deep insights into the quantum properties of certain black holes. In this paper we shall focus on 1/2 BPS-states which in the heterotic picture correspond to Dabholkar-Harvey states \cite{Dabholkar:1990yf}, i.e. with zero excitations in the right-moving sector but arbitrary excitations in the left-moving sector. In the type IIA picture these are electrically charged  black holes with zero classical entropy (so called `small black holes'). In the non-orbifold situation the degeneracy $\Omega(Q)$ of 1/2 BPS-states with electric charge vector $Q\in \Gamma^{6,22}$ are famously counted by the Fourier coefficients of the discriminant function 
\be
\frac{1}{\Delta(\tau)}=\frac{1}{\eta(\tau)^{24}} = \sum_{n\in \mathbb{Z}} c(n) q^{n},
\ee
via the relation:
\be
\Omega(Q) = c(Q^2/2),
\label{1/2BPS}
\ee
where $Q^2$ is the $SO(6,22;\mathbb{Z})$-invariant inner product on $\Gamma^{6,22}$. More generally, the 1/2 BPS-states in CHL models associated with orbifolds of some symmetry $g$ are captured by the \emph{fourth helicity supertrace} \cite{Kiritsis:1997hj}
\be
B^{\text{CHL}[g]}_4:= \text{Tr}_g(-1)^F J_3^4 q^{L_0} \bar{q}^{\bar{L}_0},
\ee
where $J_3$ is the third component of the massive little group in 4 dimensions. This can be viewed as a generating function of the 1/2-BPS index. 
%It is convenient to assemble these into a generating function 
%\be 
%B_4^{\text{CHL[g]}}(q,\bar{q})=\sum \Omega^{\text{CHL[g]}}_4 q^{L_0} \bar{q}^{\bar{L}_0}.
%\ee
Consider now the CHL model obtained by orbifolding type IIA string theory on $K3\times S^1\times \tilde{S}^1$ by an order $N$ symmetry $g$ with Frame shape (\ref{frameshape}) and let $(T,U)$ be the 
K\"ahler and complex structure moduli of the torus $T^2=S^1\times \tilde{S}^1$. We show that the fourth helicity supertrace of this CHL model is given by
\be
B^{\text{CHL}[g]}_{4}(\tau, T, U)=\frac{3}{2\tau_2}\sum_{n|N}m(n) \Theta_{\Gamma^{2,2}}(\tau,Tn, Un),
\ee
where $\Theta_{\Gamma^{2,2}}$ is the standard Narain theta function of the lattice. It is then interesting to ask about the role of Fricke S-duality for the counting of BPS-states. We expect that there should exist 1/2 BPS-saturated couplings in the theory which are invariant under the Fricke action (\ref{Fricke}). To this end, notice that for the unorbifolded theory, there is an $R^2$-correction to the Einstein-Hilbert action with coupling $f_{R^2}$ determined by the one-loop topological amplitude \cite{Harvey:1996ir}:
\be
F_1=\frac{2}{3} \int_{SL(2, \mathbb{Z})\backslash \mathbb{H}}d^2\tau B_4(\tau, T, U),
\ee
via the relation 
\be
\frac{\partial}{\partial T} f_{R^2}=\frac{\partial}{\partial T}  F_1, 
\ee
where $F_1$ depends on both $(T,U)$ while $f_{R^2}$ is $U$-independent. Performing the (renormalized) integral, one obtains the well-known result \cite{Harvey:1996ir}
\be
f_{R^2}(T)=-\log(T_2^{24}|\Delta(T)|^{4})+const.
\ee
Notice that the same discriminant function now appears in a \emph{spacetime} context, with argument given by the K\"ahler modulus $T$ of the spacetime torus, rather than the 
world sheet parameter $\tau$ \cite{Ooguri:2004zv,Dabholkar:2004yr,Dabholkar:2005by,Dabholkar:2005dt}. We generalise this analysis to all CHL models by calculating the topological one-loop amplitude in the orbifold theory
\be
F^{\text{CHL}[g]}_{1} =\frac{2}{3} \int_{SL(2,\mathbb{Z})\backslash \mathbb{H}}d^2\tau B^{\text{CHL}[g]}_4(\tau, T, U).
\ee
We then obtain the following result for the threshold coupling of the CHL model:
\be
f^{\text{CHL}[g]}_{R^2}(T)=-\log(T_2^{\sum_{a|N} m(a)}|\eta_g(T/N)|^4)+const,
\label{fR2CHL}
\ee
where $\eta_g$ is the eta-product associated with the Frame shape (\ref{frameshape}):
\be
\eta_g(T)=\prod_{a|N}\eta(aT)^{m(a)}.
\ee
The result (\ref{fR2CHL}) generalises earlier analyses in \cite{Antoniadis:1995zn,Dabholkar:2005dt}; for example, for the frame shapes $1^{24}, 1^82^8, 2^{12}$ and $1^{-8}2^{16}$ our formula reproduces the four lines of eq. 3.9 in \cite{Dabholkar:2005dt}. Thus, the Fourier coefficients 
\be
\frac{1}{\eta_g(T)}=\sum_{n\in \mathbb{Z}}c_g(n) e^{2\pi i n T},
\ee
 capture the 1/2 BPS-degeneracies in the CHL model:
\be
\Omega_4^{\text{CHL}[g]}(Q)=c_g(Q^2/2),
\ee
generalising the result (\ref{1/2BPS}). We show that the coupling $f^{\text{CHL}[g]}_{R^2}(T)$ is invariant under the Fricke T-duality $T\to -1/(NT)$ provided that the Frame shape is balanced. Implementing string-string duality $T\leftrightarrow S$ we find that in this case the coupling is indeed invariant under Fricke S-duality $S\to -1/(NS)$.

\subsection{Connection with  moonshine}
Monstrous moonshine relates representations of the monster group $\mathbb{M}$ with modular forms, Borcherds-Kac-Moody algebras and conformal field theory. To each element $g\in \mathbb{M}$  one associates a modular form $T_g(\tau)$ (McKay-Thompson series), whose Fourier coefficients are characters of the monster group. The basic conjecture of Conway-Norton, later proven by Borcherds, was that the McKay-Thompson series $T_g(\tau)$ should be invariant under a certain genus zero subgroup $\Gamma_g\subset SL(2,\mathbb{R})$. For some elements $g$, the invariance group $\Gamma_g$ contains a Fricke involution $\tau \to -1/(N\tau)$, analogously to our Fricke S-duality (\ref{Fricke}).

The relation with moonshine goes in fact much deeper than this simple analogy. In recent years a new moonshine phenomenon has been discovered, dubbed Mathieu moonshine, which relates the representation theory of the Mathieu group $M_{24}$ with weak Jacobi forms and superstring theory on $K3$-surfaces \cite{Eguchi:2010ej,Cheng:2010pq,Gaberdiel:2010ch,Taormina:2010pf,Gaberdiel:2010ca,Eguchi:2010fg,Gaberdiel:2011fg,Govindarajan:2011em,Taormina:2011rr,Cheng:2011ay,Eguchi:2011aj,Cheng:2012tq,Gaberdiel:2012gf,Gannon:2012ck,Gaberdiel:2013nya,Taormina:2013mda,Cheng:2013wca,Harrison:2013bya,Creutzig:2013mqa,Gaberdiel:2013psa,Persson:2013xpa}. The role of the McKay-Thompson series is here played by the so called \emph{twining genera} $\phi_g(\tau, z)$, which are weak Jacobi forms with respect to subgroups of $SL(2,\mathbb{Z})$. In our previous work \cite{Persson:2013xpa} (generalising the earlier results \cite{Cheng:2010pq,2012arXiv1208.3453R}), we defined a class of infinite-products, labelled by \emph{commuting pairs} $g,h$ of elements in $M_{24}$:
\begin{align}\label{infprod} \Phi_{g,h}(\sigma,\tau,z):= p q^{\frac{1}{N}} y\prod_{(d,m,\ell)>0}\prod_{t=0}^{M-1}(1-e^{\frac{2\pi it}{M}}q^{\frac{m}{N}}y^\ell p^d)^{\hat c_{g,h}(d,m,\ell ,t)},
\end{align}
where the first product runs over
\be d,m\in\ZZ_{\ge 0}\quad \text{and} \quad \begin{cases} \ell\in\ZZ,\ \ell<0, & \text{if }m=0=d\ ,\\
 \ell\in\ZZ, & \text{otherwise.}
\end{cases}
\ee
We proved that the functions  $\Phi_{g,h}(\sigma,\tau,z)$ are Siegel modular forms with respect to certain discrete subgroups $\Gamma_{g,h}^{(2)}\subset Sp(4;\mathbb{R})$. The automorphic properties of $\Phi_{g,h}(\sigma,\tau,z)$  imply, in particular, that they satisfy the relation: 
\be\label{Sduality} 
\Phi_{g,h}(\sigma,\tau,z)=\Phi_{g,h'}(\frac{\tau}{N},N\sigma,z)\ ,
\ee
where we note that $h$ and $h'$ may not be in the same conjugacy class of $M_{24}$. For the case $(g,h)=(e,e)$ one obtains the famous Igusa cusp form $\Phi_{10}$, whose inverse is the generating function of 1/4 BPS-states in $\mathcal{N}=4$ string theory. In this case, (\ref{Sduality}) reduces to ordinary electric-magnetic duality. It was conjectured in \cite{Persson:2013xpa} that $\Phi_{g,h}^{-1}$ similarly count twisted dyons in the CHL model obtained through orbifolding by $g$. We therefore conclude that whenever these functions have a physical interpretation, the associated CHL model should exhibit a generalization of electric-magnetic dual which acts on the vector of electric-magnetic charges $(Q,P)$ in the following way:
\be
\left(\begin{array}{c} Q \\ P \end{array}\right) \longmapsto \left(\begin{array}{c} -\frac{1}{\sqrt{N}} P \\ \phantom{-}\sqrt{N} Q \end{array}\right).
\ee
This electric-magnetic duality precisely corresponds to the $N$-modularity of the associated charge lattices. Thus, the results of the present paper show that this modular property is a consequence of the fact that CHL models satisfy Fricke S-duality.
\label{sec:moonshine}

\subsection{Outline}
In section~\ref{sec:CHL} we introduce the CHL models that we shall analyse in the paper and we discuss them from the point of view of heterotic-type II duality. We also prove consistency 
of a non-geometric CHL models and present a partial classification. In section~\ref{s:Sdual} we then prove that these CHL models are invariant under Fricke S-duality and analyse the implications for heterotic-type II duality. We also show that the Fricke S-duality in fact generalise to the full Atkin-Lehner involutions. In section~\ref{s:emlattices} we perform a detailed analysis of the electric-magnetic charge lattices for certain CHL models and verify that they satisfy the N-modularity conditions that follow from Fricke S-duality. In section~\ref{sec:BPScount} we calculate the 1/2 BPS-indices (i.e. the fourth helicity supertraces) for both the type IIA and the heterotic CHL models. In the type IIA picture we integrate this index against the fundamental domain of the modular group to obtain the one-loop topological string amplitude, from which we can extract the 1/2 BPS-saturated $R^2$-coupling and verify that this is invariant under Fricke S-duality, precisely when the CHL model is self-dual. In section~\ref{sec:conclusions} we present some conclusions and suggestions for future work. Some technical details are relegated to the appendix. In appendix~\ref{s:nongeom} we generalise our analysis to non-geometric CHL models where the level-matching condition does not hold. We verify that all of our results concerning Fricke S-duality and N-modularity also hold in this more general setting. In appendix~\ref{a:hetTdual} we discuss Fricke T-duality in heterotic CHL models, and, finally, in appendix~\ref{s:lattices} we provide tables listing the symmetries and various charge lattices in the CHL models we study. 
\section{Non-geometric CHL models}
\label{sec:CHL}

In this section we introduce the CHL models we will study in this paper. To this end we begin by recalling some standard facts about heterotic-type II dualities, after which we present the 
(non-geometric) CHL models and demonstrate their consistency. We also provide a partial classification of all such models. 

\subsection{Heterotic-type II duality}
Let us first recall some basic properties of compactification of heterotic string on $T^6$. The moduli space of such compactification is
\be O(6,22,\ZZ)\backslash O(6,22)/(O(6)\times O(22))\times SL(2,\ZZ)\backslash SL(2,\RR)/U(1)\ .
\ee The first factor is the Narain moduli space, parametrizing the metric and B-field of the torus $T^6$ as well as the Wilson lines along the internal directions, while $SL(2,\RR)/U(1)$ factor represent the heterotic axio-dilaton modulus $S_{\text{het}}$. The discrete groups $O(6,22,\ZZ)$ and $SL(2,\ZZ)$ are the T-duality and S-duality groups, respectively. The low energy effective theory is a four dimensional $N=4$ supergravity theory with gauge group $U(1)^{28}$. The lattices $\Lambda_e$ and $\Lambda_m$ of electric and magnetic charges are both isomorphic to $\Gamma^{6,22}$, the unique even unimodular lattice of signature $(6,22)$
\be \Lambda_e\cong\Lambda_m\cong \Gamma^{6,22}\ .
\ee
 The lattice $\Lambda_e$ of electric charges is identified with the Narain lattice along the internal directions, and the T-duality group is the group of automorphisms of such a lattice $O(6,22,\ZZ)\cong O(\Gamma^{6,22})$. Under the action of the S-duality group, a vector of electric-magnetic charges $\left(\begin{smallmatrix}
 Q \\ P
 \end{smallmatrix}\right)
\in \Lambda_e\oplus \Lambda_m$ transforms in the fundamental representation.
% The first factor in the moduli space is the quotient of the Grassmannian parametrizing a six dimensional positive definite subspace within $\Gamma^{6,22}\otimes\RR\cong \RR^{6,22}$, modulo automorphisms of the lattice; physically, this subspace corresponds to the space of internal bosonic oscillators in the  right-moving (supersymmetric) sector of the theory.

The heterotic model has a dual description as a type IIA compactification on $K3\times T^2$. Under this duality, the type IIA moduli $S_{\text{IIA}}$, $T_{\text{IIA}}$, $U_{\text{IIA}}$, representing, respectively, the axio-dilaton  and the K\"ahler and complex structure of $T^2$, are identified with the corresponding heterotic moduli as follows
\be S_{\text{het}}\leftrightarrow T_{\text{IIA}},\qquad T_{\text{het}}\leftrightarrow S_{\text{IIA}},\qquad U_{\text{het}}\leftrightarrow U_{\text{IIA}},
\ee
where $S_{\text{het}}$ is the heterotic axio-dilaton   and   $T_{\text{het}}$ $U_{\text{het}}$ the complex structure and K\"ahler modulus of a two dimensional torus $T^2\subset T^6$ in the heterotic description. The heterotic S-duality group $SL(2,\ZZ)$ is identified in the type IIA picture with $SL(2,\ZZ)_{T_{\text{IIA}}}$, which is part of the T-duality group $O(2,2,\ZZ)\cong SL(2,\ZZ)_{T_{\text{IIA}}}\times SL(2,\ZZ)_{U_{\text{IIA}}}$ along $T^2$.  The choice of $T^2\subset T^6$ in the heterotic picture determines a splitting of the Narain lattice as $\Gamma^{2,2}\oplus \Gamma^{4,20}$.
The lattice $\Gamma^{4,20}$ corresponds in the type IIA picture to the lattice $\Gamma_{K3}\cong \Gamma^{4,20}$ of R-R charges of D0, D2, D4-branes wrapping cycles of the K3 surface. More geometrically, $\Gamma_{K3}$ is the Mukai lattice of even integral cohomology of K3, endowed with the Mukai pairing.

\bigskip

\subsection{Generalities on CHL models}\label{s:CHLgeneral}

%\subsubsection{Basic setup}

In this paper, we discuss certain four dimensional CHL models with $16$ supersymmetries. These models are obtained by orbifolding the heterotic string on $T^6\equiv T^4\times S^1\times \tilde S^1$ (or, equivalently, of type IIA on $K3\times S^1\times \tilde S^1$) by a cyclic group of symmetries preserving all space-time supersymmetries. 

We focus on cyclic groups generated by symmetries $\hat g$ of finite order $\hat N$ of the form $\hat g=(g,\delta)$. Here, $g$ is an element of the T-duality group $g\in O(6,22,\ZZ)$ that fixes a sublattice $\Gamma^{2,2}$ of $\Gamma^{6,22}$ and preserves all space-time supersymmetries. The latter condition implies  one has to restrict to the family of heterotic models in the moduli space such that $g$ acts trivially on the left-moving (supersymmetric) sector; this family is non-empty if and only if $g$ fixes pointwise a six dimensional positive definite subspace of $\Gamma^{6,22}\otimes \RR$. We also assume that $g$ exists in some point in the moduli space where the gauge group is the generic $U(1)^{28}$. In the heterotic picture, the $g$-fixed $\Gamma^{2,2}\subset \Lambda_e$ is the Narain lattice of a torus  $T^2\subset T^6$ and $\delta\in \Gamma^{2,2}\otimes \RR$ is a (generalized) shift of finite order (modulo $\Gamma^{2,2}$) along a circle $S^1\subset T^2$.
% From a perturbative heterotic CFT point of view, this is an element of the $U(1)_L\times U(1)_R$ world-sheet symmetry group associated with the left- and right-moving $u(1)$ currents of the sigma model on $S^1$. The elements of this group are parametrized by $\delta\in (\Gamma_{S^1} \otimes\RR)/\Gamma_{S^1}$. 

In the type IIA description, $g$ is a symmetry of the non-linear sigma model on K3 that preserves the $\N=(4,4)$ superconformal algebra and the spectral flow. The latter condition ensures that all space-time supersymmetries are preserved. All such symmetries $g$ have been classified in \cite{K3symm}. They correspond to elements of $O(\Gamma^{4,20})$ that fix a four-dimensional positive definite subspace of $\Gamma^{4,20}\otimes \RR$. With each such symmetry one can associate an element in the Conway group $Co_0$, the group of automorphisms of the Leech lattice, that fix a sublattice of rank (at least) four. The conjugacy class of this element is uniquely determined by the condition that it has the same eigenvalues as $g$ in the $24$-dimensional representation.

\bigskip

%\subsubsection{Effective action and dualities}

Let us consider a heterotic CHL model given by the orbifold
\be \frac{T^4\times S^1}{\ZZ_{\hat N}}\times \tilde S^1\ .
\ee where $\ZZ_{\hat N}$ is generated by a symmetry $\hat g=(g,\delta)$ of order $\hat N$, with $g\in O(4,20)\subset O(6,22)$ of order $N$ satisfying the conditions above and $\delta\in \Gamma^{6,22}\otimes \RR$ a shift along the $S^1$ circle by $1/\hat N$ of a period. We always choose $\hat N$ to be a multiple  of $N$ (see sections \ref{s:levelmatch} and \ref{s:CHLclass}).
%The symmetry $g$ acts  on the space $O(4,20)/(O(4)\times O(20))$ of local deformations of the K3 model by a $O(4,20,\ZZ)$ transformation that preserves a subspace of $\RR^{4,20}$ of  signature $(4,d)$, for some $0\le d\le 20$. 
The orbifold projection eliminates $20-d$ $U(1)$ gauge fields from the untwisted sector and, because of the shift along the $S^1$, the twisted sectors contain no massless states. Therefore, the moduli space for this compactification is locally of the form
\be \frac{O(6,d+2)}{O(6)\times O(d+2)}\times \frac{SU(2)}{U(1)}\ ,\qquad  0\le d\le 20\ .
\ee All such models admit three complex moduli $S_{\text{het}}$, $T_{\text{het}}$ and $U_{\text{het}}$, corresponding, respectively, to the axio-dilaton, and the K\"ahler and the complex structure of $S^1\times \tilde S^1$.
The duality group $SL(2,\ZZ)\times O(\Gamma^{6,22})$ of the original (unorbifolded) toroidal theory is also broken to a subgroup.
In particular, the $SL(2,\ZZ)$ S-duality factor is broken to
\be \Gamma_1(\hat N):=\{ \left(\begin{smallmatrix}
a & b \\ c & d
\end{smallmatrix}\right)\in SL(2,\ZZ)\mid a\equiv 1, c\equiv 0\mod \hat N  \}\ .
\ee  This is clear in the type IIB picture, where $SL(2,\ZZ)$ acts on the basis of the first homology group of the torus $T^2$ and $\Gamma_1(\hat N)$ is the subgroup that preserves the shift $\delta$ along $S^1$.

\bigskip

The electric and magnetic charges of a dyonic state form a doublet of $8+d$ dimensional vectors $(Q,P)$.  The standard Dirac-Zwanziger-Schwinger-Witten quantization condition implies that the electric and magnetic charges take values in dual lattices $Q\in \Lambda_e$ and $P\in \Lambda_m$ of signature $6,2+d$
\be \Lambda_e=\Lambda_m^*\ .
\ee The lattices $\Lambda_e$ and $\Lambda_m$ for a generic orbifold symmetry $\hat g$ are described in section \ref{s:emlattices}. 
The T-duality group $O(\Gamma^{6,22})$ is broken to the subgroup $C(\hat g)$ of elements commuting with $\hat g=(\delta,g)$.

Under the action of $\Gamma_1(\hat N)\subseteq SL(2,\ZZ)$, the electric and magnetic charges and the heterotic axio-dilaton transform as
\be  S_{\text{het}}\to \frac{aS_{\text{het}}+b}{cS_{\text{het}}+d}\qquad\qquad \begin{pmatrix}
Q\\ P
\end{pmatrix}\to \begin{pmatrix}
d & -b\\ -c & a
\end{pmatrix}\begin{pmatrix}
Q\\ P
\end{pmatrix}\ .
\ee
The full duality group of a CHL model is in general larger than the direct product $\Gamma_1(\hat N)\times C(\hat g)$. Indeed, suppose that, for some $a$ coprime with $\hat N$, the symmetry $g\in O(\Gamma_{K3})\subset O(\Gamma^{6,22})$ and its power $g^a$  are conjugated 
\be g=hg^ah^{-1}\ ,
\ee
for some $h\in O(\Gamma^{6,22})$ fixing $\delta$.  Then, $(\delta,g)$ and $(a\delta,g)$  give rise to the same CHL model, since the orbifold group $\langle (\delta,g)\rangle =\langle (a\delta,g^a)\rangle$ and the group $\langle(a\delta,g)\rangle$ are related by conjugation within $O(\Gamma^{6,22})$. Thus, the duality group $\Gamma_1(N)\times O(\Lambda_e)$ can be extended by an element \be (\left(\begin{smallmatrix}
a & b \\ c & d
\end{smallmatrix}\right),h)\in SL(2,\ZZ)\times O(\Gamma^{6,22})\ ,\ee where $c\equiv 0\mod \hat N$.

\medskip

All  pairs $(g,\delta)$ are classified, up to conjugation within $O(\Gamma^{6,22})$ in section \ref{s:CHLclass}. Each inequivalent CHL model can be characterised by the eigenvalues of $g\in O(\Gamma^{4,20})$ in the defining $24$-dimensional representation. As every $g$ acts rationally in such a representation, its characteristic polynomial has the form
\be \chi_g(t):=\det\nolimits_{\bf 24}(t-g)=\prod_{a|N} (t^a-1)^{m(a)}\ ,
\ee where $N$ is the order of $g$ and $m(a)\in \ZZ$ are integers such that 
\be \sum_{a|N} am(a)=24\ .\ee When all $m(a)$ are non-negative, then $g$ acts by a permutation (in a suitable basis) with cycle shape \be\label{Frame} \prod_{a|N} a^{m(a)}\ .\ee It is useful to adopt the same formal product \eqref{Frame}  also when some of the $m(a)$ are negative; this symbol is known as (generalized) Frame shape of $g$.

\subsection{Consistency and level matching}\label{s:levelmatch}

In order for the model to be consistent, the level matching condition has to be satisfied, namely the eigenvalues of $L_0-\bar L_0$ in each $g^r$-twisted sector should take values in $\ZZ/\hat{N}\ZZ$, where $\hat N$ is the order of the symmetry. For $\hat g=(g,\delta)$, with $g\in O(\Gamma^{6,22})$ of order $N $ and $\delta\in \Gamma^{6,22}\otimes \RR$ fixed by $g$, this means 
\be\label{levelmatch} \Delta E_g+\frac{\delta^2}{2}\in \frac{1}{\hat N}\ZZ\ ,
\ee where  
$ \Delta E_g:= E_{g,L}-E_{g,R}$  is the level mismatch for the $g$-twisted ground state in the heterotic model and $\hat N$ is the least common multiple of the orders of $g$ and $\delta$. In general,
\be\label{Elev2} \Delta E_g\equiv \frac{\E_g}{N\lambda}\mod \frac{1}{N}\ZZ\ ,\ee where $\lambda$ is a divisor of $N$ and $\E_g\in \ZZ/\lambda\ZZ$ is coprime to $\lambda$. If $g$ has Frame shape $\prod_{a|N} a^{m(a)}$, then the corresponding level mismatch $\Delta E_g$ for heterotic strings on $T^6$ is easily computed to be
\be\label{Elevel} \Delta E_g\equiv \frac{\E_g}{N\lambda}\equiv  -\frac{1}{24}\sum_{a|N}m(a)(a-\frac{1}{a})\equiv \frac{1}{24}\sum_{a|N}\frac{m(a)}{a}\equiv \frac{1}{24N}\sum_{a|N} m(N/a)a\mod \frac{1}{N}\ .
\ee
%This expression corresponds to the level of a $g$-twisted ground state in a theory with $24$ free bosonic oscillators. 
In the type IIA picture, $g$ is interpreted as a symmetry of the corresponding non-linear sigma model on K3. In appendix \ref{s:genlatticeem} we will show that  string-string duality implies that the level mismatch $\Delta E_g'$ in the type IIA frame satisfies
\be \Delta E_g'\equiv  \frac{\E'_g}{N\lambda} \mod \frac{1}{N}\ZZ
\ee where $\lambda$ is the same as in the heterotic frame and $\E'_g\in \ZZ/\lambda\ZZ$ is coprime relative to $\lambda$ (but not necessarily equal to $\E_g$).
%Throughout the paper, we assume that this is true in general. A classification of all possible symmetries $g$ and the corresponding level mismatch $E_g$ is discussed in section \ref{s:CHLclass}.

\bigskip

A standard construction of the CHL model consists in choosing $\delta$ of the same order $N$ as $g$ (that is $N\delta\in \Gamma^{6,22}$), so that $\hat N=N$ \cite{Chaudhuri:1995ee}. With this choice, the level matching condition \eqref{levelmatch} requires $\delta^2\equiv -\frac{2\E_g}{N\lambda}\mod \frac{1}{N}\ZZ$, so that $\delta$ can be a null vector only when $\lambda=1$. In this paper, we choose $\delta$ to have order $N\lambda$, so that the order of $(g,\delta)$ is
\be \hat N=N\lambda\ .
\ee In this case, \eqref{levelmatch} requires $\delta^2\equiv 0\mod \frac{1}{N\lambda}\ZZ$. Since the action of the shift symmetry on the original (unorbifolded) theory is only defined by $\delta$ modulo vectors in $\Gamma^{6,22}$, we can choose $\delta$ to be null. The two constructions ($\delta$ of order $N$ or $\delta$ of order $N\lambda$) are physically equivalent, in the sense that they lead to the same CHL model at different points in the moduli space. 
However, the second choice has the advantage that a null $\delta$ admits a geometric interpretation as a shift along one of the circles of $S^1$ both in the type IIA and the heterotic frame. 

%We want to focus on symmetries $\hat g=(g,\delta)$ with left- right-symmetric $U(1)_L\times U(1)_R$ transformations, i.e. with $\delta_L=\delta_R$. These symmetries correspond to the product of a symmetry $g$ of the corresponding $\N=(4,4)$ $c=6$ superconformal field theory (a K3 model) and a shift along the circle $S^1$. In this case, \eqref{levelmatch} reduces to
%\be\label{levelmatch2} E_g\in \frac{1}{\hat N}\ZZ\ .
%\ee where $\hat N$ is the order of $\hat g$. For each given $g$, this condition can be satisfied simply by defining $\delta$ to be a $1/\hat N$ period along $S^1$, where $\hat N$ is the minimum integer such that $\hat NE_g\in \ZZ$. Notice that $\hat N$ might be strictly larger than the order $N$ of the symmetry $g$. This happens exactly when $NE_g\not\in \ZZ$, i.e. when the orbifold of the K3 model $\C$ by $g$ is not consistent, because the level-matching condition is not satisfied. Level-matching is restored in the enlarged CFT $\C=\C_{K3}\otimes \C_{S^1}$, where the action of $g$ is accompanied by a shift $\delta$ of $1/\hat N$ period along $S^1$. The relation between $\hat N$ and $N$ is, in general,
%\be \hat N=N\lambda\ ,
%\ee where $\lambda$ is a divisor of $N$. For all the symmetries $g$ that we are considering, $\lambda$ is also a divisor of $24$, which leads to various simplifications in the calculations in the next sections.
%
%

\subsection{Classification}\label{s:CHLclass}

%{\bf ??? Where should we put this? Maybe back to section2? }

We are considering four dimensional CHL models obtained by orbifolding symmetry $\hat g=(\delta,g)$ where $g\in O(\Gamma^{6,22})$ fixes a sublattice $\Gamma_{S^1}\oplus \Gamma_{\tilde S^1}\cong \Gamma^{2,2}$ and $\delta\in \Gamma_{S^1}\otimes\RR$ has finite order modulo $\Gamma_{S^1}$. Each CHL model can be labeled by the corresponding pair $(\delta,g)$. However, this labelling is highly redundant, in the sense that there are infinitely many pairs $(\delta,g)$ leading to the same CHL model, i.e. to the same connected component in the moduli space. A first step toward reducing this redundancy is to choose $\delta\in \Gamma^{6,22}$ to have order $\hat N=N\lambda$; different choices for the order lead to the same set of CHL models. As discussed in section \ref{s:CHLgeneral}, with this choice, the level-matching condition \eqref{levelmatch}
implies that $\delta^2\in \frac{1}{\hat N}\ZZ$ and, upon shifting $\delta$ by vectors in $\Gamma^{6,22}$, we can assume that $\delta^2=0$.

Whenever two pairs $\hat g=(\delta,g)$ and $\hat g'=(\delta',g')$ are related by a $O(\Gamma^{6,22})$ transformation, the corresponding CHL models belong to same component of the moduli space. An obvious necessary condition for $\hat g=(\delta,g)$ and $\hat g'=(\delta',g')$ to be conjugated is that $\delta$ and $\delta'$ have the same order $\hat N=\hat N'$ and norm $\delta^2=(\delta')^2$. Furthermore, up to a replacement of $\hat g$ by $\hat g^a$, for some $a$ coprime to $\hat N$, we can assume that $\delta$ and $\delta'$ are $1/\hat N$ fractions of some primitive null vectors $u,u'\in \Gamma^{6,22}$. It is well known that there is always a $O(\Gamma^{6,22})$ transformation that relates such $u,u'$ and therefore $\delta$ and $\delta'$. Thus, without loss of generality, if $\delta$ and $\delta'$ have the same order, we can assume that $\delta=\delta'$. It remains to classify all possible choices of $g$ and $g'$ up to conjugation within the subgroup of $O(\Gamma^{6,22})$ stabilizing $\delta=\delta'$. Since $\delta\in \Gamma_{S^1}\otimes \RR$, the stabilizer of $\delta$ contains in particular $O(\Gamma^{5,21})$, where $\Gamma^{5,21}$ is the orthogonal complement of $\Gamma_{S^1}$ in $\Gamma^{6,22}$.

We also assumed that both $g$ and $g'$ fix some sublattice of $\Gamma^{5,21}$ isomorphic to $\Gamma^{1,1}$. Thus, both $g$ and $g'$ can be interpreted as symmetries of some non-linear sigma model on K3 and, in particular, as elements of $O(\Gamma_{K3})$, with $\Gamma_{K3}\cong \Gamma^{4,20}$.  As explained in \cite{K3symm}, each symmetry $g$ of a K3 model is associated with an element (that, with abuse of language, we still denote by $g$) in the Conway group $Co_0$, the group of automorphisms of the Leech lattice $\Lambda$, fixing a sublattice of rank at least four. Let $\Lambda^g$ (respectively, $\Gamma_{K3}^g$) be the $g$-fixed sublattice within the Leech lattice $\Lambda$ (resp., within $\Gamma_{K3}$) and $\Lambda_g=(\Lambda^g)^\perp\cap \Lambda$  (respectively,  $\Gamma_{K3,g}=(\Gamma_{K3}^g)^\perp\cap \Gamma_{K3}$) its orthogonal complement. Then, the isomorphism class of the sublattice $\Lambda_g\cong \Gamma_{K3,g}$ depends only on the conjugacy class of $g$ in $Co_0$ \cite{K3symm}. Notice that two sublattices $\Lambda_g$ and $\Lambda_{g'}$ of rank at least $4$, corresponding to two different $Co_0$ classes are not isomorphic. It follows that two symmetries $g,g'\in O(\Gamma^{6,22})$ associated with distinct $Co_0$ classes cannot be related by a $O(\Gamma^{6,22})$ conjugation.

On the other hand, consider two symmetries $g$, $g'$ associated with the same conjugacy class in $Co_0$ and fixing a sublattice of rank at least four in $\Lambda$. These symmetries act on the same way on isomorphic abstract lattices $\Lambda_g\cong \Lambda_{g'}$. Now, $\Lambda_g$ can always be primitively embedded in $\Gamma_{K3}\cong \Gamma^{4,20}$. This means that $\Lambda_g$ can be primitively embedded in $\Gamma^{5,21}=\Gamma^{1,1}\oplus \Gamma_{K3}$, in such a way that its orthogonal complement contains $\Gamma^{1,1}$. By theorem 1.14.2 of \cite{Nikulin}, it follows that such a primitive embedding is unique up to conjugation in $O(\Gamma^{5,21})$. We conclude that $g$ and $g'$ associated with the same $Co_0$ class are also related by a $O(\Gamma^{5,21})$ conjugation. Notice that this in \emph{not} necessarily true if we restrict to conjugation within $O(\Gamma_{K3})$ rather than $O(\Gamma^{5,21})$: it might happen that elements in different $O(\Gamma_{K3})$ classes are associated with the same $Co_0$ class. 

To summarize, the distinct CHL models can be labeled simply by the $Co_0$ class associated with $g$. Equivalently, they can be labeled by the Frame shape of $g$, since distinct $Co_0$ classes  (at least the ones fixing a four dimensional sublattice of $\Lambda$) have different Frame shapes. The relevant Frame shapes are listed in tables \ref{t:case1}, \ref{t:case2}, \ref{t:case3} in appendix \ref{s:lattices}. The Frame shape determines both $N$ (the order of $g$) and $\hat N=N\lambda$ (via \eqref{Elevel}), and we take $\hat N\delta$ to be any primitive null vector in $\Gamma^{6,22}$ fixed by $g$. In the following sections, we will see that  this labelling is still, in some sense, redundant: there exist non-trivial dualities relating CHL models associated with different Frame shapes.

\section{Fricke S-duality and $N$-modularity}\label{s:Sdual}

The equivalence between type IIA string theory compactified on $(K3\times S^1)/\ZZ_{\hat N}\times \tilde S^1$ and heterotic string theory on $(T^4\times S^1)/\ZZ_{\hat N}\times \tilde S^1$ can be explained using the standard adiabatic argument of \cite{VafaWitten95}. We will use this argument to discuss the strong coupling limit of the heterotic CHL model. This will lead us to discover new Fricke-type S- and T-dualities. In this section we derive these new dualities and analyse their implications for heterotic-type II duality and electric-magnetic duality.

\subsection{String-string duality and Fricke S-duality}

 For definiteness, we restrict to a point in the moduli space where, in the type IIA picture, the circles $S^1$ and $\tilde S^1$ are orthogonal and the B-field and Wilson lines along $S^1$ and $\tilde S^1$ vanish. We denote by $R$ and $\tilde R$ the radii of the circles in the type IIA picture. In this section, we consider  symmetries $\hat g=(\delta,g)$ where  $g$ is a symmetry of order $N$ of the internal K3 model such that the $g$-twisted sectors satisfy the level-matching condition (i.e. $\lambda=1$ in the language of the previous section), while the case $\lambda>1$ is left to appendix \ref{s:STdualGen}. Then, $\delta$ is simply a shift of $2\pi R/N$ along the circle $S^1$. Let us first take the weak coupling limit of heterotic string theory, with all the other moduli fixed. In the type IIA picture this corresponds to the limit of large volume of $S^1\times \tilde S^1$, by the usual exchange of the axio-dilaton heterotic modulus with the type IIA K\"ahler modulus.  For very large $R$, $\tilde R$, a local observer in the type IIA picture sees a six dimensional space-time. The orbifold construction implies that the winding $w$  and momentum $m$ around $S^1$  are quantized as 
\be w=\frac{nR}{N},\qquad \qquad m=\frac{k}{R},\qquad  \qquad\qquad n,k\in\ZZ\ .
\ee  We label the states in this six-dimensional orbifold theory by
\be |w=\frac{nR}{N},m=\frac{k}{R},\chi_{n,k}\rangle\ ,
\ee where $\chi_{n,k}$ denotes a state  in the $g^n$-twisted sector of the internal K3 model $\C$, with $g$-eigenvalue $e^{\frac{2\pi i k}{N}}$.
As $R\to \infty$, local excitations carry winding number $w=0$ and the spectrum of momenta $m$ approaches a continuum. Such excitations are associated with states $\chi_{0,k}$ of the untwisted sector of the internal K3 model with any possible $g$-eigenvalue. Therefore, in the large $R$ limit, the local physics is simply described by type IIA string theory compactified on the internal K3 model $\C$. By string-string duality, the same physics can be equivalently described by heterotic string compactified on $T^4$. The orbifold construction has a `global' effect: as a local observer moves by $2\pi R/N$ around the $S^1$ direction, the states get mapped into themselves up to a $g$ transformation acting on the internal K3 model. More precisely, a shift by $2\pi R/N$ multiplies the eigenstates of momentum $m=k/R$ by $e^{\frac{2\pi ik}{N}}$. This is the same as the action of $g$ on the internal state $\chi_{0,k}$, since this is a $g$-eigenstate with the same eigenvalue. The same `global' effect occurs in the heterotic string compactified on $\frac{T^4\times S^1}{\ZZ_{\hat N}}$, and one can then argue that the two orbifold models are equivalent \cite{VafaWitten95}.

\bigskip

Now, let us consider the strongly coupled limit of the heterotic string on $\frac{T^4\times S^1}{\ZZ_{\hat N}}\times \tilde S^1$, with all the other moduli being fixed. In the type IIA picture, this means that both radii $R$ and $\tilde R$ of $S^1$ and $\tilde S^1$ become infinitesimal, with their ratio being fixed.  It is useful to consider T-duality both in the $S^1$ and $\tilde S^1$ directions, leading again to a compactification of a six dimensional type IIA theory along two circles $ {S'}^1$ and $\tilde {S'}^1$ of very large radii\footnote{Here and in the following, we set the type IIA string length to $1$}
\be\label{Tdual} R'=\frac{N}{R}\qquad \tilde R'=\frac{1}{\tilde R}\ .
\ee Notice the non-standard factor $N$ in \eqref{Tdual}: one can check that upon setting $n'=k$ and $k'=n$ the action on the Narain left- and right-moving momenta
\be (p_L,p_R)\equiv \frac{1}{\sqrt2} (m+w,m-w)=\frac{1}{\sqrt2}\bigl(\frac{k}{R}+\frac{nR}{N},\frac{k}{R}-\frac{nR}{N}\bigr) \qquad\qquad n,k\in\ZZ\ , \ee 
is the expected one
\be (p_L,p_R)\to (p'_L,p'_R)=(p_L,-p_R)\ ,
\ee
thus assuring that this is indeed a symmetry of the string theory. As in the previous case, in the limit $R'\to \infty$ a local observer will see local excitations with vanishing winding $w'=0$ and spectrum of momenta $m'=k'/{R'}$, $k'\in \ZZ$ approaching a continuum. However, the $\hat g$ symmetry acts differently in this T-dual picture. Namely, the states with $w'=0$ and $k'$ quanta of momentum are tensored with $g$-invariant states $\chi_{k',0}$ in the $g^{k'}$-twisted sector of the internal K3 model $\C$. Thus, in the large $R'$ limit, the local physics appears as type IIA compactified on an internal CFT whose spectrum consists of $g$-invariant twisted and untwisted states. But this is simply the orbifold $\C'$ of the internal K3 model by $g$. Since we assumed that $\lambda=1$, $\C'$ is again a consistent $\N=(4,4)$ superconformal theory with central charge $c=6$, and can be either a non-linear sigma model on K3 or on a torus $T^4$. As we move around the circle $S^1$ by a shift $2\pi R'/N$, a state $|w'=0,m',\chi_{k',0}\rangle$ of momentum $m'=k'/{R'}$ picks up a phase $e^{\frac{2\pi ik'}{N}}$. One can think of this transformation as the action on $\chi_{k',0}$ of the quantum symmetry $\Q$: by definition, $\Q$ is the symmetry of the internal orbifold CFT $\C'$ that multiplies each state in the $g^{k'}$-twisted sector by $e^{\frac{2\pi ik'}{N}}$. Therefore, this model is equivalently described by a type IIA theory compactified on $\frac{\C'\times {S'}^1}{\ZZ_{N}}\times \tilde {S'}^1$, where $\C'$ is the $g$-orbifold of the original K3 model and the $\ZZ_{N}$-symmetry $\hat{g}'=(\Q,\delta')$ acts by the quantum symmetry $\Q$ on $\C'$ and by a shift by $2\pi R'/N$ along ${S'}^1$. 

\bigskip

To summarize, we have shown that type IIA on $\frac{\C\times T^2}{\ZZ_N}$, where $\C$ is a non-linear sigma model on K3 and $\ZZ_N$ is generated by $(\delta,g)$, is dual to type IIA on $\frac{\C'\times T^2}{\ZZ_N}$, where $\C'$ is the orbifold of $\C$ by $g$ and $\ZZ_N$ is generated by $(\delta',\Q)$. The duality acts on the type IIA axio-dilaton and on the K\"ahler and complex structure moduli of $T^2$ by a \emph{Fricke involution}
\be\label{FrickeTdual} S_{\text{IIA}}\to S_{\text{IIA}}\qquad T_{\text{IIA}}\to -\frac{1}{NT_{\text{IIA}}}\qquad U_{\text{IIA}}\to -\frac{1}{NU_{\text{IIA}}}\ .
\ee At the special points in the moduli space where
\be T_{\text{IIA}}=i\frac{R_{\text{IIA}}\tilde R_{\text{IIA}}}{N}\qquad U_{\text{IIA}}=i\frac{R_{\text{IIA}}}{N\tilde R_{\text{IIA}}}\ ,
\ee eq.\eqref{FrickeTdual} reduces to \eqref{Tdual}. Furthermore, the moduli determining the non-linear sigma model $\C$ on K3 are mapped to the moduli determining the orbifold model $\C'$. A similar T-duality was noticed before in  \cite{Vafa:1997mh}, where ALE spaces were considered, rather than K3 surfaces.

\bigskip

If the orbifold theory $\C'$ can be interpreted again as a non-linear sigma model on K3, then this is dual to a weakly-coupled heterotic string compactified on ${T^6}/{\ZZ_{N}}$. Furthermore, if $\hat g'$ is in the same $O(6,22,\ZZ)$ conjugacy class as $\hat g$, then this duality relates two points in the moduli space of the same heterotic CHL model. 
Using the standard correspondence between type IIA and heterotic moduli, we obtain the following transformations
\be\label{hetmodSdual} T_{\text{het}}\to T_{\text{het}}\qquad S_{\text{het}}\to -\frac{1}{NS_{\text{het}}}\qquad U_{\text{het}}\to -\frac{1}{NU_{\text{het}}}\ ,
\ee
We refer to this transformation as (heterotic) \emph{Fricke S-duality}.

\bigskip

When $\C'$ is not a non-linear sigma model on K3, then the strong coupling dual of the heterotic CHL model is a genuinely different theory. More precisely, since, in this case, $\C'$ is necessarily a non-linear sigma model on a torus $T^4$, the S-dual theory is type IIA on $\frac{T^4\times {S'}^1}{\ZZ_{N}}\times \tilde {S'}^1$.  

Notice that type IIA on $\frac{T^4\times {S'}^1}{\langle\hat g' \rangle}\times \tilde {S'}^1$ has no dual heterotic description. It is natural to ask whether there exists any string theory related to this model by an exchange of the dilaton and the K\"ahler moduli $S\leftrightarrow T$. In fact, this is the case, as shown in \cite{Sen:1995ff}: the dual theory is again a compactification of type IIA on $\frac{T^4\times {S'}^1}{\langle\hat g'' \rangle}\times \tilde {S'}^1$, where $\hat g''$ acts by a shift along ${S'}^1$ as well as by a symmetry of the non-linear sigma model on $T^4$ acting only on the right-moving oscillators. The symmetry $\hat g''$ breaks all spacetime supersymmetries coming from right-moving sector of type IIA superstring, but preserves all the left-moving ones, thus giving rise to a four dimensional $\N=4$ theory. Following \cite{Sen:1995ff}, we call such a theory of $(4,0)$ type.  This should be contrasted with the properties of the $\hat g'$ symmetry, which breaks half of the spacetime supersymmetries in both the left- and the right-moving sector; therefore, type IIA on $\frac{T^4\times {S'}^1}{\langle\hat g' \rangle}\times \tilde {S'}^1$ is of $(2,2)$ type.

\bigskip

Invariance of the four-dimensional low energy equations of motion implies that the electric and magnetic charges of the two S-dual models are related by
\be\label{emcharges}  \begin{pmatrix}
Q'\\ P'
\end{pmatrix}= \begin{pmatrix}
0 & -\frac{1}{\sqrt{N}}\\ \sqrt{N} & 0
\end{pmatrix}\begin{pmatrix}
Q\\ P
\end{pmatrix}=\begin{pmatrix}
-\frac{1}{\sqrt{N}}P\\ \sqrt{N} Q
\end{pmatrix}
\ .\ee 
Let $\Lambda_e'$ and $\Lambda_m'$ denote the electric and magnetic charges in the dual theory. By \eqref{emcharges},
\be\label{duallatte}  \Lambda_e'=\Lambda_m(\frac{1}{N})\ ,
\ee and dually
\be\label{duallattm} \Lambda_m'\cong \Lambda_e(N)
\ee
 where $L(n)$ denotes a lattices $L$ with the quadratic form rescaled by $n$ (i.e., each vector is rescaled by $\sqrt{n}$). If a theory is self-dual under the Fricke S-duality, using the relation $\Lambda_e\cong \Lambda_m^*$, we obtain the
 \be \Lambda_e^*\cong \Lambda_e(N)
 \ee or, equivalently,
  \be \Lambda_m\cong \Lambda_m^*(N)\ .
 \ee
  A lattice $L$ satisfying the condition $L\cong L^*(N)$ is called $N$-modular. Although all integral lattices of dimension $2$ are $N$-modular, for higher dimensional lattices $N$-modularity is a very special property. Thus, the fact that the lattice of magnetic charges $\Lambda_m$ of a $\ZZ_N$ self-dual CHL model is an integral $N$-modular lattice is a very non-trivial prediction of S-duality. We will verify this prediction by explicitly computing the lattices $\Lambda_e$ and $\Lambda_m$ for several CHL models.
 
When a CHL model is not self-dual,  eqs.\eqref{duallatte} and \eqref{duallattm} 
imply
\be \Lambda_e'= \Lambda_m(\frac{1}{N})\qquad \Lambda_m'= \Lambda_e(N)\ ,
\ee but we cannot argue that $\Lambda_e'\cong \Lambda_e$ and $\Lambda_m'\cong \Lambda_e$, and hence $\Lambda_e$ and $\Lambda_m$ need not be $N$-modular in this case. We will verify that, in all cases where $\C'$ is a torus model, $N$-modularity does not hold.

\subsection{Self-duality and the Witten index}
\label{s:quantsymm}

From the above discussion, it is clearly important to determine which CHL models are self-dual under Fricke S-duality. This can be determined simply in terms of the Frame shape $\prod_{a|N} a^{m(a)}$ of $g$, by analyzing the associated Witten index as we now show.

Let $\C$ be a non-linear sigma model on K3 and $g$ be a symmetry of $\C$  of order $N$ preserving the $\N=(4,4)$ superconformal algebra. We will assume that the orbifold of $\C$ by $g$ is a consistent CFT $\C'$; in our previous notation, this means that $\lambda=1$ and $\hat N=N$ (the case $\lambda>1$ is treated in appendix \ref{a:quantsymm}). It follows that the orbifold CFT $\C'$ is either a non-linear sigma model on K3 or on a four dimensional torus $T^4$. Furthermore, $\C'$ has a symmetry $\Q$ of order $N$ (the quantum symmetry) that acts by multiplication by $e^{\frac{2\pi i r}{N}}$ on the $g^r$-twisted states.  

Let $\prod_{a|N} a^{m(a)}$, $m(a)\in \ZZ$, be the cycle shape associated to $g$. By \eqref{Elevel}, the condition that the orbifold is consistent, i.e $\lambda=1$, is equivalent to
\be\label{consorbif} \sum_{a|N}m(N/a)a\in 24\ZZ\ .\ee The cycle shape determines the eigenvalues of $g$ on the $24$ dimensional space Ramond-Ramond ground states of $\C$. In this section, we will prove that the cycle shape describing the action of $\Q$ on the RR ground states of the orbifold theory $\C'$ is $\prod_{a|N} a^{m(N/a)}$. This statement needs a refinement when $\C'$ is a torus model; we will be more precise below.

In general, the topology of the target space of a $\N=(4,4)$ superconformal field theory $\C$ with central charge $c=6$ can be distinguished by the Witten index\footnote{Of course, the Witten index can be defined even for theories with a smaller superconformal algebra, in particular for $\N=(2,2)$ theories.}
\be \I^\C(\tau.\bar\tau)=\Tr_{RR}((-1)^{F+\tilde F}q^{L_0-\frac{c}{24}}\bar q^{\bar L_0-\frac{c}{24}})=\begin{cases}24 & \text{if $\C$ is a K3 model}\\ 0 & \text{if $\C$ is a torus model}\ ,\end{cases}
\ee where $q=e^{2\pi i\tau}$. The Witten index is the specialization to $z=0$ of the elliptic genus $\phi(\tau,z)$. By standard arguments, the only non-vanishing contribution to this trace comes from the RR ground states, i.e. states with conformal weights $(h,\bar h)=(\frac{c}{24},\frac{\bar c}{24})=(\frac{1}{4},\frac{1}{4})$, so that the Witten index is actually a constant and it equals the Euler characteristic of the target space.

If the theory has a symmetry $g$ one can consider a $g$-twined version of the Witten index (the specialization to $z=0$ of the twining elliptic genus $\phi_g(\tau,z)$)
\be \I^\C_g:=\Tr_{RR}(g(-1)^{F+\tilde F}q^{L_0-\frac{c}{24}}\bar q^{\bar L_0-\frac{c}{24}})\ ,
\ee which is again a constant. In the case we are considering, where $\C$ is a K3 model and $g$ has cycle shape $\prod_{a|N} a^{m(a)}$, the twining index $\I^\C_g$ is simply the trace of $g$ on its $24$-dimensional representation (the RR ground states in K3 models have all positive fermion number $(-1)^{F+\tilde F}$), so that
\be\label{firstIg} \I^\C_{g}=\sum_{a|N} m(a) \sum_{k=1}^a e^{\frac{2\pi i k}{a}}=m(1)\ ,\ee which is just the dimension of the fixed subspace of $g$. More generally, for any power $g^j$ we have
%It is easy to check that,  for a cycle shape $\prod_{a|N}a^m(a)$, the multiplicity of a given eigenvalue $e^{\frac{2\pi i x}{N}}$, $x=1,\ldots, N$, is 
%\be \sum_{y|(x,N)} m(N/y)\ .
%\ee In fact, each cycle of length $a$ contributes to this multiplicity if and only if $y=N/a$ is a divisor of $x$, so that $e^{\frac{2\pi i x}{N}}=e^{\frac{2\pi i x/y}{a}}$. Thus, we obtain the formula
\be\label{secondIg} \I^\C_{g^j}=
%\sum_{x=1}^N e^{\frac{2\pi i xj}{N}}\sum_{y|(x,N)} m(N/y)=
\sum_{u|(j,N)} m(u)u\ ,
\ee where the right hand side is the dimension of the fixed subspace of $g^j$. 
%equality follows from \eqref{firstIg} by noticing that the cycle shape of $g^j$ has the form $1^{\sum_{u|(j,N)} m(u)u}\cdot\ldots$. 
Notice that, once the twined indices $\I^\C_{g^j}$ are known for all $j=1,\ldots,N$, using \eqref{secondIg} one can reconstruct the cycle shape of $g$.

\medskip

Similarly, one can consider the twined Witten indices $\I^{\C'}_{\Q^k}$, $k=1,\ldots,N$, associated with the action of the quantum symmetry $\Q$ on the RR ground states of the orbifold theory $\C'$.  Using standard orbifold techniques, this index can be computed  in terms of the Frame shape of $g$ (see appendix \ref{a:quantsymm})
\be\label{Qindex} \I^{\C'}_{\Q^k}=\sum_{n|(k,N)}m(N/n)n\ .
\ee
By specializing to the case $k=N$, we obtain the Witten index of the orbifold theory $\C'$
\be \I^{\C'}=\sum_{n|N}m(N/n)n\ .
\ee Thus, the condition \eqref{consorbif} of consistency of the orbifold $\C'$ implies that the elliptic genus of the orbifold theory is a multiple of $24$.  Now, we see that, if \eqref{consorbif} is satisfied, then the only two possibilities are $\I^{\C'}=0$ for a torus model and $\I^{\C'}=24$ for a K3 model. This can be also verified a posteriori for all possible cycle shapes of Conway elements.

If $\I^{\C'}=24$, then $\C'$ is a K3 model, and by comparing this formula with \eqref{secondIg}, we conclude that $\Q$ must correspond to a cycle shape $\prod_{a|N}a^{m(N/a)}$. In particular, $\Q$ has the same cycle shape as $g$ if and only if the cycle shape is `balanced' with balancing number $N$, i.e.
\be\label{balance} m(a)=m(N/a)\ .
\ee
If \eqref{balance} holds, then \eqref{consorbif} is automatically satisfied and $\C$ is a K3 model.

When $\C'$ is a torus model, i.e. $\I^{\C'}=0$, the cycle shape $\prod_{a|N}a^{m(N/a)}$ determines the eigenvalues of $\Q(-1)^{F+\tilde F}$ on the RR ground states. Notice that, in this case, the fermion number has to be taken into account, as it acts non-trivially on the RR ground states.

%To conclude, we have shown that the Witten index of the orbifold theory $\C'$ is given by
%%In appendix \ref{a:quantsymm}, we show that the Witten index of the orbifold theory $\C'$ is given by
%\be \sum_{a|N}m(N/a)a
%\ee and the Frame shape of the quantum symmetry is $\prod_{a|N}a^{m(N/a)}$. 

To summarise, the above analysis shows that there are three possible cases to take into account:
\begin{itemize}
\item[\bf \emph{Case 1:}] If the Frame shape is \emph{balanced} with balancing number $N$, i.e. if
\be m(N/a)=a\ ,
\ee for all $a|N$, then 
\be \sum_{a|N}m(N/a)a=24\ ,\ee
the orbifold $\C'$ is again a K3 model and the quantum symmetry $\Q$ has the same Frame shape as $g$. Thus, in this case, the CHL model is self-dual and the lattice $\Lambda_m$ is $N$-modular.
\item[\bf \emph{Case 2:}] If the Frame shape is not balanced, but
\be \sum_{a|N}m(N/a)a=24\ ,
\ee then the orbifold $\C'$ is again a K3 model, but Frame shape $\prod_{a|n}a^{m(N/a)}$ of the quantum symmetry $\Q$ is different from $g$. Thus, in this case, S-duality relates two different CHL models. In particular, if $\Lambda_m$ and $\Lambda_m'$ are the lattices of magnetic charges in these two models, we have the relation
\be \Lambda_m^*\cong\Lambda'_m(N)\ .
\ee
\item[\bf \emph{Case 3:}] If
\be \sum_{a|N}m(N/a)a=0\ ,
\ee then the orbifold $\C'$ is a torus model. Thus, the CHL model is dual to type IIA on $(T^4\times S')/\ZZ_N\times \tilde S'$.
\end{itemize}

\subsection{Heterotic T-duality}\label{s:hetTdual}

The analysis of T-duality in the type IIA frame of CHL models can be repeated without essential modifications to describe T-duality in the heterotic frame (equivalently, S-duality in the type IIA frame). Under a change in the moduli
\be T_{\text{het}}\to -\frac{1}{NT_{\text{het}}}\ ,\qquad U_{\text{het}}\to -\frac{1}{NU_{\text{het}}}\ , \qquad S_{\text{het}}\to S_{\text{het}}\ ,
\ee the heterotic CHL model is mapped to another four dimensional $\N=4$ string model. 

In analogy with the type IIA T-duality, one can argue that the six dimensional decompactification limit of the T-dual heterotic model is simply the orbifold of heterotic string on $T^4$ by the symmetry $g\in O(\Gamma_{K3})$. In appendix \ref{a:hetTdual}, we show that, in \emph{all} cases, this is simply the same heterotic CHL model we started from, possibly at a different point in the moduli space. Thus, all CHL models are self-dual under heterotic T-duality.

\subsection{Atkin-Lehner dualities}
\label{s:AtkinLehner}

The Fricke S-duality $S\to -1/NS$ can be generalized to dualities acting on the heterotic axio-dilaton by Atkin-Lehner involutions. Let $e$ be an exact divisor of $N$ (denoted by $e\| N$), i.e $e| N$ and $\gcd(e,N/e)=1$.  An Atkin-Lehner involution is an element of $SL(2,\RR)$ of the form\footnote{The normalization factor $\frac{1}{\sqrt{e}}$ is sometimes omitted in the definition, since it is irrelevant in the action of $W_e$ as a fractional linear transformation on the upper half plane.}
\be\label{ALdef} W_e=\frac{1}{\sqrt{e}}\begin{pmatrix}
ae & b\\ Nc & de
\end{pmatrix}\qquad \qquad a,b,c,d,e,N\in \ZZ, \quad ade^2-Nbc=e,\quad e\|N\ .
\ee Atkin-Lehner involutions are elements of the normalizer of $\Gamma_0(N)$ within $SL(2,\RR)$ and they have order $2$ modulo $\Gamma_0(N)$, i.e. $W_e^2\in \Gamma_0(N)$. For $e=N$, one recovers the Fricke involution $W_N=\frac{1}{\sqrt{N}}\left(\begin{smallmatrix}
0 & -1\\ N & 0
\end{smallmatrix}\right)$.
Let us focus on a CHL model in the type IIA frame (similar arguments extend to the heterotic frame). One can easily verify that, at least in the perturbative limit, the model is invariant under the transformation
\be\label{ALmoduli} T_{\text{IIA}}\to T'_{\text{IIA}}=\frac{ae T_{\text{IIA}}+b}{NcT_{\text{IIA}}+de}\qquad U_{\text{IIA}}\to U'_{\text{IIA}}=\frac{ae U_{\text{IIA}}+b}{NcU_{\text{IIA}}+de}\qquad S_{\text{IIA}}\to S'_{\text{IIA}}=S_{\text{IIA}}
\ee
and 
\be\label{ALtransf} \begin{pmatrix}
\tilde n & n/N\\ -k & \tilde k
\end{pmatrix}\to \begin{pmatrix}
\tilde n' & n'/N\\ -k' & \tilde k'
\end{pmatrix}=\frac{1}{e} \begin{pmatrix}
de & -c\\ -bN & ae
\end{pmatrix}\begin{pmatrix}
\tilde n & n/N\\ -k & \tilde k
\end{pmatrix}\begin{pmatrix}
de & -b\\ -Nc & ae
\end{pmatrix}\ ,
\ee where $n,k\in \ZZ$  (respectively, $\tilde n,\tilde k\in \ZZ$) are the units of winding and momentum along $S^1$ (resp., $\tilde S^1$).
In particular, \eqref{ALtransf} is an automorphism of the lattice of winding-momenta and, for each charge in such a lattice, the corresponding left- and right-moving conformal weights $\frac{1}{2}(p_L^2+\tilde p_L^2)$ and $\frac{1}{2}(p_R^2+\tilde p_R^2)$ are invariant under \eqref{ALmoduli} and \eqref{ALtransf}. We will refer to this kind of transformations as a type IIA Atkin-Lehner T-duality.

\medskip

Let us consider the limit $T'_{\text{IIA}}\to \infty$, where the dual theory  decompactifies to a six dimensional model. In this limit, the only states with finite mass are the ones with zero winding $n',\tilde n'$ around the dual torus $S'^1\times \tilde S'^1$. This condition is equivalent to
\be\label{zerowind}
%\begin{cases}w'=0\\ \tilde w'=0
%\end{cases}\Leftrightarrow\quad
\begin{cases} \tilde n'= c^2 \frac{N}{e}\tilde k+c d k-c d n+d^2 e \tilde n=0\\
 n'= -a c N \tilde k+a d e n-b c \frac{ N}{e}k-b d N \tilde n=0
\end{cases}\Leftrightarrow\quad
\begin{cases}ck=-de\tilde n\\ dn=c\frac{N}{e}\tilde k
\end{cases}\ .
\ee By the condition $ade^2-Nbc=e$ in the definition \eqref{ALdef} it follows that $\gcd(c,e)=1$ and $\gcd(d,N/e)=1$, so that \eqref{zerowind} implies
\be k\in e\ZZ\qquad\qquad n\in \frac{N}{e}\ZZ\ .
\ee
Therefore, in the $T'_{\text{IIA}}\to \infty$ limit, the theory contain only states in the $(g^{N/e})^r$-twisted sectors, $r=1,\ldots, e$ and that are invariant under $g^{N/e}$. Thus, the model is a compactification of type IIA on the orbifold CFT
\be \C'=\C/\langle g^{N/e}\rangle\ .
\ee For large but finite $T'$, as a local observer moves around the circle $S'$ by $1/N$-th of a period, a state with  $k'$ units of momentum picks a phase
\be e^{\frac{2\pi i k'}{N}}=e^{\frac{2\pi i}{N}(a c N \tilde k+a d e k-\frac{N}{e}bc n+b d N \tilde n)}=e^{\frac{2\pi i}{N}(ade k-\frac{N}{e}bc n)}\ .
\ee This phase corresponds to the action of $g':=g^{ade}\Q^{-\frac{N}{e}bc}$ on the states of $\C'=\C/\langle g^{N/e}\rangle$, where $\Q$ is the quantum symmetry acting by a phase on the twisted sectors of the orbifold CFT $\C'$ and $g$ (with slight abuse of notation) is the symmetry of $\C'$ induced by the symmetry $g$ of $\C$. Notice that $\Q$ has order $e$ and $g$ has order $N/e$ in $\C'$, so that, by the condition $ade-\frac{N}{e}bc=1$, we obtain \be g^{ade}=g^{1+\frac{N}{e}bc}=g\qquad\qquad \Q^{-\frac{N}{e}bc}=\Q^{1-ade}=\Q\ .\ee Furthermore, since $\gcd(e,N/e)=1$, it follows that $g'=g\Q$ has order $N$. We conclude that the T-dual theory is type IIA compactified on the orbifold
\be \frac{\C'\times S'}{\langle (\delta',g')\rangle}\times \tilde S'\qquad \text{with } \C'=\C/\langle g^{N/e}\rangle,\ g'=g\Q\ ,
\ee and $\delta'$ is a shift of order $N$ along $S'$. For $e=N$, i.e. the Fricke involution, the induced $g$ has order $1$ in $\C'=\C/\langle g\rangle$, so that $g'=\Q$ is simply the quantum symmetry.

\medskip

A cumbersome calculation along the lines of \ref{a:quantsymm} shows that if $g$ has Frame shape $\prod_{a|N}a^{m(a)}$, then $g'$ has Frame shape
\be \prod_{a|N} a^{m(\frac{ea}{\gcd(a,e)^2})}\ .
\ee This formula reduces to $\prod_{a|N} a^{m(\frac{N}{a})}$ for the Fricke involution $W_N$. A case by case analysis shows that whenever $g$ has a balanced Frame shape (case 1), then $\C'$ is a K3 model and $g'$ has the same Frame shape as $g$. Therefore, the CHL model in case 1 are self-dual under \emph{all} Atkin-Lehner involutions. On the contrary, the Atkin-Lehner involutions relate the CHL models in case $3$ to the ones in case 2, according to the following duality diagrams:
$$ \begin{tikzcd}
K3(1^{-4} 2^5 3^4 6^1) \color{red}{\arrow[leftrightarrow,color=blue]{r}{W_6}}\arrow[leftrightarrow,color=blue]{d}{W_2}\arrow[leftrightarrow,color=blue]{dr}[description]{W_3} & %\frac{T^4\times S^1}{\ZZ_6}\times \tilde S^1
T^4(1^1 2^4 3^5 6^{-4})\\[20pt] 
K3(1^5 2^{-4} 3^1 6^4)\arrow[leftrightarrow,color=blue]{r}{W_6}\arrow[leftrightarrow,crossing over,color=blue]{ru}[description]{W_3} & K3(1^4 2^1 3^{-4} 6^5) \arrow[leftrightarrow,color=blue]{u}{W_2}
\end{tikzcd}
%\ee
%\be 
\qquad \qquad\begin{tikzcd}
K3(1^{-2} 2^3 5^2 10^1) \arrow[leftrightarrow,color=blue]{r}{W_{10}}\arrow[leftrightarrow,color=blue]{d}{W_2}\arrow[leftrightarrow,color=blue]{dr}[description]{W_5} & 
%\frac{T^4\times S^1}{\ZZ_{10}}\times \tilde S^1
T^4(1^1 2^2 5^3 10^{-2})\\[20pt] 
K3(1^3 2^{-2} 5^1 10^2)\arrow[leftrightarrow,color=blue]{r}{W_{10}}\arrow[leftrightarrow,crossing over,color=blue]{ru}[description]{W_5} & K3(1^2 2^1 5^{-2} 10^3) \arrow[leftrightarrow,color=blue]{u}{W_2}
\end{tikzcd}
$$
$$ \begin{tikzcd}
K3(1^{-2} 2^2 3^2 4^1 12^1) \arrow[leftrightarrow,color=blue]{r}{W_{12}}\arrow[leftrightarrow,color=blue]{d}{W_4}\arrow[leftrightarrow,color=blue]{dr}[description]{W_3} & 
%\frac{T^4\times S^1}{\ZZ_{12}}\times \tilde S^1
T^4(1^1  3^1 4^2 6^2 12^{-2})\\[20pt] 
K3(1^1 2^2 3^1 4^{-2} 12^2)\arrow[leftrightarrow,color=blue]{r}{W_{12}}\arrow[leftrightarrow,crossing over,color=blue]{ru}[description]{W_3} & K3(1^2 3^{-2} 4^1 6^2 12^1) \arrow[leftrightarrow,color=blue]{u}{W_4}
\end{tikzcd}
$$
$$ \begin{tikzcd}
K3(1^{-2} 2^4 3^{-2} 6^4) \arrow[leftrightarrow,color=blue]{r}{W_2,W_{6}}\arrow[leftrightarrow,loop left,color=blue]{}{W_3} & 
%\frac{T^4\times S^1}{\ZZ_{12}}\times \tilde S^1
T^4(1^4 2^{-2} 3^4 6^{-2})\arrow[leftrightarrow,loop right,color=blue]{}{W_3}
\end{tikzcd}
$$
$$ \begin{tikzcd}
K3(1^{-1} 2^{-1} 3^3 6^3) \arrow[leftrightarrow,color=blue]{r}{W_3,W_{6}}\arrow[leftrightarrow,loop left,color=blue]{}{W_2} & 
%\frac{T^4\times S^1}{\ZZ_{12}}\times \tilde S^1
T^4(1^3 2^3 3^{-1} 6^{-1})\arrow[leftrightarrow,loop right,color=blue]{}{W_2}
\end{tikzcd}
$$
In these diagrams, the notation $X(\prod_{a|N} a^{m(a)})$, with $X\in \{K3,T^4\}$, denotes a type IIA model on $\frac{X\times T^2}{\ZZ_N}$, where $\ZZ_n$ is generated by $(\delta,g)$ and $g$ has Frame shape $\prod_{a|N} a^{m(a)}$.

\medskip

Under string-string duality, these Atkin-Lehner involutions are mapped to S-duality transformations in the heterotic frame
\be S_{\text{het}}\to S'_{\text{het}}=\frac{ae S_{\text{het}}+b}{NcS_{\text{het}}+de}\qquad U_{\text{het}}\to U'_{\text{het}}=\frac{ae U_{\text{het}}+b}{NcU_{\text{het}}+de}\qquad T_{\text{het}}\to T_{\text{het}}'=T_{\text{het}}\ .
\ee Each node in the diagrams above admits a heterotic dual (or, more generally, a $(4,0)$ dual) and these heterotic (or $(4,0)$) models are related to one each other by a network of Atkin-Lehner S-dualities.

The Atkin-Lehner S-duality $W_e$ acts on the electric-magnetic charges by
\be \begin{pmatrix}
Q \\ P
\end{pmatrix}\to  \begin{pmatrix}
Q' \\ P'
\end{pmatrix}=\frac{1}{\sqrt{e}}\begin{pmatrix}
de & -b\\ -Nc & ae
\end{pmatrix} \begin{pmatrix}
Q \\ P
\end{pmatrix}\ .
\ee  As a consequence, the electric and magnetic lattices $\Lambda_e,\Lambda_m$ and $\Lambda_e',\Lambda_m'$ of two CHL models related by $W_e$ must be related as follows
\be \Lambda_e'\cong {\rm span}_\ZZ \Bigl(\sqrt{e}\Lambda_e\cup \frac{1}{\sqrt{e}}\Lambda_m\Bigr)\subset \Lambda_m\otimes \RR
\ee and
\be \Lambda_m'\cong {\rm span}_\ZZ \Bigl(\frac{N}{\sqrt{e}}\Lambda_e\cup \sqrt{e}\Lambda_m\Bigr)\subset \Lambda_m\otimes \RR\ .
\ee 
We have verified these relations in several examples.

\medskip

Finally, we note that an analogous Atkin-Lehner involution is expected to act as T-duality in the heterotic frame or, equivalently, as S-duality in the type IIA frame.

\section{Lattice of electric-magnetic charges}\label{s:emlattices}

The lattice of electric-magnetic charges $\Lambda_{em}$ of a CHL models contains the sublattices $\Lambda_e$ and $\Lambda_m$  of purely electric and purely electric charges
\be\label{emlattices} \Lambda_e\oplus \Lambda_m\subseteq \Lambda_{em}\ .
\ee By convention, purely electric charges are the ones carried by perturbative heterotic states. Correspondingly, the electric fields are given by the metric and B-field with one leg along the compactified $T^6$ and one along the space-time directions, as well as the the $U(1)^{16}$ group coming from the spontaneously broken $E_8\times E_8$ gauge fields in $10$ dimensions. In the original unorbifolded model, namely heterotic on $T^6$, one has $\Lambda_e\cong \Lambda_m\cong \Gamma^{6,22}$ and \eqref{emlattices} is simply an equality $\Lambda_{em}=\Lambda_e\oplus \Lambda_m$. The purely magnetic charges are carried by NS5 branes wrapping a 5-dimensional torus within $T^6$, by KK monopoles with asymptotic circle along $T^6$ and by magnetic monopoles for the heterotic gauge group of rank $16$ (in ten dimensions $E_8\times E_8$ or $SO(32)$).

\medskip

In a generic CHL model, both $\Lambda_e$ and $\Lambda_m$ are lattices of signature $(6,2+d)$, $d=0,\ldots,20$, and they can be thought as embedded in a real space $\RR^{6,2+d}$ with metric $L$ (roughly speaking, $L$ is the metric appearing in the low energy effective action in the kinetic and $\theta$-angle terms for the $U(1)$ gauge fields; the electric charges then live in a space $\RR^{6,d+2}$ that is dual to the space of $U(1)$ gauge fields; by classical electric-magnetic duality of the low energy equations of motion, the magnetic charges live in an isomorphic $\RR^{6,d+2}$ space).  Any two elements $(P_1,Q_1), (P_2,Q_2)\in \Lambda_{em}$ must satisfy the Dirac quantization condition
\be\label{DiracQuant} P_1 L Q_2 -P_2 L Q_1 \in \ZZ\ .
\ee This implies
\be\label{emDirac} \Lambda_e\oplus \Lambda_m\subseteq \Lambda_{em}\subseteq \Lambda_e^*\oplus \Lambda_m^*
\ee and
\be\label{eminclude} \Lambda_e\subseteq \Lambda_m^*\qquad \Lambda_m\subseteq \Lambda_e^*\ ,
\ee for the lattices of purely magnetic and purely electric charges (the two inclusion are actually equivalent). 

\medskip

In this section, we will show that for $\lambda=1$ we always have an equality
\be \Lambda_m=\Lambda_e^*\ ,
\ee which, by \eqref{emDirac}, implies
\be\label{emsplit} \Lambda_{em}=\Lambda_e\oplus\Lambda_m=\Lambda_e\oplus\Lambda_e^*\ .
\ee The situation is a little more complicated when $\lambda>1$, i.e. when the rotation $g$ within the symmetry $\hat g=(\delta,g)$ does not satisfy the level-matching condition by itself, as will be shown in appendix \ref{s:genlatticeem}. In this case the natural choice of electric and magnetic charge lattices does not yield a Lagrangian decomposition, as in (\ref{emsplit}). Although such a decomposition always exists---indeed for  any  lattice with bilinear form \eqref{DiracQuant}  there is  a choice of maximal isotropic sublattices $\Lambda_e$ and $\Lambda_m$ for which \eqref{emsplit} holds---this is not necessarily the natural choice in the heterotic string frame and, most importantly, it might not be the choice where the S-duality group and space parity transformation have the standard action. We deal with this situation in appendix \ref{s:genlatticeem}.

\subsection{Purely electric charges}

The lattice of electric charges of the CHL model can be most conveniently described in the heterotic frame. Let $u,u^*$ denote two null vectors generating $\Gamma_{S^1}$, so that
\be u\cdot u=u^*\cdot u^*=0\qquad u\cdot u^*=1\qquad \Gamma_{S^1}=\langle u, u^*\rangle\ .
\ee  We can consider $u$ as representing a unit of winding along $S^1$ and $u^*$ as a unit of momentum. Let $g$ be a symmetry of order $N$ of the K3 model in the dual type IIA picture, so that $g$ acts by automorphisms of $\Gamma_{K3}\cong \Gamma^{4,20}$ and fixes $\Gamma_{S^1}$ and $\Gamma_{\tilde S^1}$. The symmetry $g$ fixes a sublattice $\Gamma_{K3}^g\subseteq \Gamma_{K3}$ of signature $(4,d)$, $0\le d\le 20$. In this section, we assume that both in the heterotic and in the type IIA picture, a $g$-twisted ground state satisfies the level matching condition, i.e. $\lambda=1$. This means that the shift $\delta$ can be taken to have the same order $N$ as the symmetry $g$, for example \be \delta:=u/N \ ,
\ee representing a shift along $S^1$ of $1/ N$-th of a period. The case $\lambda>1$ is considered in appendix \ref{s:pureelectgen}.

% Then, we take the orbifold by a symmetry $\hat g=(\delta,g)$ of order $N$, for a suitable $\lambda|N$, where 
%\be \delta:=u/\hat N\ ,
%\ee represents a shift along $S^1$ of $1/\hat N$-th of a period.

We can write a basis of perturbative states in the parent (unorbifolded) heterotic string picture as a tensor product
\be |\N;(\tilde p,p,v)\rangle=\N |(\tilde p,p,v)\rangle \ .
\ee  where $|(\tilde p,p,v)\rangle$ is a lowest level state with Narain winding-momentum  
\be (\tilde p,p,v)\in \Gamma_{\tilde S^1}\oplus\Gamma_{S^1}\oplus\Gamma_{K3}\ ,
\ee and $\N$ is a combination of left- and right-moving oscillators. 
% $|N\rangle$ denotes a state obtained by acting on the vacuum with $24$ left-moving (bosonic) oscillators, $|\tilde p\rangle$ denotes a lowest level state with Narain winding-momentum $\tilde p\in \Gamma_{\tilde S^1}$ along $\tilde S^1$,
%$|p=wu+nu^*\rangle$ with $w,m\in \ZZ$ denote a lowest level state with winding $w$ and momentum $m$ along $S^1$, and $ |v\rangle$ denotes the lowest level state with winding-momenta $v\in \Gamma_{K3}$. We drop the dependence on the right-moving oscillators, which are $\hat g$-invariant and do not affect the electric charge associated with a given state. The symmetry $\hat g$ acts by
%\be \hat g|N_L\rangle =|g(N_L)\rangle,\qquad \hat g|wu+nu^*\rangle=e^{\frac{2\pi i n}{N}}|wu+nu^*\rangle\qquad \hat g|v\rangle= |g(v)\rangle\ .
%\ee 
%
%
%\medskip
%
%
The orbifold projection excludes $20-d$ right-moving bosonic oscillators, so that the gauge group of the low energy effective field  theory is reduced from $U(1)^{28}$ to $U(1)^{8+d}$ and the state $|\N;(\tilde p,p,v)\rangle$ has electric charge
\be \tilde p+p+\Pi_g(v)\in \RR^{4,4+d}\ ,
\ee with respect to this reduced gauge group. Here,
\be \Pi_g(v):=\frac{1}{N}\sum_{i=1}^N g^i(v)\in \Gamma_{K3}^g\otimes \RR\ ,\qquad v\in \Gamma_{K3}\ ,\ee
is the projection of $v\in \Gamma_{K3}$ onto the $g$-invariant subspace $\Gamma_{K3}^g\otimes \RR$. Since $\Gamma_{K3}$ is self-dual, it follows that \be \Pi_g(\Gamma_{K3})=(\Gamma_{K3}^g)^*\ .\ee For each electric charge vector $\tilde p+p+\Pi_g(v)$, one can find a configuration of oscillators such that the corresponding state is $\hat g$-invariant. 
% Thus, \emph{before} the projection onto the $g$-invariant space, the lattice spanned by the electric charges of the untwisted states with respect to the reduced $U(1)^{8+d}$ gauge group is
%\be\label{untwtotal} \Pi_g(\Gamma_{K3})\oplus \Gamma_{S^1} \oplus \Gamma_{\tilde S^1}\cong (\Gamma_{K3}^g)^*\oplus \Gamma_{S^1} \oplus \Gamma_{\tilde S^1}\ .
%\ee
%\medskip
%We can take a basis of $\hat g$-eigenvectors in the complex vector space spanned by the states $|N_L\rangle$ and similarly for the space spanned by $|v\rangle$, $v\in \Gamma_{K3}$. Over each of these two spaces, the set of eigenvalues for $\hat g$ is the set of $N$-th roots of unity. Therefore, in order to obtain a $\hat g$-invariant state, also the eigenvalue of the state $|wu+nu^*\rangle$ must be a $N$-th root of unity. %, which implies $n\equiv 0\mod \lambda$. 
We conclude that the lattice of electric charges for the untwisted sector is
\be\label{untwproj} \Lambda_e^{U}=(\Gamma_{K3}^g)^*\oplus \Gamma_{S^1} \oplus \Gamma_{\tilde S^1}\ .
\ee In the $\hat g$-twisted sector, \emph{before} the projection over the $\hat g$-invariant states, the possible $U(1)^{8+d}$ charges form a translate of the lattice \eqref{untwproj} by $\delta=u/ N$. After projecting onto the $\hat g$ invariant subspace, we expect only a subset of these charges to survive, and this subset should form a translate of the lattice \eqref{untwproj}, in order to have a consistent OPE between untwisted and twisted states.
%The reason is that the space of $\hat g$-invariant $\hat g$-twisted states should be an irreducible module over the algebra of $g$-invariant untwisted fields.  If $\hat g$ acts by $e^{-\frac{2\pi i\E_g}{N\lambda}}$ on the $\hat g$-twisted ground state, where $\E_g\in \ZZ/N\lambda \ZZ$ is defined by  
%\be \frac{\E_g}{N\lambda}\equiv E_g\mod \ZZ\ ,
%\ee in terms of the ground state $(L_0-\bar L_0)$-eigenvalue $E_g$,
%then the $\hat g$-invariant $\hat g$-twisted states span a set of charges
%\be\label{twproj} \Lambda_e^{\hat g-\text{twisted}}=\delta+\E_g u^*+ \Lambda_e^{U}=\frac{u}{N\lambda}+\E_gu^*+ \Lambda_e^{U}\ .
%\ee 
More generally, since the spectrum of the orbifold is generated by the OPE of $\hat g$-invariant $\hat g$-twisted fields, we obtain that the lattice of electric charges is
%\be\label{eleclatt} \Lambda_e=(\Gamma_{K3}^g)^*\oplus \langle \frac{u}{N\lambda}+\E_g u^*, \lambda u^*\rangle \oplus \Gamma_{\tilde S^1}\ .
%\ee 
\be\label{electLattsimpl} \Lambda_e=(\Gamma_{K3}^g)^*\oplus \langle \frac{u}{N}, u^*\rangle \oplus \Gamma_{\tilde S^1}\ ,\qquad \text{for }\lambda=1\ .
\ee  
We stress that this result is only valid when $\lambda=1$, i.e. when the level matching condition is satisfied. 

\subsection{Purely magnetic charges}

Let us now determine the lattice $\Lambda_m$ of purely magnetic charges. Once again, we will focus on the case $\lambda=1$, postponing the discussion of the general case to section \ref{s:genlatticeem}.
%For $\lambda=1$, \eqref{eleclatt} simplifies to
%\be\label{electLattsimpl} \Lambda_e=(\Gamma_{K3}^g)^*\oplus \langle \frac{u}{N}, u^*\rangle \oplus \Gamma_{\tilde S^1}\ .
%\ee 
By the Dirac quantization condition, the lattice of purely magnetic charges $\Lambda_m$ must be contained in the dual of the lattice \eqref{electLattsimpl}, that is 
%\be
%\Lambda_m\subseteq \Lambda_e^* \Gamma_{K3}^g\oplus \langle N\lambda u^*, \frac{u}{\lambda}-N\E_gu^*\rangle \oplus \Gamma_{\tilde S^1}\ .
%\ee
%In order to give a more precise statement, one needs to consider the possible candidates for purely magnetic charges. 
%
%\subsection{Case $\lambda=1$} Let us start from the case $\lambda=1$, for which
\be\label{magnincl} \Lambda_m\subseteq \Lambda_e^*=\Gamma_{K3}^g\oplus \langle N u^*, u\rangle \oplus \Gamma_{\tilde S^1}\ ,\qquad \text{for }\lambda=1\ .
\ee We will now argue that the CHL model contains non-perturbative states whose charges generate the lattice $\Lambda_e^*$, so that this inclusion is actually an equality. It is useful to split the problem into two parts: we consider separately the states that carry magnetic charges within $\langle N u^*, u\rangle \oplus \Gamma_{\tilde S^1}$ and the states that carry charges within $\Gamma_{K3}^g$. The former are related to the four $U(1)$ gauge fields arising from metric and B-field along $S^1\times \tilde S^1$ and are dual to the winding and momentum along $S^1\times \tilde S^1$. The latter are related to the remaining $U(1)^{4+d}$ gauge group, which survives in the decompactification limit where the volume of $S^1\times \tilde S^1$ is large and the theory becomes effectively six-dimensional (heterotic string on $T^4$ or, equivalently, type IIA on K3). 

Let us start from the latter. In the original unorbifolded theory, the magnetic charges under these $4+d=24$ $U(1)$ gauge fields generate the lattice $\Gamma_{K3}\cong \Gamma^{4,20}$. After orbifolding by the symmetry $\hat g=(\delta, g)$, only the states invariant under $g\in O(\Gamma_{K3})$ should be included in the spectrum (the shift $\delta$ has no effect on this set of charges). The magnetic charges of these $g$-invariant  states generate a lattice $\Gamma_{K3}^g\subset \Lambda_m$, which must therefore be contained as a sublattice of $\Lambda_m$  in the CHL model. 

%In principle, one could imagine that there might be new states in the CHL model, carrying a magnetic charge which is a fraction of the one of these $g$-invariant configurations.
%However, eq.\eqref{magnincl} excludes this possibility: $\Gamma_{K3}^g$ is the finest possible lattice compatible with the Dirac quantization condition.

Now, let us consider the four $U(1)$ gauge fields given by the metric and B-field with one leg along $S^1\times \tilde S^1$ and one leg in an uncompactified direction. The associated electric charges correspond to momenta $m,\tilde m$ and windings $w,\tilde w$ along, respectively, $S^1$ and $\tilde S^1$ and generate the lattice $\langle \frac{u}{N}, u^*\rangle \oplus \Gamma_{\tilde S^1}$. Explicitly, the lattice is spanned by vectors
\be  (\begin{matrix}
m, \tilde m , w , \tilde w
\end{matrix})\in \ZZ\oplus \ZZ\oplus \frac{1}{N}\ZZ\oplus \ZZ\ ,\ee
with standard quadratic form
\be\label{quadform} (m,\tilde m,w,\tilde w)^2=2mw+2\tilde m\tilde w\ . \ee
 Notice that,  in our conventions, both the charges and the moduli are normalized as in the original unorbifolded model.
Let us denote by $M,\tilde M,W,\tilde W$ the dual magnetic charges. 
 In the unorbifolded theory, one unit of charge $M$, dual to momentum $m$ along $S^1$, is carried by a NS5-brane wrapping $T^4\times \tilde S^1$. The orbifold projection preserves a configuration of $N$ such branes, periodically localized at distance $\delta$ along $S^1$. Therefore, in the CHL model, $M$ is quantized in units of $N\ZZ$. One unit of charge $\tilde M$, dual to the momentum $\tilde m$ along $\tilde S^1$, is carried by an NS5 brane wrapped along $T^4$ and wrapped once along $S^1/\ZZ_N$, so that $\tilde M\in \ZZ$. One unit of charge $W$, dual to the winding $w$ along $S^1$, is carried by a KK monopole with asymptotic circle $S^1/\ZZ_N$, so that $W\in \ZZ$. Finally, one unit of charge $\tilde W$ is carried in the unorbifolded theory by a KK monopole with asymptotic circle $\tilde S^1$. This configuration is not affected by the symmetry $\hat g$, so it also appears in the CHL model, where it carries the same charge. The magnetic charges of these non-perturbative states span a lattice
\be  (\begin{matrix}
M , \tilde M , W , \tilde W
\end{matrix})\in N\ZZ\oplus \ZZ\oplus \ZZ\oplus \ZZ\ ,\qquad \qquad (M , \tilde M , W , \tilde W)^2=2MW+2\tilde M\tilde W\ ,\ee 
which is isomorphic to  $\langle N u^*, u\rangle \oplus \Gamma_{\tilde S^1}$.
 
 \medskip 
 
 To summarize, we have identified in the CHL model a set of states whose magnetic charges generate a lattice
\be \Gamma_{K3}^g\oplus \langle N u^*, u\rangle \oplus \Gamma_{\tilde S^1}\ .
\ee By \eqref{magnincl}, this is the finest possible lattice compatible with the Dirac quantization condition, so that $\Lambda_m=\Lambda_e^*$ and \be \Lambda_{em}=\Lambda_e\oplus\Lambda_m=\Bigl((\Gamma_{K3}^g)^*\oplus \langle \frac{u}{N}, u^*\rangle \oplus \Gamma_{\tilde S^1}\Bigr)\oplus \Bigl(\Gamma_{K3}^g\oplus \langle N u^*, u\rangle \oplus \Gamma_{\tilde S^1}\Bigr)\ ,\qquad \text{for }\lambda=1\ . \ee

\medskip

For $\lambda>1$, the argument showing that $\Gamma_{K3}^g\subset \Lambda_m$ goes through without essential modifications. However,  as explained in appendix \ref{s:genlatticeem}, the analysis of the charges carried by NS5-branes and KK monopoles is more subtle in this case. The outcome is that the inclusion $\Lambda_m\subset \Lambda_e^*$ is strict for $\lambda>1$ and, as a consequence, $\Lambda_{em}$ is not simply the direct sum of the purely electric and purely magnetic sublattices.

%The S-duality group $SL(2,\ZZ)$ of heterotic string becomes, in the type IIB picture, the group of modular transformations on the complex torus $S^1\times \tilde S^1$. The orbifold procedure breaks this group to the subgroup $\Gamma_1(N\lambda)$ that preserves a shift $\delta$ of order $N\lambda$ along $S^1$. This reduced S-duality group acts on the electric-magnetic charges by
%\be \begin{pmatrix}
%Q\\ P
%\end{pmatrix}\mapsto\begin{pmatrix}
%a & b\\ c & d
%\end{pmatrix}\begin{pmatrix}
%Q\\ P
%\end{pmatrix}\qquad \begin{pmatrix}
%a & b\\ c & d
%\end{pmatrix}\in \Gamma_1(N\lambda)\ .
%\ee By consistency, the lattice $\Lambda_e\oplus \Lambda_m$ must be closed under these transformations or, equivalently,
%\be N\lambda\Lambda_e\subseteq \Lambda_m\subseteq \Lambda_e\ .
%\ee It is easy to verify that both inclusions are satisfied. Since $\Lambda_e=\Lambda_m^*$, the second inclusion implies that $\Lambda_m$ is integral (in fact, it is also even). When $\lambda=1$, for example for the geometric symmetries, we have $\E_g=0$ and the lattices of electric and magnetic charges become
%\be \Lambda_e=(\Gamma_{K3}^g)^*\oplus \langle \frac{u}{N}, u^*\rangle \oplus \Gamma_{\tilde S^1}\ ,
%\ee and
%\be \Lambda_m=\Gamma_{K3}^g\oplus \langle N u^*, u\rangle \oplus \Gamma_{\tilde S^1}\ .
%\ee
%

\subsection{Fricke S-duality and $N$-modular lattices}\label{s:Nmodular}

In the previous section, we derived the lattice $\Lambda_{em}$ of electric-magnetic charges of  a CHL model, in the case $\lambda=1$. Suppose that the CHL model is in case 1 of section \ref{s:Sdual}, i.e. the symmetry $g$ has balanced cycle shape.  In this case, the Fricke S-duality connects two different points in the moduli space of the same CHL model. In particular, the moduli $S_{\text{het}},T_{\text{het}},U_{\text{het}}$ of the dual theories are related as in \eqref{hetmodSdual} and the electric and magnetic charges as in \eqref{emcharges}.
This implies that the lattice of magnetic charges $\Lambda_m$ of a $\ZZ_N$ self-dual CHL model is an integral $N$-modular lattice
\be \Lambda_m^*(N)\cong \Lambda_m\ .
\ee
This prediction can be verified in detail. By \eqref{electLattsimpl}, we have
\be \Lambda_m^*(N)=(\Gamma_{K3}^g)^*(N)\oplus \langle \frac{u}{\sqrt{N}}, \sqrt{N}u^*\rangle \oplus \Gamma_{\tilde S^1}(N)\ .
\ee The summands $\langle \frac{u}{\sqrt{N}}, \sqrt{N}u^*\rangle$ and $\Gamma_{\tilde S^1}(N)$ are $\ZZ^2$ lattices with quadratic forms, respectively
\be \begin{pmatrix}
0 & 1 \\ 
1 & 0 \end{pmatrix}\qquad \qquad \begin{pmatrix}
 0 & N\\
 N & 0
\end{pmatrix}\ .
\ee On the other hand, the summands $\langle N u^*, u\rangle$, $ \Gamma_{\tilde S^1}\subset \Lambda_m$ are $\ZZ^2$ lattices with quadratic forms, respectively
\be \begin{pmatrix}
0 & N \\ 
N & 0 \end{pmatrix}\qquad \qquad \begin{pmatrix}
 0 & 1\\
 1 & 0
\end{pmatrix}\ .\ee
Therefore, the lattices $\langle \frac{u}{\sqrt{N}}, \sqrt{N}u^*\rangle\oplus \Gamma_{\tilde S^1}(N)$ and $\langle N u^*, u\rangle$, $ \Gamma_{\tilde S^1}\subset \Lambda_m$ are isomorphic, as expected, but the isomorphism requires the exchange of the winding-momentum charges related to the circle $S^1$ with the ones related to $\tilde S^1$. This is compatible with the non-trivial action of the S-duality transformation \eqref{hetmodSdual} on the complex structure modulus $U_{\text{het}}$.

$N$-modularity of the summands $\langle N u^*, u\rangle$, $ \Gamma_{\tilde S^1}$ actually holds for every CHL model (with $\lambda=1$), independently of the model being self-dual or not. What is peculiar to self-dual CHL models is the $N$-modularity property of the orthogonal complements
\be (\Gamma_{K3}^g)^*(N)\cong \Gamma_{K3}^g\ .
\ee We have verified this property directly by deriving the quadratic forms of the lattices $\Gamma_{K3}^g$ for all possible Frame shapes of $g$ (see tables \ref{t:case1},\ref{t:case2},\ref{t:case3} in appendix \ref{s:lattices}). It turns out that the lattice $\Gamma_{K3}^g$ is $N$-modular if and only if the Frame shape of $g$ is balanced (table \ref{t:case1}). This conclusion extends to the cases $\lambda>1$, as described in section \ref{s:genlatticeem}. A highly non-trivial example is provided by the Frame shape $1^82^8$, which corresponds on the type IIA side to a symplectic automorphism of order $2$ of the K3 manifold (Nikulin involution), and on the $E_8\times E_8$ heterotic side to the involution exchanging the two $E_8$ factors. In this case, one obtains
\be \Gamma_{K3}^g= \Gamma^{4,4}\oplus E_8(-2)\ ,
\ee  and 
\be (\Gamma_{K3}^g)^*= \Gamma^{4,4}\oplus E_8(-1/2)\ .
\ee S-duality, in this case, predicts the existence of an isomorphism
\be \Gamma^{4,4}(2)\oplus E_8(-1)\cong \Gamma^{4,4}\oplus E_8(-2)\ .
\ee This isomorphism can be verified by first checking that the lattices on both sides are in the same genus, and then noting that the genus has only one isomorphism class (as follows from theorem~1.14.2 of \cite{Nikulin}).

In case 2, S-duality relates pairs of distinct CHL models, related to symmetries $g, g'\in O(\Gamma_{K3})$ of the same order but different Frame shape, and we have (table \ref{t:case3})
\be (\Gamma_{K3}^g)^*(N)\cong \Gamma_{K3}^{g'}\ .
\ee

Finally,  for case 3, i.e. when the orbifold of the type IIA K3 model by $g$ is a sigma model on $T^4$, one has
\be (\Gamma_{K3}^g)^*(N)\cong (\Gamma^{4,4})^{g'}\ ,
\ee where $\Gamma^{4,4}$ is the lattice of even D-brane charges on a non-linear sigma model on $T^4$ and $g'\in O(\Gamma^{4,4})$ is a symmetry of this model preserving the (small) $\N=(4,4)$ superconformal algebra \cite{Volpato:2014zla} (see table \ref{t:case2} in appendix \ref{s:lattices}).

\section{BPS-state counting}
\label{sec:BPScount}

In this section we shall calculate the 1/2 BPS-indices in arbitrary CHL models, both from the type IIA and the heterotic perspective. In $\mathcal{N}=4$ string theories the 1/2 BPS-states are counted by the index
\be
\Omega_4:= \text{Tr}(-1)^F J_3^{4}, 
\ee
where $F$ is the spacetime fermion number and $J_3$ is the Cartan generator of the massive little group in four dimensions. Due to the 
insertion of $J_3^4$ this trace is sensitive only to 1/2 BPS-states. The 1/4 BPS-states only contribute to the index $\Omega_6$ which 
 has an insertion of $J_3^6$. More generally, one can define the $n$:th index $\Omega_n$ by insertion of $J_3^{n}$. By symmetry 
reasons this vanishes for $n$ odd. 
 
We will  calculate certain   one-loop topological amplitudes and show that they receive contributions precisely from the fourth helicity supertraces, and, furthermore, that they are invariant under Fricke S-duality precisely when the CHL models are self-dual.

\subsection{The type II helicity supertrace}
% Let us introduce the generating function for these helicity supertraces
% \be
% B_n(q, \bar{q})=\sum \Omega_n q^{L_0}\bar{q}^{\bar{L}_0}.
% \ee
It is convenient to introduce a generating function, called the \emph{helicity string partition function} in \cite{Kiritsis:1997hj}:
\be
Z( \upsilon, \bar{ \upsilon})=\text{Tr}(-1)^F e^{2\pi i  \upsilon J_3^{R}} e^{2\pi i \bar{ \upsilon} J_3^{L}} q^{L_0} \bar{q}^{\bar{L}_0}, 
\ee
where $( \upsilon, \bar{ \upsilon})$ are chemical potentials for the left- and right-moving spacetime helicities $(J_3^{R}, J_3^L)$. From the generating function $Z( \upsilon, \bar{ \upsilon})$ one can compute the associated helicity supertraces  $B_n$ can then 
be computed by taking derivatives of the partition function: 
\be B_n=\left(\frac{1}{2\pi i} \frac{\partial}{\partial \upsilon}+\frac{1}{2\pi i} \frac{\partial}{\partial \bar\upsilon}\right)^n Z(\upsilon,\bar \upsilon)\Big|_{\upsilon=\bar \upsilon=0}\ .
\ee 
%where 
%\be Q:=\frac{1}{2\pi i} \frac{\partial}{\partial \upsilon}\qquad \bar Q:=\frac{1}{2\pi i} \frac{\partial}{\partial \bar\upsilon}
%\ee and $Z(\upsilon,\bar \upsilon)$ is the helicity supertrace generating function
%\be  Z(\upsilon,\bar \upsilon)=STr(q^{L_0}\bar q^{\bar L_0}e^{2\pi i\upsilon \lambda_L}e^{2\pi i\bar \upsilon \lambda_R})
%\ee
%and $\lambda_L$ and $\lambda_R$ are the contributions to space-time helicity from left- and right-moving.
The $B_n$'s should then be viewed as generating functions of the BPS-indices $\Omega_n$. For type IIA compactified on K3$\times T^2$, we have
\be\label{ZIIA} Z^{\text{IIA}}(\upsilon,\bar \upsilon)=\frac{\vartheta_1(\upsilon/2)^2\bar\vartheta_1(\bar\upsilon/2)^2  }{\eta^2\bar\eta^2}\frac{\xi(\upsilon)\bar\xi(\bar\upsilon)}{\tau_2\eta^2\bar\eta^2}C%\bigl[{}^1_1{}^1_1\bigr]
(\upsilon/2,\bar\upsilon/2)\frac{\Theta_{\Gamma^{2,2}} }{\eta^2\bar\eta^2}\ ,
\ee 
where we defined
\be \xi(\upsilon)=\prod_{n=1}^\infty \frac{(1-q^n)^2}{(1-e^{2\pi i \upsilon} q^n)(1-e^{-2\pi i \upsilon} q^n)}\ ,
\ee 
and $\Theta_{\Gamma^{2,2}}$ is the usual Narain theta function of the lattice.
%The uncompactified Minkowski space $R^{1,3}$ extends along the directions 0--3, the torus $T^2$ along 4--5 and the K3 manifold along 6--9 and we made a light-cone quantization of superstrings along the 0,1 directions. 
The contributions to $Z^{\text{IIA}}$ are as follows:
\begin{itemize}
\item $\frac{\xi(\upsilon)\bar\xi(\bar\upsilon)}{\tau_2\eta^2\bar\eta^2}$ 
 comes from the scalar fields along the two uncompactified transverse directions.
 \item $C%\bigl[{}^1_1{}^1_1\bigr]
 (\upsilon/2,\bar\upsilon/2)$ comes from the internal $\N=(4,4)$ SCFT with central charge $(c,\bar c)=(6,6)$ (non-linear sigma model on K3). For $\bar \upsilon=0$, this is just the elliptic genus of K3.
\item $\frac{\Theta_{\Gamma^{2,2}} }{\eta^2\bar\eta^2}$ comes from the oscillators and winding-momenta along $T^2$.
\item $\frac{\vartheta_1(\upsilon/2)^2}{\eta^2}$ and its complex conjugate come from left- and right-moving fermions along the two uncompactified transverse directions and along the torus $T^2$.
\end{itemize}
Since $\vartheta _1(0)=0$, the only non-vanishing contributions to the fourth helicity trace $B_4$  come from terms where all four derivatives act on the theta functions. Using
\be \vartheta_1'(0)=2\pi \eta^3\ ,\qquad \xi(0)=1\ ,\qquad C%\bigl[{}^1_1{}^1_1\bigr]
(0,0)=24
\ee we obtain the following expression for the fourth helicity supertrace:
\be
\begin{split}
 B_4&=\left(\frac{1}{2\pi i} \frac{\partial}{\partial \upsilon}+\frac{1}{2\pi i} \frac{\partial}{\partial \bar\upsilon}\right)^4 Z^{\text{IIA}}(\upsilon,\bar \upsilon)\Big|_{ \upsilon=0}
 \nonumber \\
 &= 6\left(\frac{1}{2\pi i} \frac{\partial}{\partial \upsilon}\right)^2\left(\frac{1}{2\pi i} \frac{\partial}{\partial \bar\upsilon}\right)^2 Z^{\text{IIA}}(\upsilon,\bar \upsilon)\Big|_{ \upsilon=0}
 \nonumber \\
 &=\frac{3}{2} C%\bigl[{}^1_1{}^1_1\bigr]
 (0,0)\frac{\Theta_{\Gamma^{2,2}} }{\tau_2}
 %\nonumber \\ & 
 =36\frac{\Theta_{\Gamma^{2,2}} }{\tau_2}\ .
 \end{split}\ee
 
 \subsection{Helicity supertrace for type IIA CHL models}
 Let us now generalize the previous analysis to CHL models in type IIA string theory. We are interested in the case where the orbifold group $G$ is cyclic of order $\hat N$, with generator $\hat g$ acting by a shift $\delta$ of a $1/\hat N$ period along one of the cycles of $T^2$ and by a symmetry $g$ of the K3 sigma model. Then, the generating function $Z^{\text{IIA}}_{\text{CHL}[\hat g]}$ is obtained  from \eqref{ZIIA}  by the usual orbifold formula
\be Z^{\text{IIA}}_{\text{CHL}[\hat g]}(\upsilon,\bar \upsilon)=\frac{\vartheta_1(\upsilon/2)^2\bar\vartheta_1(\bar\upsilon/2)^2  }{\eta^6\bar\eta^6\tau_2}\xi(\upsilon)\bar\xi(\bar\upsilon)\frac{1}{|\hat N|}\sum_{r,s=1}^{\hat N}C_{g^r,g^s}%\bigl[{}^1_1{}^1_1\bigr]
(\upsilon/2,\bar\upsilon/2)\Theta_{\Gamma^{2,2}} \bigl[{}^r_s\bigr]\ ,
\label{CHLpartfun}
\ee 
where 
\be \Theta_{\Gamma^{2,2}} \bigl[{}^r_s\bigr]:=\sum_{(p_L,p_R)\in r\delta+\Gamma^{2,2}} e^{2\pi is \delta \cdot (p_L,p_R)} q^{\frac{p_L^2}{2}}
\bar q^{\frac{p_R^2}{2}}\ .
\ee
Here we used the fact that the orbifold group acts only on the winding-momenta along $T^2$ and on the internal K3 sigma model. The fourth helicity trace is then evaluated to
\be B_4^{\text{CHL}[\hat g]}=\frac{3}{2\tau_2}\frac{1}{\hat N}\sum_{r,s=1}^{ \hat N}C_{g^r,g^s}%\bigl[{}^1_1{}^1_1\bigr]
(0,0)\Theta_{\Gamma^{2,2}} \bigl[{}^r_s\bigr]\ .
\ee  We have
\be C_{g^r,g^s}%\bigl[{}^1_1{}^1_1\bigr]
(0,0) = \Tr_{\bf 24} (g^{\gcd(r,s,N)})\ ,
\ee where ${\bf 24}$ denotes the $24$ dimensional representation of R-R ground states in the K3 sigma model. Therefore, if $g$ has Frame shape $\prod_{a|N} a^{m(a)}$, we obtain
\be \Tr_{\bf 24} (g^d)=\sum_{a|d}m(a)a
\ee for all  $d|\hat N$. Assembling the pieces we find that the fourth helicity supertrace of the CHL model becomes 
\begin{align} B_4^{\text{CHL}[\hat g]}= &\frac{3}{2\tau_2}\frac{1}{\hat N}\sum_{d|\hat N} \Tr_{\bf 24} (g^d)\sum_{\substack{r,s=1\\ \gcd(r,s,\hat N/d)=1}}^{ \hat N/d}\Theta_{\Gamma^{2,2}} \bigl[{}^{dr}_{ds}\bigr]\\= &\frac{3}{2\tau_2}\frac{1}{\hat N}\sum_{d|\hat N} \Tr_{\bf 24} (g^d)\sum_{l|\frac{\hat N}{d}}\mu(l)\sum_{r,s=1}^{ \hat N/ld}\Theta_{\Gamma^{2,2}} \bigl[{}^{ldr}_{lds}\bigr]\\
\stackrel{n:=ld}{=}&\frac{3}{2\tau_2}\frac{1}{\hat N}\sum_{n|\hat N}\Bigl(\sum_{d|n} \Tr_{\bf 24} (g^d)\mu(n/d)\Bigr)\sum_{r,s=1}^{ \hat N/n}\Theta_{\Gamma^{2,2}} \bigl[{}^{nr}_{ns}\bigr]
\\
=&\frac{3}{2\tau_2}\frac{1}{\hat N}\sum_{n|\hat N}\Bigl(\sum_{d|n} \mu(n/d)\sum_{a|d}m(a)a\Bigr)\sum_{r,s=1}^{ \hat N/n}\Theta_{\Gamma^{2,2}} \bigl[{}^{nr}_{ns}\bigr]
\end{align} where $\mu$ is the M\"obius function.
 Using the property
\be \sum_{x|y}\mu(x)=\begin{cases}1 &\text{if }y=1\ ,\\ 0 &\text{otherwise}\ ,\end{cases}
\ee
the divisor sum simplifies:
\be \sum_{d|n} \mu(n/d)\sum_{a|d}m(a)a\stackrel{t:=n/d}{=}\sum_{a|n}m(a) a\sum_{t|\frac{n}{a}}\mu(t)=m(n)n
\ee so that
\be B_4^{\text{CHL}[\hat g]}=\frac{3}{2\tau_2}\sum_{n|\hat N}m(n)\frac{n}{\hat N}\sum_{r,s=1}^{ \hat N/n}\Theta_{\Gamma^{2,2}} \bigl[{}^{nr}_{ns}\bigr]
\ee
Finally, we note that, if $T,U$ are the K\"ahler and complex moduli of $T^2$ in the original (unorbifolded) model, then
\be \frac{n}{\hat N}\sum_{r,s=1}^{ \hat N/n}\Theta_{\Gamma^{2,2}} \bigl[{}^{nr}_{ns}\bigr](T,U)=\Theta_{\Gamma^{2,2}}(Tn/\hat N,Un/\hat N)\ ,
\ee because the LHS is simply (up to a factor $1/|\eta|^4$)  the partition function for a bosonic sigma model on a torus $T^2$ with moduli $(T,U)$, orbifolded by a shift of order $\hat N/n$ along one of the cycles; this orbifold is again a sigma model on $T^2$, with the $T$ and $U$ moduli divided by $\hat N/n$. Notice that, in our conventions the K\"ahler and complex structure moduli $T_{\text{IIA}}$ and $U_{\text{IIA}}$ of $T^2$ in the CHL model are related to $T,U$ via
\be T_{\text{IIA}}\equiv \frac{T}{\hat N}\qquad U_{\text{IIA}}\equiv \frac{U}{\hat N}\ .
\ee
Therefore the final result for the helicity supertrace is
\be B_4^{\text{CHL}[\hat g]}=\frac{3}{2\tau_2}\sum_{n|\hat N}m(n)\Theta_{\Gamma^{2,2}}(nT_{\text{IIA}},nU_{\text{IIA}}).
\ee

\subsection{Topological amplitudes and Fricke S-duality}
Our aim is now to test the Fricke S-duality in the context of 1/2 BPS-state counting. Recall that the fourth helicity supertrace determines the one-loop topological amplitude according to the formula:
\be
F_1=\frac{2}{3}\int_{\F} d^2\tau\, B_4 \ .
\ee  
where $\mathcal{F}$ is a fundamental domain for $SL(2,\mathbb{Z})\backslash \mathbb{H}$. The $T_{\text{IIA}}$-dependent part of this amplitude determines the coupling $f_{R^2}$ associated with the $R^2$-correction term in the low-energy effective action. Explicitly the relation between $F_1$ and $f_{R^2}$ is \cite{Gregori:1997hi}
\be \partial_{T_{\text{IIA}}} f_{R^2} = \partial_{T_{\text{IIA}}} F_1\ .
\ee 
If Fricke S-duality holds, then the associated coupling $f^{\text{CHL}[\hat g]}_{R^2}$ for CHL models should be invariant under $T_{\text{IIA}}\to -1/(\hat{N} T_{\text{IIA}})$ whenever the model is self-dual. Let us now verify this prediction. 
Let $g\in G$ be determined by the Frame shape
\be
\prod_{a|\hat{N}} a^{m(a)}.
\ee
Then, using the well-known fact the (renormalized) integral of the theta series is
\be
\int_{\F} \frac{d^2\tau}{\tau_2} \Theta_{\Gamma^{2,2}}(T,U)=-\log(\im T\,|\eta(T)|^4)+\ldots \ee 
where the ellipsis denote terms independent of $T$, we obtain for the CHL models
\be f^{\text{CHL}[\hat g]}_{R^2}(T_{\text{IIA}})=-\log\prod_{a|\hat N} \Bigl(\im T_{\text{IIA}}\,a|\eta(aT_{\text{IIA}})|^4\Bigr)^{m(a)}=-\log (\im T_{\text{IIA}}^{24}\,|\eta_g(T_{\text{IIA}})|^4)+const,
\label{R2couplingCHL}
\ee
where $\eta_g$ is the eta-product
\be
\eta_g(T_{\text{IIA}})=\prod_{a|N}\eta(aT_{\text{IIA}})^{m(a)}.
\ee
Under the Fricke T-duality transformation $T_{\text{IIA}}\to -1/(\hat N T_{\text{IIA}})$, the $R^2$ coupling transforms according to
\be f_{R^2}(-1/(\hat{N}T_{\text{IIA}}))=-\log\prod_{a|\hat N} \Bigl(\im T_{\text{IIA}} \,\frac{\hat N}{a}|\eta(T_{\text{IIA}}\frac{\hat N}{a})|^4\Bigr)^{m(a)}=-\log\prod_{a|\hat N} \Bigl(\im T_{\text{IIA}}\,a|\eta(aT_{\text{IIA}})|^4\Bigr)^{m(\frac{\hat N}{a})}\ ,
\ee 
indeed verifying that the coupling is invariant if and only if the Frame shape of $\hat g$ is balanced
\be m(\hat N/a)=m(a)\ ,
\ee which is exactly the condition for the model to be self-dual.

%By the OSV-conjecture \cite{Ooguri:2004zv,Dabholkar:2004yr,Dabholkar:2005dt}, the topological string amplitude captures the degeneracies of BPS-states. From the result (\ref{R2couplingCHL}) we may therefore extract the generating function of 1/2 BPS-degeneracies: 
%\be
%\frac{1}{\eta_g(T_{\text{IIA}})}=\sum_{Q^2/2\in \mathbb{Z}}\Omega_4^{\text{CHL}[g]}(Q)e^{\pi i T_{\text{IIA}} Q^2},
%\label{BPSchl}
%\ee
%For $g=e$ this reduces to the well-known counting of Dabholkar-Harvey states as reviewed in the introduction. 

\subsection{The heterotic helicity supertrace}

Let us now analyse the counting of BPS-states from the heterotic perspective. The generating function of helicity supertraces for heterotic strings on $T^6$ is \cite{Kiritsis:1997hj}
\be Z^{\text{het}}(\upsilon,\bar\upsilon)=
\frac{\vartheta_1(\upsilon/2)^4  }{\eta^{12}\bar\eta^{24}}\frac{\xi(\upsilon)\bar\xi(\bar\upsilon)}{\tau_2}\Theta_{\Gamma^{6,22}}\ ,
\ee and the fourth helicity supertrace is\footnote{This corrects a typo in eq. (G.9) of \cite{Kiritsis:1997hj} which lacks the factor $\Theta_{\Gamma^{6,22}}$.}
\be B_4=\left(\frac{1}{2\pi i} \frac{\partial}{\partial \upsilon}+\frac{1}{2\pi i} \frac{\partial}{\partial \bar\upsilon}\right)^4 Z^{\text{het}}(\upsilon,\bar \upsilon)_{\rvert \upsilon=0}=\left(\frac{1}{2\pi i} \frac{\partial}{\partial \upsilon}\right)^4Z^{\text{het}}(\upsilon,\bar \upsilon)_{\rvert \upsilon=0}=\frac{3}{2\tau_2} \frac{\Theta_{\Gamma^{6,22}}}{\bar\eta^{24}}\ .
\ee 

Let us consider the case where $T^6$ is the product of two orthogonal tori $T^2\times T^4$, with no Wilson lines along $T^2$. The lattice of winding momenta  decomposes accordingly as
\be \Gamma^{6,22}\cong \Gamma^{2,2}\oplus \Gamma^{4,20}\ .
\ee A CHL model is given by a orbifold by a symmetry $\hat g$ acting by a $1/\hat N$ period shift $\delta\in \Gamma^{2,2}\otimes \RR$ along one of the circles of $T^2$ and by an automorphism $g$ on the lattice $\Gamma^{4,20}$. The generating function is given by
\be Z^{\text{het}}_{\text{CHL}[g]}(\upsilon,\bar\upsilon)=
\frac{\vartheta_1(\upsilon/2)^4  }{\eta^{12}}\frac{\xi(\upsilon)\bar\xi(\bar\upsilon)}{\tau_2}\frac{1}{\hat N}\sum_{r,s=0}^{\hat N-1}\Theta_{\Gamma^{2,2}}\bigl[{}^{r}_{s}\bigr]\frac{\Theta_{\Gamma^{4,20}}^{(r,s)}}{\bar\eta_{g^r,g^s}}\ ,
\label{genhetchl}
\ee where
%\be Z_{0,0}=\frac{\Theta_{\Gamma^{4,20}}}{\eta^4\bar\eta^{20}}
%\ee
\be \Theta_{\Gamma^{4,20}}^{(0,s)}=\Theta_{(\Gamma^{4,20})^{g^s}}
\ee
and when $(r,s)\equiv (lc,ld)$, with $c,d,l\in \ZZ$ and $c,d$ coprime, we have
\be \Theta_{\Gamma^{4,20}}^{(lc,ld)}(\tau,\bar\tau)=(c\tau+d)^{-2}(c\bar \tau+d)^{-10}\Theta_{\Gamma^{4,20}}^{(0,l)}\Bigl(\frac{a\tau+b}{c\tau+d},\frac{a\bar\tau+b}{c\bar\tau+d}\Bigr)\ ,
\ee for some integers $a,b$ such that $\left(\begin{smallmatrix}
a & b\\ c& d
\end{smallmatrix}\right)\in SL(2,\ZZ)$. In the denominator of (\ref{genhetchl}) we also defined the generalized eta product which in general takes form \cite{Persson:2013xpa}:
 \be\label{etagh} \eta_{g,h}(\tau)=q^{\frac{1}{N\lambda}}\prod_{i=1}^{24}\prod_{n=0}^\infty (1-e^{\frac{2\pi it_i}{M}}q^{\frac{r_i}{N}+n})\ ,\qquad\qquad  gh=hg\ ,
 \ee
where $M$ is the order of $h$, and the integers $0<r_1,\ldots,r_{24}\le N$ and $0<t_1,\ldots,t_{24}\le N$ are such that $\{(e^{2\pi i\frac{r_i}{N}},\,e^{2\pi i\frac{t_i}{M}})\}$ is the set of simultaneous eigenvalues of $g$ and $h$. Taking derivatives of $Z^{\text{het}}_{\text{CHL}[g]}(\upsilon,\bar\upsilon)$ we obtain the fourth helicity supertrace
\be B^{\text{CHL}[g]}_{4, \text{het}}=\frac{3}{2\tau_2}\frac{1}{\hat N}\sum_{r,s=0}^{\hat N-1}\Theta_{\Gamma^{2,2}}\bigl[{}^{r}_{s}\bigr]\frac{\Theta_{\Gamma^{4,20}}^{(r,s)}}{\bar\eta_{g^r,g^s}}\ .
\ee
%We notice  the interesting fact that the generalised theta function $\eta_{g,h}(\tau)$ here appears as a function of the \emph{worldsheet} parameter, while its $(g,h)=(g,e)$ specialisation previously appeared as the \emph{spacetime} generating function of 1/2 BPS-states (\ref{BPSchl}). This  can be understood by combining standard duality arguments. 
Integrating $ B^{\text{CHL}[g]}_{4, \text{het}}$ against the fundamental domain of $SL(2,\mathbb{Z})$ should now yield the heterotic topological 1-loop amplitude \cite{Antoniadis:1997zt}
\be
\tilde{F}_1=\frac{2}{3}\int_{\mathcal{F}}  B^{\text{CHL}[g]}_{4, \text{het}},
\ee
which in particular depends on $(T_{\text{het}}, U_{\text{het}})$ but is independent of $S_{\text{het}}$. Therefore, this coupling is 1-loop exact on the heterotic side. If we expand this in the large-volume limit $\im T_{\text{het}}\to \infty$ and map to the type IIA picture via $(U_{\text{het}} \leftrightarrow U_{\text{IIA}}, T_{\text{het}} \leftrightarrow S_{\text{IIA}})$ we obtain  the one-loop IIA counterpart of $\tilde{F}_1$ supplemented with an infinite series of exponentially suppressed terms of order $\mathcal{O}(e^{-\im S_{\text{IIA}}})$ due to D-instantons. Finally, performing the mirror map $U_{\text{IIA}} \leftrightarrow T_{\text{IIA}}$ (see \cite{Antoniadis:1997zt}) then reproduces precisely the coupling $f^{\text{CHL}[\hat g]}_{R^2}(T_{\text{IIA}})$ in (\ref{R2couplingCHL}). Thus, the sequence of dualities just described ensures that the counting of 1/2 BPS-states in CHL via the fourth helicity supertrace is indeed consistent with heterotic-type II duality and Fricke S-duality.

\section{Conclusions}
\label{sec:conclusions}
In this paper we have analysed a large class of non-geometric CHL models and showed that they exhibit larger symmetries than previously expected. These models are in particular invariant under Fricke involution $S\to -1/(NS)$ which lies outside of the $SL(2,\mathbb{Z})$-symmetry of the original unorbifolded model. We showed that this leads to new non-trivial heterotic-type II dualities and gives rise to strong constraints on the associated electric and magnetic charge lattices. In particular, for self-dual CHL models these lattice are required to be $N$-modular. We also demonstrated that Fricke S-duality is compatible with the counting of 1/2 BPS-states by calculating the relevant helicity supertraces and verified the invariance of certain BPS-saturated topological couplings.

As mentioned in the introduction, our results  give a physical interpretation of the modular properties of the class of Siegel modular forms $\Phi_{g,h}$ constructed in \cite{Persson:2013xpa} in the context of Mathieu moonshine. It is therefore natural to speculate about the physical interpretation of these Siegel modular forms. It is known that in cases when $g$ is a geometric symmetry of the target space the reciprocals $1/\Phi_{g, h}$ are generating functions of certain (twisted) 1/4 BPS-states in CHL models. One can also relate the $\Phi_{g,h}$ to the generalised eta-products via the following limit
\be \lim_{z\to 0}\frac{\Phi_{g,h}(\sigma,\tau,z)}{(2\pi iz)^2}=\eta_{g,h}(\tau)\eta_{g,h'}(N\sigma),
\ee
Since we have seen in this paper that the generalised eta-products appear in certain topological one-loop couplings involving the fourth helicity supertraces, it is natural to speculate that the $\Phi_{g,h}$ would similarly appear in some topological couplings involving the \emph{sixth helicity supertrace} 
\be
B_6^{\text{CHL}[g]}:=\text{Tr}_g(-1)^{F} J_3^6 q^{L_0} \bar{q}^{\bar{L}_0},
\ee
which indeed receives contributions from 1/4 BPS-states. Schematically, this coupling should take the form \cite{Kawai:1995hy}
\be
\int_{\mathcal{F}} B_6^{\text{CHL}[g]} = \text{log} \left((\text{det}\im \Omega)^{w_{g,h}} |\Phi_{g,e}(T, U, V)|^2\right),
\ee
where $V$ is a Wilson line modulus, $\Omega =  \left(\begin{smallmatrix}
T & V \\ V & U
\end{smallmatrix}\right)$ is the genus 2 period matrix and $w_{g,h}$ is the weight of the corresponding Siegel modular form. This would generalise the ``threshold integrals'' previously calculated by Sen and collaborators for a class of CHL models \cite{David:2006ji,David:2006yn,Sen:2010ts}.  

A related observation is the following. It has been conjectured that the negative reciprocal of the weight 10 Igusa cusp form $\Phi_{10}$ is the generating function for Gromov-Witten invariants on $K3\times T^2$  \cite{Katz:1999xq,2014arXiv1404.6698P,2014arXiv1411.1514O}. By standard string theory dualities this can be interpreted as counting BPS-states in type II string theory on $K3\times T^2$. Moreover, since $\Phi_{e,e}$ is precisely the Igusa cusp form it is natural to conjecture that the Siegel modular forms $\Phi_{g,e}$ are generating functions for Gromov-Witten invariants on the CHL orbifold $(K3\times S^1)/\mathbb{Z}_N \times \tilde{S}^1$. It would be very interesting to verify this by  generalising the original analysis of Katz-Klemm-Vafa \cite{Katz:1999xq} to our class of CHL models.

The Fricke involutions appear in the context of topological string amplitudes for certain families of non-compact Calabi-Yau threefolds. In this setup, the involutions can be interpreted as dualities exchanging the expansions of the amplitude at two different cusps in the moduli space \cite{Alim2014}. It is tantalizing to conjecture a relationship between this construction and the present paper, but the details of this connection are not clear to us.

As mentioned in the introduction the Fricke involution also plays a role in the context of Monstrous moonshine, where some of the McKay-Thompson series $T_g(\tau)$ exhibit invariance under $\tau \to -1/(N\tau)$ for $N$ the order of $g\in \mathbb{M}$. In fact, Tuite has proposed the this ``Fricke property'' holds the key for understanding the elusive genus zero property of monstrous moonshine, namely the fact that all McKay-Thompson series are hauptmoduln for the modular groups $\Gamma_g\subset SL(2,\mathbb{R})$ under which the $T_g(\tau)$'s are invariant \cite{Tuite1995}. In short, the argument is that this genus zero property is equivalent to the statement that only for Fricke elements $g\in \mathbb{M}$ does the orbifold $V^{\natural}/\left<g\right>$  of the monster CFT $V^{\natural}$  give back the same theory, while orbifolding by non-Fricke elements instead yields the Leech lattice CFT. Tuite was able to demonstrate this for a large class of elements of the monster. This observation  is very similar to what we find in the present paper. Namely, it is precisely when the orbifold $\mathcal{C}'=\mathcal{C}/\left<g\right>$ of the K3 sigma model $\mathcal{C}$ is again a K3-sigma model, that the CHL model is self-dual with respect to Fricke S-duality $S\to -1/(NS)$. Thus, it would be very interesting to see if our techniques can be applied to monstrous moonshine, and perhaps shed some light on the relation between genus zero and Fricke properties. 

\acknowledgments

We are grateful to all organizers and participants of the workshop `(Mock) modularity, Moonshine, and String Theory' and, in particular, Miranda Cheng, John Duncan, Matthias Gaberdiel, Terry Gannon, Jeff Harvey, Shamit Kachru, Sameer Murthy, Natalie Paquette and Max Zimet for helpful discussions. We are grateful to Boris Pioline for pointing out some mistakes in a first version of the paper. We thank Murad Alim and Jie Zhou for bringing the paper \cite{Alim2014} to our attention and for interesting discussions about possible connections with our work.
DP thanks SITP Stanford and both of us thank the Perimeter Institute for hospitality while this work was being finalised.

\medskip

\noindent {\it This research was supported in part by Perimeter Institute for Theoretical Physics. Research at Perimeter Institute is supported by the Government of Canada through Industry Canada and by the Province of Ontario through the Ministry of Economic Development \& Innovation.}

%\paragraph{Note added.} This is also a good position for notes added
%after the paper has been written.

\appendix

\appendix

\section{Non-geometric CHL models: general case}\label{s:nongeom}

In this section, we generalize the analysis of CHL models to the most general case, where the level matching condition for a $g$-twisted sector is not satisfied.

\subsection{Lattice of purely electric charges}\label{s:pureelectgen}

The calculation lattice of purely electric charges is similar to the $\lambda=1$ case. In this case, $\delta=\frac{u}{\hat N}$ is a shift of order $\hat N=N\lambda$. In the untwisted sector, the lattice of electric charges before the projection onto $g$-invariant states is again
\be(\Gamma_{K3}^g)^*\oplus \langle u,u^*\rangle \oplus \Gamma_{\tilde S^1}\ .
\ee However, in this case $\hat g$ acts on a vector with winding-momentum $p=nu+ku^*\in \Gamma_{S^1}$ along $S^1$ by multiplication by  a $\hat N$-th root of unity
\be \hat g|nu+ku^*\rangle=e^{\frac{2\pi i k}{N\lambda}}|nu+ku^*\rangle\ ,\ee
while the $\hat g$-eigenvalues on  oscillators and winding-momenta along $\Gamma_{K3}$ are $N$-th roots of unity. It follows that the $\hat g$ invariant states the units of momenta $k$ along $S^1$ must be multiple of $\lambda$, so that the lattice of electric charges in the untwisted sector is
\be\label{untwproj2} \Lambda_e^{U}=(\Gamma_{K3}^g)^*\oplus \langle u,\lambda u^*\rangle \oplus \Gamma_{\tilde S^1}\ .
\ee 
The electric-charges in the $\hat g$-twisted sector, \emph{before} the projection over the $\hat g$-invariant states, form a translate of the lattice \eqref{untwproj2} by $\delta=u/ \hat N$. 
%After projecting onto the $\hat g$ invariant subspace, we expect only a subset of these charges to survive, and this subset should form a translate of the lattice \eqref{untwproj}, in order to have a consistent OPE between untwisted and twisted states.
%The reason is that the space of $\hat g$-invariant $\hat g$-twisted states should be an irreducible module over the algebra of $g$-invariant untwisted fields.  
The symmetry $\hat g$ acts by $e^{2\pi i(L_0-\bar L_0)}=e^{2\pi i\Delta E_g}$ on the $\hat g$-twisted ground state, where $\Delta E_g\in \frac{1}{N\lambda}\ZZ$ satisfies (see eq.\eqref{Elevel}) $\Delta E_g\equiv \frac{\E_g}{N\lambda}\mod \frac{1}{N}\ZZ$
%\be\label{levelmismatch} \Delta E_g\equiv \frac{\E_g}{N\lambda}\equiv  \frac{1}{24N}\sum_{a|N}m(N/a)a\mod \frac{1}{N}\ZZ\ ,
%\ee 
for a suitable $\E_g\in \ZZ/\lambda\ZZ$ coprime to $\lambda$.
Then the $\hat g$-invariant $\hat g$-twisted states span a set of charges
\be\label{twproj} \Lambda_e^{\hat g-\text{twisted}}=\delta-\E_g u^*+ \Lambda_e^{U}=\frac{u}{N\lambda}-\E_gu^*+ \Lambda_e^{U}\ .
\ee 
By taking the OPE of $\hat g$-invariant $\hat g$-twisted fields, we obtain the full lattice of purely electric charges in the general case
%\be\label{eleclatt} \Lambda_e=(\Gamma_{K3}^g)^*\oplus \langle \frac{u}{N\lambda}+\E_g u^*, \lambda u^*\rangle \oplus \Gamma_{\tilde S^1}\ .
%\ee 
\be\label{electLatt} \Lambda_e=(\Gamma_{K3}^g)^*\oplus \langle \frac{u}{N\lambda}-\E_gu^*, \lambda u^*\rangle \oplus \Gamma_{\tilde S^1}\ .
\ee  
The Dirac quantization condition requires the lattice $\Lambda$ of purely magnetic charges to be contained in the dual $\Lambda_e^*$ of the purely electric lattice.    Based on the experience with the $\lambda=1$ case, one would naively expect an equality
\be\label{Lambdamnaive} \Lambda_m=\Lambda_e^*= \Gamma_{K3}^g\oplus \langle N\lambda u^*,\frac{u}{\lambda}+N\E_gu^*\rangle \oplus \Gamma_{\tilde S^1}\qquad \text{(naive)}\ .
\ee The argument suggesting that $\Lambda_m$ must contain a primitive sublattice $\Gamma_{K3}^g$ is still valid for $\lambda>1$. The inclusion of the summand $\langle N\lambda u^*,\frac{u}{\lambda}+N\E_gu^* \rangle \oplus \Gamma_{\tilde S^1}$, however, is more subtle. String-string duality exchanges NS5-branes wrapping $S^1\times K3$ (or its orbifold) with fundamental strings winding $S^1$. Via this duality, eqs.\eqref{electLatt} and \eqref{Lambdamnaive} provide the lattice of winding and momenta around $S^1$ and $\tilde S^1$ for the type IIA fundamental string in the orbifold. One can check that this lattice is compatible with an orbifold  by a symmetry $\hat g=(\delta,g)$ in the type IIA frame only if the $g$-twisted sector of the K3 sigma model satisfies the level-matching condition. However, the only case where the level-mismatch of an element with $\lambda>1$ has been computed in the type IIA picture  ($g$ of Frame shape $2^{12}$) seems to contradict this conclusion. This lattice of magnetic charges in this CHL, therefore, cannot be described by \eqref{Lambdamnaive}. Furthermore, the non-perturbative type IIA charges seem to break the perturbative T-duality. These issues suggest that the naive guess \eqref{Lambdamnaive} might be wrong and that the lattice of electric magnetic charges is more complicated in this case. 

The adiabatic argument and the analysis of perturbative states in type IIA model seem to suggest that string-string duality and type IIA S-duality should hold also in the case $\lambda>1$. In section \ref{s:genlatticeem}, we will therefore \emph{assume} that these dualities hold and derive the most general form of the lattice $\Lambda_{em}$ compatible with these assumptions.

\subsection{Heterotic S-duality and T-duality}\label{s:STdualGen}

The properties of the CHL models under the T-dualities in type IIA and heterotic frame, described in section \ref{s:Sdual}, generalize to the case where $\lambda>1$. The heterotic and type IIA frame are very similar and can be treated together. Let us generically denote by $\C$ the internal CFT in a six dimensional compactification of string theory. For heterotic strings compactified on $T^4$, $\C$ is a $\N=(4,0)$ superconformal field theory with central charges $(c,\tilde c)=(6,16)$. In type IIA compactified on K3, $\C$ denotes a $\N=(4,4)$ SCFT with central charges $(c,\tilde c)=(6,6)$. Let us consider a symmetry $g\in O(\Gamma_{K3})$ of $\C$, preserving the superconformal algebra, such that the level-matching condition for the $g$-twisted sector is not satisfied. We define the four dimensional CHL model as the orbifold of our string theory (heterotic or type IIA) on $\C\times S^1\times \tilde S^1$ by a symmetry $\hat g=(g,\delta)$, where $\delta$ has order $\hat N=N\lambda$. Once again, $R$ and $\tilde R$ are the radii of $S^1$ and $\tilde S^1$ and, for simplicity, we consider a point in the moduli space where the K\"ahler and the complex moduli of the torus $S^1\times\tilde S^1$ are 
\be T=i\frac{R\tilde R}{N\lambda}\qquad U=i\frac{R}{N\lambda\tilde R}
\ee and there are no Wilson line along $S^1\times \tilde S^1$. 

The left- and right-moving momenta along $S^1$ are quantized as (see appendix \ref{s:pureelectgen})
\be (p_L,p_R)=\frac{1}{\sqrt2}\Bigl(\frac{k\lambda-n\E_g}{R}+\frac{nR}{N\lambda },\frac{k\lambda-n\E_g}{R}-\frac{nR}{N\lambda }\Bigr)\ ,\qquad n,k\in\ZZ
\ee  and are tensored with states in the $\hat g^n$-twisted sector of $\C$ with $\hat g$-eigenvalue $e^{2\pi i\frac{k\lambda-n\E_g}{N\lambda}}$. In the limit $R,\tilde R\to \infty$ we recover the original six dimensional theory with internal CFT $\C$. The T-dual picture is obtained by
\be R\to R'=\frac{N\lambda }{R}\qquad \begin{pmatrix}
k\\ n
\end{pmatrix} \to \begin{pmatrix}
k'\\ n'
\end{pmatrix}=\begin{pmatrix}
\E_g & \frac{1-\E_g^2}{\lambda}\\ \lambda & -\E_g
\end{pmatrix}\begin{pmatrix}
k\\ n
\end{pmatrix}
%\begin{cases} m'=-m\E_g+n\frac{1-\E_g^2}{\lambda}\\ n'=m\lambda+n\E_g\end{cases}
\ee (together with the analogous transformation along $\tilde S^1$), so that
\be (p_L,p_R)\to (p'_L,p'_R)=(p_l,-p_R)\ ,
\ee as appropriate for a T-duality.
Since $\lambda|24$ and $(\E_g,\lambda)=1$, it follows that $\E_g^2\equiv 1\mod \lambda$ for all $g$, so that $k',n'$ are integers, as expected.
 
  In the limit $R'\to\infty$, local excitations satisfy
\be\label{orbiproj} n'=k\lambda-n\E_g=0\ ,
\ee so that the limit enforces a projection onto the $\hat g$-invariant states. In fact, since $\E_g$ and $\lambda$ are coprime, the condition \eqref{orbiproj} can only be satisfied if $n$ is a multiple of $\lambda$. Define $a=0,\ldots,N/\lambda-1$ and $b=0,\ldots, \lambda-1$, such that 
\be n\equiv a\lambda+bN\mod N\lambda\ ,\ee 
and denote by $\Hh_n$ the $g^n$-twisted sector of the internal CFT $\C$.   Since $g$ has order $N$ on the original untwisted theory $\C$, the $\hat g^{a\lambda+bN}$-twisted sectors, for $a$ fixed and $b=0,\ldots, \lambda-1$, are all isomorphic as vector spaces
\be\label{twistisom} \Hh_{a\lambda}\stackrel{\cong}{\longrightarrow} \Hh_{a\lambda+N}\stackrel{\cong}{\longrightarrow} \ldots\stackrel{\cong}{\longrightarrow} \Hh_{a\lambda+bN}\stackrel{\cong}{\longrightarrow} \ldots\stackrel{\cong}{\longrightarrow}\Hh_{a\lambda+N(\lambda-1)}\ .
\ee One can think of this isomorphism as being generated by the OPE with the vertex operator $\V_N$ associated with the unique $g^N$-twisted ground state, which has conformal dimension $0$. 
However, the action of $g$ on the twisted sectors $\Hh_{a\lambda+bN}$ depends on \emph{both} $a$ and $b$. To see this, first notice that by definition $g^{N+1}$ acts by $e^{2\pi i(N+1)(L-\bar L_0)}$ on the $g$-twisted sector $\Hh_1$, while it acts by $e^{2\pi i(L-\bar L_0)}$ on $\Hh_{N+1}$. Since on $\Hh_1$ and $\Hh_{N+1}$ the eigenvalues of $L_0-\bar L_0$ take value in $\frac{\E_g}{N\lambda}+\frac{1}{N}\ZZ$, and since the OPE by $\V_N$ maps states of $\Hh_1$ to states of the same spin on $\Hh_{N+1}$, we obtain that $g^{N+1}(\V_N)=e^{-\frac{2\pi i\E_g}{\lambda}}\V_N$. Since $g$ has order $N\lambda$ on the twisted sectors, it follows that $g^\lambda(\V_N)=g^{(N+1)\lambda}(\V_N)=\V_N$ and since $\lambda$ divides $N$ we finally get
\be g(\V_N)=e^{-\frac{2\pi i\E_g}{\lambda}}\V_N\ .
\ee
  Thus, if we denote by $\Hh_n(g=\zeta)$ the $g$-eigenspace of $\Hh_n$ with $g$-eigenvalue $\zeta$, we have
\be \Hh_{a\lambda}(g=\zeta)\stackrel{\cong}{\longrightarrow} \ldots\stackrel{\cong}{\longrightarrow} \Hh_{a\lambda+bN}(g=e^{-\frac{2\pi i b\E_g}{\lambda}}\zeta)\stackrel{\cong}{\longrightarrow} \ldots\stackrel{\cong}{\longrightarrow}\Hh_{a\lambda+N(\lambda-1)}(g=e^{-\frac{2\pi i (\lambda-1)\E_g}{\lambda}}\zeta)\ .
\ee Therefore, the direct sum of the $g$-invariant subspaces of the $g^n$-twisted sectors is
\be \bigoplus_{n=0}^{\hat N-1} \Hh_n(g=1)\cong \bigoplus_{a=0}^{N/\lambda-1}\Bigl(\bigoplus_{b=0}^{\lambda-1} \Hh_{a\lambda}(g=e^{\frac{2\pi i \E_g b}{\lambda}})\Bigr)\cong \bigoplus_{a=0}^{N/\lambda-1} \Hh_{a\lambda}(g^\lambda=1)\ .
\ee The latter space is the spectrum of the orbifold $\C'$ of $\C$ by the symmetry $g^\lambda$. This is a consistent SCFT with the same world-sheet supersymmetry as $\C$; in particular, the level-matching is satisfied (while the orbifold by $g$ is inconsistent).  Thus, the local physics in this case is described by type IIA (or heterotic) superstring theory compactified on the internal CFT $\C'$. 

As we move along $S'^1$ by a shift of $2\pi R'/(N\lambda)$, a state with zero winding $n'=0$ and momentum $k'=n/\lambda$ picks up a phase $e^{\frac{2\pi i k'\lambda}{N\lambda}}=e^{\frac{2\pi i n/\lambda}{N}}$. This is equivalent to a symmetry $g'$ of order $N$ in the orbifold CFT $\C'$, acting by $e^{\frac{2\pi i a}{N}}g^{\E_g^{-1}}$ on the $g^{a\lambda}$-twisted sector ($\E_g^{-1}\in \ZZ/\lambda\ZZ$ is determined by $\E_g\E_g^{-1}\equiv 1\mod \lambda$). Therefore, the model is actually a type IIA (or heterotic) string theory compactified on  $\frac{\C'\times {S'}^1}{\langle g'\rangle}\times \tilde {S'}^1$.

For type IIA T-duality (i.e., heterotic S-duality), if $\C'$ is a K3 model and $g'$ is in the same duality class as the original $g$, the corresponding CHL is self-dual. In fact, one can show (see appendix \ref{a:quantsymm}) that whenever $\lambda>1$, the Frame shape of $g$ is balanced with balancing number $\hat N$ and the model is self-dual. 

In the heterotic case, as explained in appendix \ref{a:hetTdual}, the T-dual theory is the same CHL model, possibly at a different point in the moduli space. 

The effect of these dualities on non-perturbative states and the lattice of electric-magnetic charges is considerably more complicated for $\lambda>1$ and is considered in appendix \ref{s:genlatticeem}.

\subsection{Witten index of the quantum symmetry}

\label{a:quantsymm}

Let $\C$ be a non-linear sigma model on K3, $g$ be a symmetry preserving the $\N=(4,4)$ superconformal algebra and let us assume that the orbifold theory $\C'=\C/\langle g\rangle$ is consistent, i.e. the twisted sectors satisfy the level-matching condition ($\lambda=1$).  In this section, we compute the twined Witten index of the quantum symmetry $\Q$.
By definition of $\Q$, this is given by
\be \I^{\C'}_{\Q^k}=\sum_{t=1}^N e^{\frac{2\pi i kt}{N}} {\rm sdim}(\Hh^{\langle g\rangle}_{g^t})\ ,
\ee where $\Hh_{g^t}$ denotes the space of $g^t$-twisted Ramond-Ramond ground states (i.e. with conformal weights $(\frac{1}{4},\frac{1}{4})$), $\Hh^{\langle g\rangle}_{g^t}$ is its $g$-invariant subspace and  ${\rm sdim}\Hh^{\langle g\rangle}_{g^t}$ is the superdimension ($\ZZ_2$-graded by $(-1)^{F+\tilde F}$)  of $\Hh^{\langle g\rangle}_{g^t}$. By the standard orbifold construction,  ${\rm sdim}\Hh^{\langle g\rangle}_{g^t}$ is given by
\be {\rm sdim}\Hh^{\langle g\rangle}_{g^t}=\frac{1}{N}\sum_{v=1}^N \I^{\C}_{g^t,g^v}\ ,
\ee where 
\be \I^{\C}_{g^t,g^v}:=\Tr_{RR,\Hh_{g^t}}(g^v(-1)^{F+\tilde F}q^{L_0-\frac{c}{24}}\bar q^{\bar L_0-\frac{c}{24}})\ ,
\ee is the $g^t$-twisted $g^v$-twined Witten index. The index $\I^{\C}_{g^t,g^v}$ is a specialization to $z=0$ of the twisted twining elliptic genus $\phi_{g^t,g^v}(\tau,z)$. In turn, the twisted-twining elliptic genus $\phi_{g^t,g^v}(\tau,z)$ can be obtained from the untwisted one $\phi_{1,g^{(t,v,N)}}(\tau,z)$ by a modular transformation. By specializing to $z=0$, the dependence on $\tau$ drops and we have simply the identity
\be \I^{\C}_{g^t,g^v}=\I^{\C}_{g^{(v,t,N)}}\ .
\ee Using \eqref{secondIg}, we can now prove \eqref{Qindex}:
\be \begin{split}
\I^{\C'}_{\Q^k}=&\sum_{t=1}^N e^{\frac{2\pi i kt}{N}}\frac{1}{N}\sum_{v=1}^N \I^{\C}_{g^t,g^v}=\sum_{t=1}^N e^{\frac{2\pi i kt}{N}}\frac{1}{N}\sum_{v=1}^N\I^{\C}_{g^{(v,t,N)}}\\
=& \sum_{t=1}^N e^{\frac{2\pi i kt}{N}}\frac{1}{N}\sum_{v=1}^N\sum_{u|(t,v,N)} m(u)u\\
=&\sum_{t=1}^N e^{\frac{2\pi i kt}{N}}\sum_{u|(t,N)}m(u)\frac{u}{N}\sum_{v=1}^{N}\delta(v \bmod u)\\
=&\sum_{t=1}^N e^{\frac{2\pi i kt}{N}}\sum_{u|(t,N)}m(u)=
\sum_{u|N}m(u)\sum_{t=1}^N e^{\frac{2\pi i kt}{N}}\delta(t\bmod u)\\
=&\sum_{u|N}m(u)\sum_{l=1}^{N/u} e^{\frac{2\pi i kl}{N/u}}=\sum_{u|N}m(u)\frac{N}{u}\delta(k\bmod N/u)\\
=&\sum_{n|(k,N)}m(N/n)n\ .
\end{split}\ee

Finally, let us consider the analysis of the Witten index for case where $\lambda> 1$. In this case, the orbifold of $\C$ by $g$ is not consistent and $\lambda$ is the smallest power such that $g^\lambda$ gives rise to a consistent orbifold. Failure of the level matching condition implies that the twining genus $\phi_g(\tau,z)$ is modular up to a phase. This implies that the twining Witten index $\I^\C_g=\phi(\tau,0)$ must vanish. The same argument applies to $g^k$, for any $k$ with $(\lambda,k)\neq \lambda$
\be\label{Ikvanish} \I^\C_{g^k}=0\qquad \text{if }(\lambda, k)\neq \lambda\ .
\ee As a consequence, a cycle of length $a$ in the cycle shape of $g$ can have non-zero multiplicity  $m(a)\neq 0$ only if $a$ is a multiple of $\lambda$. To show this, consider by absurd that there is some cycle of length $b$ such that $m(b)\neq 0$ and $(b,\lambda)<\lambda$. Without loss of generality, we can assume that $b$ is the smallest such cycle. Then, we have the contradiction
\be  0=\I^\C_{g^b}=\sum_{u|(b,N)} u m(u)=bm(b)\ ,
\ee where the last equality follows because all $u<b$ with $m(u)\neq 0$ are multiple of $\lambda$ and therefore cannot divide $b$. 

Let $\prod_{a|\frac{N}{\lambda}} a^{\tilde m(a)}$ be the cycle shape of $g^\lambda$. Since all cycle lengths in the cycle shape of $g$ are multiple of $\lambda$, we have the relation
\be\label{tildea} \tilde m(a)=\lambda m(\lambda a)\ ,
\ee which allows us to reconstruct the cycle shape of $g$ when the cycle shape of $g^\lambda$ is known. Since the orbifold by $g^\lambda$ is consistent, then \eqref{consorbif} holds and
\be \I^{\C'}:= \sum_{a|\frac{N}{\lambda}}\frac{N}{\lambda a}\tilde m(a)\in  \{0,24\}
\ee where $\I^{\C'}$ is the Witten index of the orbifold $\C'=\C/\langle g^\lambda \rangle$. Using \eqref{tildea} we have
\begin{align}
\sum_{a|\frac{N}{\lambda}}\frac{N}{\lambda a}\tilde m(a)= \sum_{a|\frac{N}{\lambda}}\frac{N}{\lambda a} \lambda m(\lambda a)=\lambda \sum_{b|N}\frac{N}{b}m(b).
\end{align} so that
\be \sum_{b|N}\frac{N}{b}m(b)=\frac{\I^{\C'}}{\lambda}\ .
\ee By assumption, \eqref{consorbif} is not satisfied by $g$, so that $\I^{\C'}\neq 0$, $\C'$ is a K3 model and
\be \sum_{b|N}\frac{N}{b}m(b)=\frac{24}{\lambda}\ .
\ee
As explained in appendix \ref{s:STdualGen}, we are interested in a symmetry $g'$ in the orbifold theory $\C'$ that acts by
\be e^{\frac{2\pi i a}{N}}g^{\E_g^{-1}}\ ,
\ee on the $g^\lambda$-invariant $g^{\lambda a}$-twisted sector $\Hh_{g^{a\lambda}}^{\langle g^\lambda\rangle}$. Here, $\E_g^{-1}\in \ZZ/\lambda\ZZ$ is determined by $\E_g^{-1} \E_g\equiv 1\mod \lambda$. Such an element always exist because $(\E_g,\lambda)=1$. Furthermore, $g$ has order $\lambda$ on $\Hh_{g^{a\lambda}}^{\langle g^\lambda\rangle}$, so that $g^{\E_g^{-1}}$ is well-defined. The standard orbifold formula gives
\be I_{g'^k}^{\C'}=\sum_{a=1}^{N/\lambda} e^{\frac{2\pi i a k}{N}}\frac{\lambda}{N}\sum_{v=1}^{N/\lambda}\I^{\C}_{g^{\lambda a},g^{\lambda v+\E_g^{-1}k}}\ .
\ee By \eqref{Ikvanish}, $I_{{g'}^k}^{\C'}$ vanishes unless $k$ is a multiple of $\lambda$, i.e.
\be I_{g'^k}^{\C'}=0\qquad \text{if }(\lambda, k)\neq \lambda\ .
\ee By the same reasoning as for $g$, we obtain that the cycle shape $\prod_{a|N} a^{m'(a)}$ of $g'$ contains only cycles with length multiple of $\lambda$ and that the cycle shape $\prod_{a|N} a^{\tilde m'(a)}$ of ${g'}^\lambda$  is given by the analogue of \eqref{tildea}
\be \tilde m'(a)=\lambda m'(\lambda a)\ .
\ee The twining indices $\I^{\C'}_{g'^{j\lambda}}$, $j=1,\ldots,N/\lambda$ are given by
\be I_{g'^{j\lambda}}^{\C'}=\sum_{a=1}^{N/\lambda} e^{\frac{2\pi i a j\lambda}{N}}\frac{\lambda}{N}\sum_{v=1}^{N/\lambda}\I^{\C}_{g^{\lambda (a,v+\E_g^{-1}j,N/\lambda)}}=\sum_{a=1}^{N/\lambda} e^{\frac{2\pi i a j\lambda}{N}}\frac{\lambda}{N}\sum_{v=1}^{N/\lambda}\I^{\C}_{g^{\lambda (a,v,N/\lambda)}}\ .
\ee
By the same computation as in the $\lambda=1$ case, we obtain
\be I_{g'^{j\lambda}}^{\C'}=\sum_{n|(j,N/\lambda)}\tilde m(\frac{N}{\lambda n})n\ .\ee
This implies that
\be \tilde m'(n)=\tilde m(\frac{N}{\lambda n})\ ,
\ee that is
\be m'(\lambda n)=m(N/n)\ ,
\ee or, equivalently,
\be m'(a)=m(\frac{N\lambda}{a})\ .
\ee

\subsection{Lattice of electric-magnetic charges}\label{s:genlatticeem}

In this section, we determine the lattice of electric-magnetic charges in a CHL model with $\lambda>1$ under the assumption that string-string duality and heterotic S-duality (or, equivalently, type IIA T-duality along $S^1\times \tilde S^1$) hold. As stressed in section \eqref{s:pureelectgen}, one expects the lattice of purely magnetic charges $\Lambda_m$ to contain a summand of the form $\Gamma_{K3}^g$, corresponding to the $\hat g$-invariant states of the original unorbifolded theory. By \eqref{electLatt} and by the Dirac quantization condition, this is the finest possible lattice of magnetic charges with respect to the $U(1)^{4+d}$ gauge group associated with the gauge fields descending from the  ${4+d}$ six dimensional vector multiplets. 

Let us focus on the lattice of electric and magnetic charges with respect to the remaining $U(1)^4$ gauge fields, associated with the metric and B-field along $S^1\times \tilde S^1$.

As explained in section \ref{s:pureelectgen}, the analysis of the spectrum of perturbative heterotic string leads to a lattice of purely electric charges of the form
\be\label{hetlattice} \begin{pmatrix}
 m \\
 \tilde m\\
 w\\
 \tilde w\end{pmatrix}\in\{\begin{pmatrix}
 \lambda & -\E_g & 0 & 0\\
 0 & 0 & 1 & 0\\ 
 0 & 1/N\lambda & 0 & 0\\
 0 & 0 & 0 & 1
\end{pmatrix}\begin{pmatrix}
 a_1 \\
 a_2\\
 a_3\\
 a_4\end{pmatrix},\ a_1,a_2,a_3,a_4\in\ZZ\}\ ,
\ee with standard quadratic form \eqref{quadform}. Here, $\E_g/N\lambda\mod \frac{1}{N}\ZZ$ is the level-mismatch \eqref{Elevel} for a heterotic string in the $g$-twisted sector, where $g\in O(\Gamma_{K3})$ is the pure rotational part of $\hat g=(\delta,g)$. Without loss of generality, we can assume that $\E_g\in \ZZ/\lambda \ZZ$ and $\lambda$ be coprime; since $\lambda$ is always a divisor of $24$, this is equivalent to
\be\label{coprime2} \gcd(\E_g,\lambda)=1\qquad \Leftrightarrow \qquad \E_g^2\equiv 1\mod \lambda\ .
\ee
Similarly, in type IIA, the momenta and winding along the torus $S^1$ and $\tilde S^1$ are quantized as follows
\be\label{IIAlattice} \begin{pmatrix}
 m' \\
 \tilde m'\\
 w'\\
 \tilde w'\end{pmatrix}\in\{\begin{pmatrix}
 \lambda & -\E'_g & 0 & 0\\
 0 & 0 & 1 & 0\\ 
 0 & 1/N\lambda & 0 & 0\\
 0 & 0 & 0 & 1
\end{pmatrix}\begin{pmatrix}
 a_1' \\
 a_2'\\
 a_3'\\
 a_4'\end{pmatrix},\ a_1',a_2',a_3',a_4'\in\ZZ\}
\ee
where, $\E'_g/N\lambda\mod \frac{1}{N}\ZZ$ is the level-mismatch for a type IIA $g$-twisted sector in a non-linear sigma model on K3. We do \emph{not} assume, a priori, that $\E_g'$ and $\lambda$ are coprime -- we will derive this property as a consequence of string-string duality, T-duality and S-duality.

  Let $m$, $\tilde m$, $w$, $\tilde w$, $M$, $\tilde M$, $W$, $\tilde W$ denote, respectively, momentum around $S^1$, and $\tilde S^1$, winding around $S^1$ and $\tilde S^1$, NS5 brane wrapping $\tilde S^1$ and $S^1$ and KK monopole with asymptotic circle $S^1$ and $\tilde S^1$ in the heterotic string frame. Let us impose the Dirac quantization condition and assume that these charges be compatible with string-string duality and heterotic T-duality along $S^1\times \tilde S^1$ (corresponding to S-duality in the type IIA frame). We find that the most general lattice of electric-magnetic charges satisfying these constraints and correctly reproducing the perturbative spectra \eqref{hetlattice} and \eqref{IIAlattice} has the following structure
\be\label{hetlattice2} {}^t\begin{pmatrix}
 m &
 \tilde m&
 w&
 \tilde w&
 \vline &
 M&
 \tilde M&
 W&
 \tilde W
\end{pmatrix}\in \Bigl\{\M_{N,\lambda,\E_g,\E_g',y}\,
  v,\ v\in\ZZ^8\Bigr\}
\ee 
 where $\M_{N,\lambda,\E_g,\E_g',y}$ is the $8\times 8$ matrix
\be\label{genemlattice} \M_{N,\lambda,\E_g,\E_g',y}:=\begin{pmatrix}
 \lambda & -\E_g & 0 & 0 &\vline & 0 & -\E_g' & y & 0\\
 0 & 0 & 1 & 0  &\vline & 0 & 0 &-\E_g\E_g'/\lambda & 0\\ 
 0 & 1/N\lambda & 0 & 0 &\vline & 0 & 0& 0& -\E_g'/N\lambda\\
 0 & 0 & 0 & 1&\vline & 0 & 0 & -\E_g'/\lambda & 0\\ \hline
 0 & 0 & 0 & 0 &\vline & N\lambda & 0 & \E_g N & 0\\
 0 & 0 & 0 & 0 &\vline & 0 & 1 & 0 & 0\\
 0 & 0 & 0 & 0 &\vline & 0 & 0 & 1/\lambda & 0\\
 0 & 0 & 0 & 0 &\vline & 0 & 0 & 0 & 1
\end{pmatrix}\ .
\ee The constraints that we are imposing are not sufficient to fix the parameter  $y\in\ZZ/\lambda\ZZ$. In order to determine its precise value a more detailed analysis of non-perturbative string states is necessary.

The analogous type IIA lattice of winding, momenta, NS5 branes and KK monopoles around $S^1$ and $\tilde S^1$ has exactly the same form with $\E_g$ and $\E_g'$ exchanged:
\be\label{IIAlattice2} {}^t\begin{pmatrix}
 m' &
 \tilde m'&
 w'&
 \tilde w'&
 \vline &
 M'&
 \tilde M'&
 W'&
 \tilde W'
\end{pmatrix}\in \Bigl\{\M_{N,\lambda,\E_g',\E_g,y}\,
  v',\ v'\in\ZZ^8\Bigr\}\ .
\ee

%Here, $\E_g\in \ZZ/\lambda\ZZ$ is the level mismatch in the heterotic string frame, $\hat \E_g\in \ZZ/\lambda\ZZ$ is the mismatch in the type IIA frame and both of them must be coprime to $\lambda$. 
 
The derivation of the matrix \eqref{genemlattice} is an easy but tedious exercise. We will skip the details, but just show that, a posteriori,  the lattice above is compatible with all expected dualities. This can be checked by comparing the BPS mass formulas
\be\label{BPSmass} M^2=\frac{|\tilde mU+\frac{m}{N\lambda}+T(N\lambda U w-\tilde w)+S(\tilde MU+\frac{M}{N\lambda})+ST(N\lambda UW-\tilde W)  |^2}{4U_2T_2S_2}
\ee for the heterotic and the type IIA frame.

The BPS mass formula is invariant under  string-string duality acting on the moduli by
\be  T'= S\qquad S'=T\qquad U'=U\ ,
\ee provided that the lattices \eqref{hetlattice2} and \eqref{IIAlattice2} are identified via the $GL(8,\ZZ)$ transformation
\be v\mapsto v'=\begin{pmatrix}
1 & 0 & 0 & 0 & 0 & 0 & 0 & 0 \\
 0 & 0 & 0 & 0 & 0 & 1 & 0 & \E_g \\
 0 & 0 & 1 & 0 & 0 & 0 & 0 & 0 \\
 0 & 0 & 0 & 0 & -1 & 0 & 0 & 0 \\
 0 & 0 & 0 & -1 & 0 & 0 & 0 & 0 \\
 0 & 1 & 0 & 0 & 0 & 0 & 0 & -\E_g' \\
 0 & 0 & 0 & 0 & 0 & 0 & 1 & 0 \\
 0 & 0 & 0 & 0 & 0 & 0 & 0 & 1
\end{pmatrix}v\ .
\ee 
The mass formula is also compatible  with heterotic T-duality (type IIA S-duality). In order to preserve the spectrum of perturbative heterotic states, the action of T-duality on the K\"ahler and complex moduli of $S^1\times \tilde S^1$ must be
\be T\to -\frac{1}{N\lambda T}\qquad U\to -\frac{1}{N\lambda U}\qquad S\to S\ .
\ee This transformation must be accompanied by an automorphism of the electric magnetic lattice such that the mass formula is invariant:
 \be
v\rightarrow
\begin{pmatrix}
\E_g  & \frac{1-\E_g ^2}{\lambda} & 0 & 0 & -y & 0 & 0 & 0 \\
 \lambda & -\E_g  & 0 & 0 & 0 & 0 & y & 0 \\
 0 & 0 & 0 & 1 & \E_g  {\E_g'}  & 0 & \frac{{-\E_g'} (1-\E_g ^2) }{\lambda} & 0 \\
 0 & 0 & 1 & 0 & {\E_g'}  & 0 & 0 & 0 \\
 0 & 0 & 0 & 0 & -\E_g  & 0 & \frac{1-\E_g ^2}{\lambda} & 0 \\
 0 & 0 & 0 & 0 & 0 & 0 & 0 & 1 \\
 0 & 0 & 0 & 0 & \lambda & 0 & \E_g  & 0 \\
 0 & 0 & 0 & 0 & 0 & 1 & 0 & 0
\end{pmatrix}v\ .
\ee
This is a $GL(8,\ZZ)$ transformation, thanks to \eqref{coprime2}.

The assumptions of string-string duality and heterotic T-duality are sufficient to fix the form of the electric-magnetic lattice as a functions of the three parameters $\E_g,\E_g',y$, with the constraint $\E_g^2\equiv 1\mod \lambda$. Let us now impose heterotic S-duality (or, equivalently, type IIA T-duality).
One can check that, in order to preserve the BPS mass formula, one needs to  act on the moduli via
\be T\to T\qquad U\to -\frac{1}{N\lambda U}\qquad S\to -\frac{1}{N\lambda S}\ .
\ee
and on the charges by
\be v\rightarrow
\begin{pmatrix}
 {\E_g'} & 0 & 0 & y & 0 & \frac{1-{\E_g'}^2}{\lambda} & 0 &
   \frac{{-\E_g} ({\E_g'}^2-1)}{\lambda} \\
 0 & {\E_g'} & 0 & 0 & 0 & 0 & 0 & 1-{\E_g'}^2 \\
 0 & 0 & 0 & -{\E_g} {\E_g'} & -1 & 0 & \frac{-{\E_g}(1-
   {\E_g'}^2)}{\lambda} & 0 \\
 0 & 0 & 0 & {-\E_g'} & 0 & 0 & \frac{{\E_g'}^2-1}{\lambda} & 0 \\
 0 & 0 & -1 & {\E_g} & 0 & 0 & 0 & 0 \\
 \lambda & -{\E_g} & 0 & 0 & 0 & -{\E_g'} & y & 0 \\
 0 & 0 & 0 & -\lambda & 0 & 0 & {\E_g'} & 0 \\
 0 & 1 & 0 & 0 & 0 & 0 & 0 & -{\E_g'}
\end{pmatrix}v\ .
\ee
This is a $GL(8,\ZZ)$ transformation if and only if
\be\label{coprime} {\E'_g}^2\equiv  1\mod \lambda\ .
\ee
We conclude that, upon imposing string-string duality, T-duality and S-duality, the level mismatch in the $g$-twisted sector in the heterotic and type IIA frames must be related
\be \gcd(\E_g,\lambda)=1\qquad \Leftrightarrow \qquad \gcd(\E_g',\lambda)=1\ .
\ee

The physical interpretation of the lattice \eqref{genemlattice} is not obvious. The form of the upper right block in \eqref{genemlattice} suggests that (in the heterotic frame) each unit of NS5 brane wrapping $S^1$ (respectively, $\tilde S^1$), must carry $ \E_g'$ units of momentum along $S^1$ (respectively, $\E_g'/N\lambda$ units of momentum along $\tilde S^1$). Furthermore (and compatibly with T-duality), each unit of KK monopole with asymptotic circle $S^1$ (respectively, $\tilde S^1$), carries also $\E_g'$ units of winding along $\tilde S^1$ (respectively, $ \E_g'/N\lambda$ units of winding along $S^1$). A similar observation holds in the type IIA picture, with $ \E_g'$ replaced by $\E_g$.  In general, it is known that, in type IIA, a NS5-brane wrapping a non-trivial 4-manifold (such as K3) times a circles $S^1$, must carry also a `geometrically induced' momentum charge along $S^1$ (by a chain of dualities, this is related to the D1-brane charge geometrically induced  on a type IIB D5 brane wrapping K3). In this case, the four manifold is an asymmetric orbifold of a K3 sigma model (or, in the heterotic picture, an asymmetric orbifold of a compactification on $T^4$), so that the precise computation of the induced charges is more complicated. This interpretation suggests that the parameter $y$ should vanish.

\medskip

Let us now consider the electric and magnetic charges under the full $U(1)^{8+d}$ gauge group. From \eqref{genemlattice}, it follows that the sublattice $\Lambda_m$ of purely magnetic charges is
\be \Lambda_m= \Gamma_{K3}^g\oplus \langle N\lambda u^*,u+N\lambda\E_gu^* \rangle \oplus \langle \lambda \tilde u^*,\tilde u+\E_g\tilde u^*\rangle\ ,
\ee where each vector $u,u^*,\tilde u,\tilde u^*$ represent, respectively, one unit of magnetic charges $W,M,\tilde W,\tilde M$.
This is a proper sublattice of the lattice $\Lambda_e^*$ thus showing that, as anticipated in section \ref{s:pureelectgen}, the naive guess \eqref{Lambdamnaive} is incompatible with all expected dualities. As a consequence, the full lattice $\Lambda_{em}$ is finer than the direct sum $\Lambda_e\oplus\Lambda_m$, i.e. that there are dyonic charges that cannot be written as sums of purely electric and purely magnetic. The last three columns of \eqref{genemlattice} are examples of such charges.
The full lattice of electric magnetic charges is given by
\be \Lambda_{em}=\M_{N,\lambda,\E_g,\E_g',y}\ZZ^8\oplus (\Gamma_{K3}^g)^*\oplus \Gamma_{K3}^g\subset \Lambda_e^*\oplus\Lambda_m^*\ ,
\ee where $\Gamma_{K3}^g\subset \Lambda_m$ is purely magnetic and its dual $(\Gamma_{K3}^g)^*\subset \Lambda_e$ is purely electric. The summand $\M_{N,\lambda,\E_g,\E_g',y}\ZZ^8\subset \Lambda_{em}$ is invariant under heterotic S-duality by construction. Notice that all $g$ such that $\lambda>1$ have balanced Frame shape, with balancing number $\hat N=N\lambda$. Therefore, the corresponding CHL models are self-dual under S-duality and the lattice $\Gamma_{K3}^g$ must be $\hat N$-modular
\be \Gamma_{K3}^g\cong (\Gamma_{K3}^g)^*(\hat N)\ .
\ee As announced in section \ref{s:Nmodular}, this prediction of S-duality can be verified for all possible Frame shapes with $\lambda>1$ (see table \ref{t:case1} in appendix \ref{s:lattices}).

%\subsection{S-duality: general case}\label{a:genSdual}
%
%
%We now describe in detail the heterotic and type IIA T-dualities in the case $\lambda>1$. We adopt the notation of section \ref{s:STdualGen}.
%
%%Finally, let us generalize our discussion to the case where $\lambda\neq 0$, i.e. the orbifold of $\C$ by $g$ is not consistent because the level-matching condition is not satisfied. In this case, we defined the orbifold by a symmetry $\hat g=(g,\delta)$, where $\delta$ has order $\hat N=N\lambda$. 

%\section{Witten index of the quantum symmetry}
%\label{a:Wittenindex}

\section{Heterotic T-duality}\label{a:hetTdual}

In this appendix we show that the orbifold of heterotic strings on $T^4$ by a symmetry $g\in O(\Gamma^{4,20})$, if consistent, is again a compactification of heterotic strings on $T^4$. The orbifold projection preserves the left-moving (supersymmetric) sector and projects out $20-d$ world-sheet currents in the right-moving (bosonic) sector, each one corresponding to a vector multiplet in the six dimensional theory. Thus, we have to show that the same number of world-sheet currents appear in the twisted sectors or, equivalently, that the six dimensional theory has still $24$ vector multiplets.
 Suppose that $g$ has Frame shape $\prod_{a|N} a^{m(a)}$. The untwisted sector contains 
\be\label{vmuntw} 4+d=\sum_{a|N} m(a) 
\ee space-time vector multiplets (we are considering a generic point in the moduli space where the gauge symmetry is not enhanced). More precisely, $4$ of these vector multiplets come from currents in the the left-moving (supersymmetric) side and $d:= \sum_{a|N} m(a)-4$ from currents in the right-moving (bosonic) side. Let us analyse the vector multiplets (or world-sheet currents) from the $g$-twisted sector. The multiplicity of these currents is the $\bar q^0$ coefficient in the anti-holomorphic part of the $g$-twisted  partition function
\be\label{gtwisted} \frac{\prod_{a|N} a^{\frac{m(a)}{2}}}{\sqrt{|(\Gamma_{K3}^g)^*/\Gamma_{K3}^g|}}\frac{1}{\eta_{g,1}(\bar\tau)}=\sqrt{\frac{\det'(1-g)}{|(\Gamma_{K3}^g)^*/\Gamma_{K3}^g|}}\prod_{a|N}\eta(\bar \tau/a)^{-m(a)}\ .
\ee The factor $1/|(\Gamma_{K3}^g)^*/\Gamma_{K3}^g|^{1/2}$ comes from the modular transformation of the theta series of $\Gamma_{K3}^g$, while $\det'(1-g)=\prod_{a|N} a^{m(a)}$ comes from the modular transformation of $\eta_{1,g}(\tau)$ (see \eqref{etagh} for the definition of $\eta_{g,h}$).  Here, $\det'(1-g)$ denotes the product of the non-zero eigenvalues of $1-g$ in the $24$-dimensional representation. In \eqref{gtwisted}, we are assuming that, at a generic point in the moduli space, the theta series of $(\Gamma_{K3}^g)^*$ provides no purely anti-holomorphic contribution.

Let us first consider the case where the Frame shape is balanced, i.e $m(a)=m(N/a)$. In this case, we have
\be \prod_{a|N} a^{m(a)}=\prod_{a|N} (N/a)^{m(a)}=N^{\sum_{a|N}m(a)} \prod_{a|N} a^{-m(a)}\ee
which implies
\be \det\nolimits '(1-g)=N^{\sum_{a|N}m(a)/2}\ .
\ee
Furthermore, heterotic S-duality implies that $\Gamma_{K3}^g$ must be $N$-modular, i.e. $\Gamma_{K3}^g\cong (\Gamma_{K3}^g)^*(N)$ and since its rank is $\sum_{a|N}m(a)$, we obtain
\be |(\Gamma_{K3}^g)^*/\Gamma_{K3}^g|=N^{\sum_{a|N}m(a)}\ .
\ee Therefore, the numerical prefactor in \eqref{gtwisted}, which counts the number of $g$-twisted ground states is simply $1$ for balanced Frame shapes. By expanding
\be
\prod_{a|N}\eta(\bar\tau/a)^{-m(a)}=\bar q^{-\frac{1}{24}\sum_{a|N} \frac{m(a)}{a}}\prod_{a|N}\prod_{n=1}^\infty \frac{1}{(1-\bar q^{\frac{n}{a}})^{m(a)}}=\bar q^{-\frac{1}{N}}(1+m(N)\bar q^{\frac{1}{N}}+\ldots)\ ,
\ee where we used the identity $\sum_{a|N} am(a)=\sum_{a|N} (N/a)\, m(a)=24$, we conclude that the $g$-twisted sector provides exactly $m(N)$ vector multiplets. Let us focus on the case where $N$ is prime. Then, the $N-1$ twisted sectors are all isomorphic to the $g$-twisted sector, each one providing $m(N)$ vector multiplets.
Using \eqref{vmuntw}, the total number of vector multiplets is
\be m(1)+m(N)+(N-1)m(N) =\sum_{a|N}a m(a)=24\ ,
\ee as expected. Thus, both the holomorphic and anti-holomorphic sectors of the theory coincide with heterotic string, so that the only consistent possibility is that the six dimensional $(1,1)$ supergravity theory arises as a compactification of heterotic strings on $T^4$.

If $N$ is not prime, we can decompose the orbifold by $g$ into several steps, where, in each step, we orbifold by a symmetry of prime order. For example, if $g$ has order $4$, we first orbifold by $g^2$, which has order $2$; $g$ induces a symmetry of order $2$ in the orbifold theory and we can divide by this symmetry. After each step, we end up with a compactification of heterotic string on $T^4$, so that the statement must be true also if we orbifold by $g$ directly.

\medskip

Let us consider now the case where the Frame shape is not balanced and $\sum_{a|N}am(N/a)=0$. By the same argument as above, we can restrict ourselves to the case where $N$ is prime. Then
\be \det\nolimits '(1-g)=N^{m(N)}\ ,
\ee so that \eqref{gtwisted} becomes
\be \frac{N^{m(N)/2}}{\sqrt{|(\Gamma_{K3}^g)^*/\Gamma_{K3}^g|}}\bar q^{-\frac{1}{24}\sum_{a|N} \frac{m(a)}{a}}\prod_{a|N}\prod_{n=1}^\infty \frac{1}{(1-\bar q^{\frac{n}{a}})^{m(a)}}=\frac{N^{m(N)/2}}{\sqrt{|(\Gamma_{K3}^g)^*/\Gamma_{K3}^g|}}\bar q^0+\ldots
\ee There are only three Frame shapes of prime order within case 3, namely $1^{-8}2^{16}$, $1^{-3}3^9$ and $1^{-1}5^5$ (see table \ref{t:case3} in appendix \ref{s:lattices}), and a case by case analysis shows that, for all such shapes 
\be \frac{N^{m(N)/2}}{\sqrt{|(\Gamma_{K3}^g)^*/\Gamma_{K3}^g|}}=m(N)\ .
\ee By the same argument as above, we conclude that the orbifold theory must be a compactification of heterotic string on $T^4$.

\medskip

It is not necessary to consider the Frame shapes in case $2$, since their order $N$ is always composite and the nature of the orbifold follows by the properties of the subgroups of $\ZZ_N$ of prime order. 

\medskip

The final step in identifying the T-dual theory with a CHL model is to determine the Frame shape of the quantum symmetry $\Q$ of the orbifold. Once again, it is sufficient to consider the case where $g$ has prime order. The previous construction shows that, among the eigenvalues of $\Q$,   $1$ appears with multiplicity $m(1)+m(N)$ and each primitive root of unity appears with multiplicity $
m(N)$. Thus, the Frame shape of $\Q$ is $1^{m(1)}N^{m(N)}$, which is the same as $g$. We conclude that heterotic T-duality maps each CHL model to itself, possibly at a different point in the moduli space.

\section{Symmetries and lattices}
\label{s:lattices}

The entries in tables \ref{t:case1}, \ref{t:case2}, \ref{t:case3} in this appendix correspond to the $42$ conjugacy classes of the Conway group that fix a sublattice of the Leech lattice $\Lambda$ of rank at least four. Each of these classes corresponds to a CHL model. For each class, we provide the order of $g\in O(\Gamma_{K3})\subset O(\Gamma^{6,22})$, the parameter $\lambda$ determined as in \eqref{Elevel} ($\lambda>1$ denotes a failure of the level matching condition in the $g$-twisted sector), the Frame shape of $g$, and the fixed lattice $\Gamma_{K3}^g$. Our notation is as follows:  $L(n)$ denotes a lattice $L$ with the quadratic form rescaled by $n$;  $A_m$, $D_m$, $E_8$ denote the root lattices of the corresponding Lie algebras and $A_m^*,D_m^*$ denote their duals (the weight lattices); $\Gamma^{n,m}$ is an indefinite even unimodular lattice of signature $n,m$. For some entries, we simply provide the quadratic form of the lattice $\Gamma_{K3}^g$. For certain Frame shapes, there are several isomorphism classes of lattices $\Gamma_{K3}^g$; we list all of them. This phenomenon is related to the fact that the class of $g$ in the Conway group does not always determine the $O(\Gamma_{K3})$ class uniquely -- in fact, even when the lattice $\Gamma_{K3}^g$ is unique, there might be different $O(\Gamma_{K3})$ classes with the same Frame shape, corresponding to the inequivalent ways of `gluing' $\Gamma_{K3}^g$ and its orthogonal complement $\Gamma_{K3,g}$ to obtain $\Gamma_{K3}\cong \Gamma^{4,20}$. Note however that, by the results of section \ref{s:CHLclass}, the full lattice of electric charges \eqref{electLatt} (and in particular $\Gamma^{1,1}\oplus \Gamma_{K3}^g$) only depends on the Frame shape of $g$, up to isomorphisms.

Fricke S-duality implies that the lattices $\Gamma_{K3}^g$ in table \ref{t:case1} must be $N\lambda$-modular $(\Gamma_{K3}^g)^*(N\lambda)\cong \Gamma_{K3}^g$. The entries of table \ref{t:case2} are grouped into S-dual pairs, whose corresponding lattices satisfy $(\Gamma_{K3}^g)^*(N)\cong \Gamma_{K3}^{g'}$. In most cases, these isomorphisms can be verified easily by noticing that $A_2$ is $3$-modular and $D_4$ is $2$-modular. The only non-trivial case, which is described in section \ref{s:emlattices}, corresponds to the Frame shape $1^82^8$ in table \ref{t:case1}.

The lattices $\Gamma_{K3}^g$ are computed as follows. The first $8$ entries of table \ref{t:case1} are taken from \cite{GarbagnatiSarti2009}, with the exception of $1^42^24^4$, where we provide a simpler (but equivalent) description. In all the other cases, our starting point is the sublattice $\Lambda^g$ of the Leech lattice fixed by the corresponding class in $Co_0$. All such lattices are described in \cite{HaradaLang1990}. From the construction in \cite{Gaberdiel:2011fg}, it follows that the orthogonal complement $\Lambda_g$ of $\Lambda^g$ in the Leech lattice  is isomorphic (up to a sign flip in the quadratic form) to the orthogonal complement  of $\Gamma_{K3}^g$ in $\Gamma_{K3}\cong \Gamma^{4,20}$. This fact can be used to determine the discriminant quadratic form of $\Gamma_{K3}^g$, and hence its genus, starting from $\Lambda^g$ (see, for example, \cite{Nikulin} for more details). Then, for a given $g$, the possible isomorphism classes of lattices $\Gamma_{K3}^g$ are exactly the classes in this genus. When $\Gamma_{K3}^g$ has signature $(4,0)$, the corresponding classes are listed in \cite{Nipp1991}. When $\Gamma_{K3}^g$ is indefinite, one can show, using Theorem 21 and Corollary 22 in Chapter 15 of \cite{ConwaySloane}, that there is only one class in the genus and we simply construct a representative for this class.

\newpage

\begin{table}[!h]\begin{center}
\newcolumntype{C}{>{$}c<{$}}
\begin{tabular}{CCC}
N|\lambda & g  &  \Gamma_{K3}^g\cong  (\Gamma_{K3}^g)^*(N\lambda)\\
\hline
1|1& 1^{24} &  \Gamma^{4,20} \\
2|1& 1^82^8 &  \Gamma^{4,4}\oplus E_8(-2) \\
3|1 & 1^63^6 & \Gamma^{2,2}\oplus \Gamma^{2,2}(3)\oplus (A_2(-1))^{\oplus 2} \\
4|1 & 1^42^24^4 &  \Gamma^{2,2}\oplus\Gamma^{2,2}(4)\oplus \ZZ(-2)^{\oplus 2} \\
5|1 & 1^45^4 &  \Gamma^{2,2}\oplus \Gamma^{2,2}(5) \\
6|1 & 1^22^23^26^2 & \Gamma^{2,2}\oplus \Gamma^{2,2}(6) \\
7|1 & 1^37^3 &  \Gamma^{1,1}\oplus \Gamma^{1,1}(7)\oplus \left[\begin{smallmatrix}
4 & 1\\ 1& 2
\end{smallmatrix}\right] \\
8|1 & 1^22^14^18^2 &  \Gamma^{1,1}\oplus \Gamma^{1,1}(8)\oplus \left[\begin{smallmatrix}
2 & 0\\ 0& 4
\end{smallmatrix}\right] \\
11|1 & 1^211^2 &  \left[\begin{smallmatrix}
4 & 2 & 1 & 1 \\
 2 & 4 & 0 & 1 \\
 1 & 0 & 4 & 2 \\
 1 & 1 & 2 & 4
\end{smallmatrix}\right],\left[\begin{smallmatrix}
2 & 0 & 1 & 0 \\
 0 & 2 & 0 & 1 \\
 1 & 0 & 6 & 0 \\
 0 & 1 & 0 & 6
\end{smallmatrix}\right],\left[\begin{smallmatrix}
2 & 1 & 1 & 1 \\
 1 & 2 & 0 & 1 \\
 1 & 0 & 8 & 4 \\
 1 & 1 & 4 & 8
\end{smallmatrix}\right]   \\[10pt]
14|1 & 1^12^17^114^1 &  \left[\begin{smallmatrix} 4 & 1 & 1 & 0 \\
 1 & 4 & 0 & 1 \\
 1 & 0 & 4 & -1 \\
 0 & 1 & -1 & 4\end{smallmatrix}\right],\left[\begin{smallmatrix}
2 & 0 & 1 & 1 \\
 0 & 2 & 1 & 1 \\
 1 & 1 & 8 & 1 \\
 1 & 1 & 1 & 8
\end{smallmatrix}\right],\left[\begin{smallmatrix}
2 & 1 & 0 & 0 \\
 1 & 4 & 0 & 0 \\
 0 & 0 & 4 & 2 \\
 0 & 0 & 2 & 8
\end{smallmatrix}\right]   \\[10pt]
15|1 & 1^13^15^115^1 &  
 \left[\begin{smallmatrix} 4 & 2 & 1 & 1 \\
 2 & 4 & -1 & 2 \\
 1 & -1 & 6 & 2 \\
 1 & 2 & 2 & 6 \end{smallmatrix}\right],
 \left[\begin{smallmatrix}
2 & 1 & 0 & 0 \\
 1 & 2 & 0 & 0 \\
 0 & 0 & 10 & 5 \\
 0 & 0 & 5 & 10
\end{smallmatrix}\right],
 \left[\begin{smallmatrix}
2 & 0 & 0 & 1 \\
 0 & 4 & 1 & 0 \\
 0 & 1 & 4 & 0 \\
 1 & 0 & 0 & 8
\end{smallmatrix}\right]   \\[10pt]
2|2 & 2^{12} & \ZZ(2)^4\oplus \ZZ(-2)^{\oplus 8}  \\
3|3 & 3^{8} &  \Gamma^{4,4}(3)  \\
4|4 & 4^6 &  \ZZ(4)^{\oplus 4}\oplus \ZZ(-4)^{\oplus 2}  \\
4|2 & 2^{4}4^4 &  D_4(2)\oplus D_4(-2)  \\
6|6 & 6^{4} &  \ZZ(6)^{\oplus 4}  \\
10|2 & 2^210^2 &  \left[\begin{smallmatrix}
6 & 4\\ 4 & 6
\end{smallmatrix}\right]^{\oplus 2},\ \ZZ(2)^{\oplus 2}\oplus \ZZ(10)^{\oplus 2}   \\
12|2 & 2^14^16^112^1  & \left[\begin{smallmatrix}
4 & 2 & 0 & 0\\ 2 & 4 & 0 & 0\\ 0 & 0 & 8 & 4\\ 0& 0& 4& 8
\end{smallmatrix}\right]  \\[10pt]
\hline && \\[-10pt]
6|2 & 2^36^3 & A_2(2)^{\oplus 2}\oplus A_2(-2) \\ 
8|4 & 4^28^2  & D_4(4) \\
\hline &&\\[-10pt]
4|1 & 1^82^{-8}4^8 &  \Gamma^{4,4}(2) \\
8|1 & 1^42^{-2}4^{-2}8^4 &  D_4(2) \\
8|2 & 2^44^{-4}8^4 &  \ZZ(4)^{\oplus 4} \\
9|1 & 1^33^{-2}9^3 &  A_2\oplus A_2(3)\\
12|1 & 1^{2}2^{-2}3^24^{2}6^{-2}12^2 &  A_2(2)\oplus A_2(2) \\
\end{tabular}\caption{Symmetries of K3 models whose Frame shape $\prod_{a|N}a^{m(a)}$ is balanced, i.e. $m(N\lambda/a)=m(a)$. We report the order $N$ of $g$, $\lambda$, and the fixed sublattice $\Gamma_{K3}^g\subset \Gamma_{K3}$. The corresponding CHL models are self-dual and the lattices $\Gamma_{K3}^g$ are $N\lambda$-modular. The first $18$ entries are the classes relevant for Mathieu moonshine; the following two classes $2^36^3$ and $4^28^2$ are relevant for other Umbral groups \cite{Cheng:2014zpa}.}\label{t:case1}\end{center}\end{table}

\newpage

\begin{table}[!h]\begin{center}
\newcolumntype{C}{>{$}c<{$}}
\begin{tabular}{CCC}
N|\lambda & g  &  \Gamma_{K3}^g \\
\hline
%4|1 & 1^82^{-8}4^8 & \Lambda_{16}(-1) & \Gamma^{4,4}(2) \\
6|1 & 1^{4}2^13^{-4}6^5 &   \Gamma^{2,2}(2)\oplus A_2(2) \\
6|1 & 1^{5}2^{-4}3^16^4 &  \Gamma^{2,2}(3)\oplus A_2 \\
%8|1 & 1^42^{-2}4^{-2}8^4 & ?& D_4(2)? \\
%8|2 & 2^44^{-4}8^4 & (\ZZ(-4)^4)^\perp & \ZZ(4)^4 \\
%9|1 & 1^33^{-3}9^3 & ?& A_2\oplus A_2(3)\\
\hline
10|1 & 1^{2}2^15^{-2}10^3  & A_4(2) \\
10|1 & 1^{3}2^{-2}5^110^2 &  A_4^*(5) \\
%12|1 & 1^{2}2^{-2}3^24^{2}6^{-2}12^2 & ? & A_2(2)\oplus A_2(2) \\
\hline
12|1 & 1^{1}2^23^14^{-2}12^2 &  A_2\oplus \ZZ(6)^{\oplus 2} \\
12|1 & 1^{2}3^{-2}4^16^212^1 &  A_2(4)\oplus \ZZ(2)^{\oplus 2} \\
\end{tabular}
\caption{Symmetries of K3 model whose Frame shape is not balanced and such that the orbifold is a K3 model. These symmetries come in pairs $g,g'$ with Frame shapes $\prod_{a|N}a^{m(a)}$ and $\prod_{a|N}a^{m(N\lambda/a)}$. The corresponding CHL models are dual to 
each other and the lattices $\Gamma^g_{K3}$ and $\Gamma^{g'}_{K3}$ are related by $(\Gamma^{g'}_{K3})^*(N\lambda)\cong\Gamma^g_{K3}$.}\label{t:case2}
\end{center}\end{table}
\begin{table}[!h]\begin{center}
\newcolumntype{C}{>{$}c<{$}}
\begin{tabular}{CCCC}
N|\lambda & g  &  \Gamma_{K3}^g & (\Gamma_{K3}^g)^*(N\lambda)\\
\hline
2|1 & 1^{-8}2^{16} &  \Gamma^{4,4}(2) & \Gamma^{4,4} \\
3|1 & 1^{-3}3^{9} &  \Gamma^{2,2}(3)\oplus A_2 & \Gamma^{2,2}\oplus A_2 \\
4|1 & 1^{-4}2^64^4 & \Gamma^{2,2}(4)\oplus\ZZ(2)^{\oplus 2} & \Gamma^{2,2}\oplus \ZZ(2)^{\oplus 2} \\
4|1 & 2^{-4}4^8 & D_4(2) & D_4 \\
5|1 & 1^{-1}5^5 & A_4^*(5) & A_4 \\
6|1 & 1^{-4}2^53^46^1 & \Gamma^{2,2}(6)\oplus A_2(2) & \Gamma^{2,2}\oplus A_2 \\
6|1 & 1^{-2}2^43^{-2}6^4 & A_2(2)^{\oplus 2} & A_2^{\oplus 2} \\
6|1 & 1^{-1}2^{-1}3^36^3 & D_4(3) & D_4 \\
8|1 & 1^{-2}2^34^18^2 & \ZZ(4)\oplus A_3^*(8) & \ZZ(2)\oplus A_3 \\
10|1 & 1^{-2}2^35^210^1 & A_4^*(10) & A_4 \\
12|1 & 1^{-2}2^23^24^112^1 & \ZZ(6)^{\oplus 2}\oplus A_2(4) & \ZZ(2)^{\oplus 2}\oplus A_2 \\
\end{tabular}\caption{The symmetries of K3 models whose orbifold is a torus. They are characterized by frame shape $\prod_{a|N}a^{m(a)}$ with $\sum_{a|N}am(N/a)=0$. The corresponding CHL models are dual to type IIA models on $(T^4\times S')/\ZZ_N\times \tilde S'$.}\label{t:case3}\end{center}\end{table}

%\subsubsection{Discriminant groups}
%
%{\bf $\ZZ_3^n$}
%When the discriminant group of an even lattice $L$ has the form $L^*/L\cong \ZZ_3^n$, the discriminant form can be written as \cite{Nikulin}
%\be\label{pure3qform} q_{a,b}:=(\frac{2}{3})^{\oplus a}\oplus (\frac{4}{3})^{\oplus b}\equiv (q_1(3))^{\oplus a}\oplus (q_{_1}(3))^{\oplus b}\ ,
%\ee with $a+b=n$. Thanks to the isomorphisms $q_{a,b}\cong q_{a+2,b-2}$, there are only $2$ inequivalent choices for $q_{a,b}$ up to isomorphisms, namely $a$ can be even or odd. These choices can be distinguished by the signature mod $8$ of the underlying lattice. For example, $L=(A_2)^{\oplus a}\oplus (A_2(-1))^{\oplus b}$ has quadratic form \eqref{pure3qform} and signature
%\be {\rm sign}(q)={\rm sign}L=2a-2b\mod 8\ .
%\ee Therefore, when $n\equiv a+b$ is even, the signature of $L$ (mod $8$) is $0$ for $a$ even and $4$ for $a$ odd, while when $n$ is odd, the signature is $6$ for $a$ even and $2$ for $a$ odd.
%
%\bigskip
%
%{\bf $4^6$}. The discriminant form $q_1(4)^4\oplus q_7(4)^2$ of $\ZZ(4)^4\oplus \ZZ(-4)^2$ is the inverse of the discriminant form $q_1(4)^6$ of $\ZZ(4)^6$.
%To prove this, we use the following two identities (\cite{Nikulin}, Proposition 1.8.2)
%\be q_\theta(4)\oplus q_{\theta'}(4)=q_{5\theta}(4)\oplus q_{5\theta'}(4)
%\ee and
%\be q_\theta(4)^2\oplus q_{\theta'}(4)=v(4)\oplus  q_{-5\theta'}(4)\qquad \text{if }\theta\equiv \theta'\mod 4\ .
%\ee 
%Thus,
%\be q_1(4)^4\oplus q_7(4)^2=v(4)\oplus q_3(4)\oplus q_1(4)\oplus q_7(4)^2=v(4)^2\oplus q_1(4)^2
%\ee 
%\be q_7(4)^6=v(4)^2\oplus q_5(4)^2=v(4)^2\oplus q_1(4)^2
%\ee
%

\bibliographystyle{JHEP}

\bibliography{Refs}

\end{document}